\documentclass[10pt]{article}
\usepackage{mathtools}
\usepackage{amsmath}
\usepackage{bm}
\usepackage{enumerate}
\usepackage{hyperref}
\usepackage{subfiles}
\usepackage{subfig}
\usepackage{epsfig}
\usepackage{epstopdf}
\usepackage{psfrag}
\usepackage{graphicx}
\usepackage{enumitem}
\usepackage{parskip}
\usepackage{xkeyval}
\usepackage[noabbrev]{cleveref}
\usepackage{url}
\usepackage[dvipsnames]{xcolor}
\usepackage[per-mode=fraction]{siunitx}
\usepackage{multirow}
\usepackage{placeins} % for float barrier
\usepackage[bbgreekl]{mathbbol}
\usepackage{pdflscape}
\usepackage{tikz} % for L-shape grid
\usepackage{boldline} % make table outer border thick
\usepackage{cleveref}
\usepackage[ruled]{algorithm2e}
\usepackage{verbatim} % for comment blocks

\usepackage[sortcites,backend=biber,bibencoding=utf8,maxnames=100,doi=false,isbn=false,url=false,eprint=false,giveninits=true,date=year,sorting=none]{biblatex}
\newif\ifbembo
\newif\ifcharter
\newif\iferewhon
\newif\iflibertine
\newif\iflibertinealt
\newif\ifpalantino
\newif\iftimesnewroman

\bembofalse
\chartertrue
\erewhonfalse
\libertinefalse
\libertinealtfalse
\palantinofalse
\timesnewromanfalse
\ifbembo
  \usepackage[p,osf]{ETbb} % osf in text, tabular lining figures in math
  \usepackage[scaled=.95,type1]{cabin} % sans serif in style of Gill Sans
  \usepackage[varqu,varl]{zi4}% inconsolata typewriter
  \usepackage[T1]{fontenc}
  \usepackage[libertine,vvarbb]{newtxmath}
  \usepackage[cal=boondoxo,bb=boondox]{mathalfa}
\fi

\ifcharter
  \usepackage[scaled=.98,p]{XCharter}
  \usepackage[scaled=1.04,varqu,varl]{inconsolata}
  \usepackage[type1]{cabin}
  \usepackage[uprightscript,xcharter,vvarbb,scaled=1.05]{newtxmath}
  \usepackage[cal=euler,scr=boondoxo,bb=boondox]{mathalpha}  % prefer bb=bboldxLight when available
\fi

\iferewhon
  \usepackage[p,osf,scaled=.98,space]{erewhon} % scaling by .98 not really necessary
  \usepackage[varqu,varl]{inconsolata} % typewriter
  \usepackage[type1,scaled=.95]{cabin} % sans serif like Gill Sans
  \usepackage[utopia,vvarbb]{newtxmath}
\fi

\iflibertine
  \usepackage[sb]{libertinus}
  \usepackage[T1]{fontenc}
  \usepackage{textcomp}
  \usepackage[varqu,varl]{zi4}% inconsolata for mono, not LibertineMono
  \usepackage{libertinust1math} % slanted integrals, by default
  \usepackage[scr=boondoxo,bb=boondox]{mathalpha} %Omit bb=boondox for default libertinus bb
\fi

\iflibertinealt
  \usepackage[sb]{libertinus}
  \usepackage[T1]{fontenc}
  \usepackage{textcomp}
  \usepackage{libertinust1math} % slanted integrals, by default
  \usepackage[scr=boondoxo,bb=boondox]{mathalpha} %Omit bb=boondox for default libertinus bb
\fi

\ifpalantino
  \usepackage[largesc]{newpxtext}
  \usepackage[vvarbb]{newpxmath}
\fi

\iftimesnewroman
  \usepackage{newtxtext} %T1 is default encoding
  \usepackage[scaled=0.95]{inconsolata} % typewriter
  \usepackage[vvarbb]{newtxmath} % vvarbb gives STIX Bbb
\fi

\usepackage[T1]{fontenc}
\usepackage[final,protrusion=true,expansion=true]{microtype}

\usepackage{titling}
\usepackage{authblk} % note: titling needs to be loaded before authblk
\pretitle{\begin{center}\bfseries\sffamily\LARGE}
\posttitle{\end{center}}
\usepackage{abstract}
\usepackage{sectsty} % note: the main alternative package, titlesec, does not work with titleref
\allsectionsfont{\sffamily}

\usepackage[left=1in,right=1in,top=1in,bottom=1in,includefoot,heightrounded]{geometry}

\makeatletter
\patchcmd{\LS@rot}{90}{-90}{}{}
\patchcmd{\endlandscape}{90}{-90}{}{}
\makeatother

\newcommand\Wi{\text{Wi}}

\newcommand{\uu}{\mathbf{u}}
\newcommand{\vun}{\mathbf{u}_\text{n}}
\newcommand{\vus}{\mathbf{u}_\text{s}}

\newcommand{\ff}{\mathbf{f}}
\newcommand{\ffn}{\ff_\text{n}}
\newcommand{\ffs}{\ff_\text{s}}

\newcommand{\viscStress}{\bbsigma}
\newcommand{\sign}{\viscStress_\text{n}}
\newcommand{\sigs}{\viscStress_\text{s}}
\newcommand{\btaun}{\bbtau_\text{n}}

\newcommand{\CC}{\bbsigma_\text{ve}}

\newcommand{\II}{\mathbb{I}}

\newcommand{\mun}{\mu_\text{n}^\text{v}}
\newcommand{\muns}{\mu_\text{n}^\text{v}}
\newcommand{\munp}{\mu_\text{n}^\text{p}}
\newcommand{\mus}{\mu_\text{s}}

\newcommand{\CFL}{C_\text{CFL}}

\newcommand{\thn}{\theta_\text{n}}
\newcommand{\ths}{\theta_\text{s}}

\newcommand{\xx}{\mathbf{x}}

\newcommand{\parens}[1]{\mathopen{}\left(#1\right)\mathclose{}}
\newcommand{\grad}{\nabla}
\newcommand{\Div}{\grad\cdot}

\newcommand{\size}[1]{\lVert#1\rVert}

\newcommand{\Gctof}{G^{\text{c}\rightarrow \text{f}}_h}
\newcommand{\Dftoc}{D^{\text{f}\rightarrow \text{c}}_h}
\newcommand{\Lapftof}{L^{\text{f}\rightarrow \text{f}}_h}
\newcommand{\interpctof}{\mathcal{J}^{\text{c}\rightarrow \text{f}}}
\newcommand{\NN}{\mathbf{N}}

\makeatletter
\def\gsh#1{%
  \vbox{\hbox{%
    \let\\\cr
    \offinterlineskip
    \valign{&\hb@xt@2\p@{\hss$##$\hss}\vskip.2ex\cr#1\crcr}%
  }\vskip-.36ex}%
}
\def\gshsym{\@ifstar\gsh@ssym\gsh@sym}
\def\gsh@sym#1#2{\mathrlap{\overset{#1}{\phantom{#2}}}#2}
\def\gsh@ssym#1#2{\overset{#1}{#2}{\vphantom{#2}}}
\makeatother
\newcommand{\tran}{^{\mkern-1.5mu\mathsf{T}}}
\graphicspath{{Graphs/}}  % Location of the graphics files

\newcommand{\todo}[1]{\textcolor{blue}{TODO: #1}}

\title{Adaptive Mesh Refinement for Two-Phase Viscoelastic Fluid Mixture Models}
\author[1,*]{Bindi M. Nagda}
\author[2,*,**]{Aaron Barrett}
\author[3,4]{Boyce E. Griffith}
\author[5]{Aaron L. Fogelson}
\author[1]{Jian Du}
\affil[1]{Department of Mathematical Sciences, Florida Institute of Technology, Melbourne, FL, USA}
\affil[2]{Department of Mathematics, University of Utah, Salt Lake City, UT, USA}
\affil[3]{Departments of Mathematics, Applied Physical Sciences, and Biomedical Engineering, University of North Carolina, Chapel Hill, NC, USA}
\affil[4]{Carolina Center for Interdisciplinary Applied Mathematics, University of North Carolina, Chapel Hill, NC, USA}
\affil[5]{Departments of Mathematics and Biomedical Engineering, University of Utah, Salt Lake City, UT, USA}
\affil[*]{These authors contributed equally to this work}
\affil[**]{To whom correspondence should be addressed: \texttt{barrett@math.utah.edu}}
\begin{document}
\maketitle

\begin{abstract}
    Multiphase flows are an important class of fluid flow and their study facilitates the development of diverse applications in industrial, natural, and biomedical systems. We consider a model that uses a continuum description of both phases in which separate momentum equations are used for each phase along with a co-incompressibility condition on the velocity fields. The resulting system of equations poses numerical challenges due to the presence of multiple non-linear terms and the co-incompressibility condition, and the resulting fluid dynamics motivate the development of an adaptive mesh refinement (AMR) technique to accurately capture regions of high stresses and large material gradients while keeping computational costs low. We present an accurate, robust, and efficient computational method for simulating multiphase mixtures on adaptive grids, and utilize a multigrid solver to precondition the saddle-point system. We demonstrate that the AMR discretization asymptotically approaches second order accuracy in $L^1$, $L^2$ and $L^\infty$ norms. The solver can accurately resolve sharp gradients in the solution and, with the multigrid preconditioning strategy introduced herein, the linear solver iterations are independent of grid spacing. Our AMR solver offers a major cost savings benefit, providing up to ten fold speedup over a uniform grid in the numerical experiments presented here, with greater speedup possible depending on the problem set-up.
    
\end{abstract}

\section{Introduction}
Multiphase flows are the simultaneous coupled flow of materials with two or more thermodynamic phases. Many practical multiphase flows have separate, non-overlapping phases that may be either miscible or immiscible, yet those length scales may be too small to efficiently resolve. Perhaps the most common example is the fluidized bed, a gas-solid system in which pressurized fluid is pumped into particles resulting in a medium that has fluid-like properties. Fluidized beds have the ability to promote high levels of contact between gases and solids, resulting in numerous applications in food processing, thermal power generation, and metallurgy \cite{philippsen2015}. Other common examples include dust storms in the atmospheric boundary layer \cite{zhang2023}, sediment transport within rivers \cite{ouda2019}, groundwater transport through porous rock \cite{moench1984}, platelet thrombus formation \cite{du2018}, cellular protrusion driven by cytoskeletal-membrane interactions \cite{mofrad2006}, DNA packaging \cite{tongu2016}, and gastric mucus secretion and storage \cite{du2023, nagda2023}. For instance, blood clots can be considered as a multiphase fluid, in which the viscoelastic thrombus phase must be porous to the viscous fluid phase in order to observe realistic platelet densities \cite{du2018,du2020}. These two phases can be resolved as non-overlapping regions at the scale of micrometers, but such descriptions become impractical when considering arterial flows for example. To obtain a tractable model, we do not explicitly track the interface between the phases. Instead, we use a homogenized material model, in which both phases are present at varying volume fractions at every point in the domain. This mixture model approach is based on separately describing each material as a continuum, with each occupying the same region in space and moving according to its own velocity field \cite{ishii2011, drew1983}. The interacting materials are called "phases" even though they are not different phases of the same material, and the resulting multiphase material is called a mixture.

Following continuum mechanics, the stresses of each phase need to be specified, along with relations for the coupling between the phases. A group of models has been proposed on the assumption that a local equilibrium is established over short spatial length scales. Depending on the choice of constitutive equations that control the degree of coupling between the phases, this model is called the two-phase mixture model \cite{ishii2011}, the algebraic-slip model \cite{pericleous1986}, the drift-flux model \cite{zuber1965}, the suspension model \cite{manninen1996}, the diffusion model \cite{ungarish1993, ishii2011}, or the local-equilibrium model \cite{johansen1990}. The two-phase mixture model used in this study follows the mathematical model derived by Ishii \cite{ishii2011}, who provided a detailed macroscopic formulation for two-phase systems. As stated at the outset, these models do not track an explicit interface between the materials. Instead, materials coexist throughout the domain in varying volume fractions. The model consists of a continuity equation and momentum equations for each phase. Each phase is described by a constitutive equation for the Cauchy stress along with an interphase drag term arising from the differences in velocity between the two phases. As the two materials can displace each other, conservation of mass is given by the co-incompressibility condition, where both velocity fields can be compressible, but the volume-averaged velocity field is incompressible. 

For multiphase flows that incorporate viscoelasticity, such as those used in thrombus formation \cite{du2018}, extensional points in the flow can lead to sharp gradients in the velocity and stress. In single phase flows, extensional points can lead to unbounded growth in the stress for some fluid models \cite{thomases2007}. While it is unclear the degree to which singular stresses arise in the multiphase models used herein, numerical difficulties in single phase models are often associated with insufficient spatial resolution to resolve the exponential profile of the stress \cite{hulsen2005}. Recent blood clotting models \cite{du2018,du2020} use multiphase models to capture both the elasticity and the porosity of the thrombus. These models require spatial resolutions that vary in time as the clot increases in size and potentially embolizes, breaking into several pieces. Further, in arterial simulations, for instance, it is necessary to use large domain sizes to ensure boundary effects do not affect the solution; however, coarse grids can effectively resolve the inlets and outlets. A similar need for high-resolution grids arises in the study of gastric mucus, particularly when modeling the flow dynamics observed during the secretion of mucus from goblet cells in the stomach lining. The process of mucus polymer swelling or collapsing in response to external stimuli occurs very rapidly and often yields interesting pattern formations. For example, when a blob of hydrated mucus polymers undergoes deswelling, the network may form a shrinking polymer annulus under certain conditions \cite{celora2023, nagda2023}, whereby there will be a high concentration of solvent on either side of the ring-like structure. A high resolution grid is therefore required to resolve the sharp change in network distribution and phase velocities across the narrow width of the polymer ring. There also is interest in developing high fidelity models of the stomach and the mucus layer lining the stomach epithelium. These models will aid in identifying the mechanisms through which the mucus layer offers a protective barrier against high stomach acidity, digestive enzymes, bacterial infections, and gastric cancer \cite{aggarwal2021, serrano2021, su2018}. Because interesting dynamics are localized over the gastric mucus layer, which is only a couple hundred microns wide, this will necessarily require high resolution over the layer to capture the sharp transitions in velocities, fluid stress, and compositions. Sufficient spatial resolution to fully resolve these flow features can lead to unreasonable computational costs when using a uniform mesh. This drawback motivates the need for an adaptive mesh refinement technique to simulate multiphase flows.  

Structured adaptive mesh refinement (AMR) was first developed by Berger and Oliger \cite{berger1984} and Berger and Colella \cite{berger1989} for simulating shock hydrodynamics. Their method utilized block-structured adaptive mesh refinement (SAMR) \cite{dubey2014}, in which the computational mesh consists of a hierarchy of levels of spatial refinement. Each level consists of several logically rectangular regions, which can be tagged for refinement across a simulation. This allows for easy grid generation and the ability to reuse uniform grid discretizations. In this way, the only communication required across levels is the filling of ``ghost cells.'' Alternatives to SAMR methods include tree-based methods, in which the generated grid is described by an octree in three spatial dimensions \cite{burstedde2011}. Completely unstructured methods \cite{lawlor2006} allow for greater flexibility in grid generation at the cost of increased complexity in discretizations. 

To our knowledge, this is the first work to develop and implement a method for simulating multiphase mixtures on adaptively refined meshes for both Newtonian and non-Newtonian fluid models. We demonstrate that the AMR multiphase discretization yields significant savings in computational cost compared to a uniformly fine discretization, while our locally refined spatial discretization retains second order accuracy of all variables. The co-incompressibility condition gives the system an additional difficulty over a single phase system. In single phase fluids, it is common to use projection-based methods \cite{bell1989, kim1985, vankan1986}, in which the incompressibility constraint is enforced by solving an additional Poisson equation. For methods that solve the full saddle point problem, the resulting linear solver can use preconditioners based on projection methods \cite{griffith2009a} or other block factorization approaches \cite{elman2008a}. For the co-incompressible condition used herein, it remains unclear how to project the velocity fields of individual phases that give fields consistent with both momentum equations. Therefore, in this work, we adapt and test a geometric multigrid based preconditioner to solve the full saddle point system on an adaptive grid. We demonstrate that the adaptive discretization can accurately resolve sharp gradients in the solution and that, with the multigrid preconditioner, the linear solver iterations are independent of grid spacing.

\section{Model Equations} \label{section 2}
In this work, we describe the multiphase material as a mixture of two fluids, the viscous solvent (s) and a viscous or viscoelastic network (n). We describe the model and numerical implementation in terms of the viscoelastic network and we detail the changes required to recover a viscous network. Full details about the derivation of the model can be found in prior works \cite{keener2011,ishii2011, wright2008, barrett2019}. 
For the two phases, the variables are the volume fractions of network $\thn\parens{\xx,t}$ and solvent $\ths\parens{\xx,t}$, the velocities of the two phases, $\vun\parens{\xx,t}$ and $\vus\parens{\xx,t}$, and the pressure $p\parens{\xx,t}$.
The dynamics of the two volume fractions are governed by the advection equations     
\begin{align} 
    \frac{\partial}{\partial t}\thn\parens{\xx,t} +\Div (\thn\parens{\xx,t} \vun\parens{\xx,t}) &= 0,
    \label{eq:net_vol_frac} \\
    \frac{\partial}{\partial t}\ths\parens{\xx,t} +\Div (\ths\parens{\xx,t} \vus\parens{\xx,t}) &= 0. 
    \label{eq:sol_vol_frac}
\end{align}
These equations, along with the identity $\thn\parens{\xx,t} + \ths\parens{\xx,t} = 1$,  yield the volume-averaged incompressibility condition,
\begin{equation} 
    \Div \parens{\thn\parens{\xx,t} \vun\parens{\xx,t} + \ths\parens{\xx,t} \vus\parens{\xx,t}} = 0. 
    \label{eq:incompressibility_a}
\end{equation}
The conservation of momentum of the two fluids yields
\begin{align}
    \rho\parens{\frac{\partial \thn\parens{\xx,t}\vun\parens{\xx,t}}{\partial t} + \grad\cdot\parens{\thn\parens{\xx,t}\vun\parens{\xx,t}\vun\parens{\xx,t}}} =&  - \thn\parens{\xx,t} \grad p\parens{\xx,t} + \Div \parens{\thn\parens{\xx,t} \sign\parens{\xx,t}} \label{eq:forcebalance_net}\\
    &- \xi
    \thn\parens{\xx,t} \ths\parens{\xx,t} \parens{ \vun\parens{\xx,t} - \vus\parens{\xx,t}} + \thn\parens{\xx,t}\ffn\parens{\xx,t}, \nonumber\\
    \rho\parens{\frac{\partial \ths\parens{\xx,t}\vus\parens{\xx,t}}{\partial t} + \grad\cdot\parens{\ths\parens{\xx,t}\vus\parens{\xx,t}\vus\parens{\xx,t}}} =& - \ths\parens{\xx,t} \grad p\parens{\xx,t} + \Div \parens{\ths\parens{\xx,t} \sigs\parens{\xx,t}} \label{eq:forcebalance_sol}\\
    &- \xi
    \thn\parens{\xx,t} \ths\parens{\xx,t} \parens{ \vus\parens{\xx,t} - \vun\parens{\xx,t}}  + \ths\parens{\xx,t}\ffs\parens{\xx,t}, \nonumber
\end{align}
in which $\rho$ is the density of each of the two phases, assumed to be equal, $\xi$ is the drag coefficient for the relative motion between the solvent and network, $\ffn\parens{\xx,t}$ and $\ffs\parens{\xx,t}$ are body forces applied to the fluids, and $\sign\parens{\xx,t}$ and $\sigs\parens{\xx,t}$ are the network and solvent stress tensors. We assume that the inertial effects are small enough so that we can neglect the convective term $\Div\parens{\theta_i\parens{\xx,t}\mathbf{u}_i\parens{\xx,t}\mathbf{u}_i\parens{\xx,t}}$ for $i=\text{n},\text{s},$ that appears in \cref{eq:forcebalance_net,eq:forcebalance_sol}. If treated explicitly, including the convective term in our numerical discretization would not effect the efficacy of the linear solver, which is the primary purpose of this study.
The viscous solvent stress tensor is
\begin{equation}\label{eq:solvent_stress}
  \sigs\parens{\xx,t} = \mus\parens{\grad\vus\parens{\xx,t} + \grad\vus\parens{\xx,t}^{\tran}} - \mus\Div\vus\parens{\xx,t}\II,
\end{equation}
in which $\mus$ is the solvent viscosity. We assume the network is a viscoelastic material based on clotting models previously derived \cite{guy2008,du2018}. In this case, the total stress consists of a viscous and viscoelastic component
\begin{align}\label{eq:network_stress}
    \sign\parens{\xx,t} &= \muns\parens{\grad\vun\parens{\xx,t} + \grad\vun\parens{\xx,t}^{\tran}} - \muns\Div\vun\parens{\xx,t}\II + \CC\parens{\xx,t}, \\
    \frac{\partial \CC\parens{\xx,t}}{\partial t} + \grad\cdot\parens{\vun\parens{\xx,t}\CC\parens{\xx,t}} &= \CC\parens{\xx,t}\cdot\grad\vun\parens{\xx,t}^{\tran} + \grad\vun\parens{\xx,t}\cdot\CC\parens{\xx,t} -\frac{1}{\lambda}\parens{\CC\parens{\xx,t} - z\parens{\xx,t}\II}, \label{eq:StressTensor}\\
    \frac{\partial z\parens{\xx,t}}{\partial t} + \grad\cdot\parens{\vun\parens{\xx,t} z\parens{\xx,t}} &= 0, \label{eq:ElasticModulus}
\end{align}
in which $\muns$ is the viscous contribution to the viscosity of the network phase, $\lambda$ is the relaxation time of the material, and $\CC\parens{\xx,t}$ is the viscoelastic stress tensor. Here, $z\parens{\xx,t}$ is a quantity that is proportional to the elastic modulus. We note that this model reduces to a viscous model if $z\parens{\xx,t} = 0$. This model is similar to the single-phase Oldroyd-B model \cite{larson1988}. Extensions to this model to incorporate shear-thinning by making the relaxation time a function of the local dynamics of the fluid are easy to incorporate. Instead, we will specialize to this simpler model for this work, which is focused on key discretization and linear solver details rather than the impacts of rheological modeling choices.

\section{Numerical Implementation}
We use a structured adaptive mesh refinement framework to discretize the governing equations. The equations are discretized on a nested hierarchy of grids consisting of locally refined rectangular patches that are organized into a sequence of levels that contain patches all with the same grid spacing. The computational domain consists of a rectangular box $\Omega$ such that the coarsest level is a uniform discretization of $\Omega$ using grid spacing $(\Delta x^0, \Delta y^0).$ The different patch levels are defined such that patch level $\ell$ uses grid spacing $(\Delta x^\ell = \frac{\Delta x^0}{r^\ell}, \Delta y^\ell = \frac{\Delta y^0}{r^\ell})$, in which $r^\ell$ is the refinement ratio between levels. For notational convenience, we describe the methodology for the case that $r^\ell = r$ is constant among all patch levels, although there is no such restriction in our implementation. We use a staggered grid, in which the velocities $\vun\parens{\xx,t}$ and $\vus\parens{\xx,t}$ are stored at cell faces, and the pressure $p\parens{\xx,t}$, volume fraction $\thn\parens{\xx,t}$, and stress tensor $\CC\parens{\xx,t}$ are stored at cell centers. A cell in patch level $\ell$ is denoted by $c_{i,j}^\ell$. We denote the center of cell $c_{i,j}^\ell$ as $\xx^\ell_{i,j}$ and the corresponding left cell face as $\xx^\ell_{i - \frac{1}{2}, j}$. We require that the space covered by patch level $\ell$, which we denote by $\Omega^\ell$, be properly nested within the next coarser level, $\Omega^\ell\subset\Omega^{\ell-1}$.

Computations using the governing multiphase equations are implemented using open-source libraries written in C++ and FORTRAN. The algorithms for our method are developed using the IBAMR open-source library \cite{ibamrREPO}, which is an adaptive and MPI parallelized version of the immersed boundary method. Structured grid components are handled using data structures provided by SAMRAI \cite{SAMRAI}, which offers a framework for block-structured adaptive mesh refinement. The resulting linear systems are solved using Krylov solvers provided by PETSc \cite{balay1998petsc}, with custom geometric multigrid preconditioners we created specifically for this study. 

\subsection{Coarse-Fine Interface Ghost Cells}\label{sec:cf_ghost_cells}
Each quantity defined on a grid patch is appended with a layer of ghost cells to allow for succinct definitions of difference stencils. Ghost cells can be of two different types: (1) a ghost cell $c_{i,j}^\ell$ can overlap a cell on a patch in the same patch level, so that $c_{i,j}^\ell\subset\Omega^\ell$, or (2) a ghost cell $c_{i,j}^\ell$ can overlap a cell on a patch in the coarser patch level, so that $c_{i,j}^\ell\subset\Omega^{\ell-1}$. In the first case, we can simply copy the values from the neighboring patch into the ghost cell $c_{i,j}^\ell$. In the second case, we employ a quadratic interpolation scheme that incorporates values from the coarse and fine patches as described by Griffith \cite{griffith2012a}. For completeness, we describe the method in detail below.

\begin{figure}[ht!]
    \begin{center}
    %\phantomsubcaption\label{fig:cf-interpolation:cell}
    %\phantomsubcaption\label{fig:cf-interpolation:side}
    %\includegraphics[width=0.45\textwidth]{Graphs/Numerics/cf_interp_cell.eps}
    %\includegraphics[width=0.45\textwidth]{Graphs/Numerics/cf_interp_side.eps}
    \subfloat[]{\epsfig{file=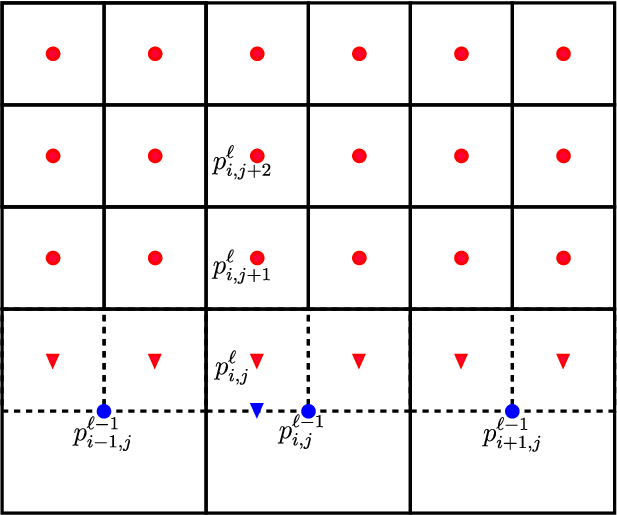, width=0.46\textwidth, height=0.38\textwidth}\label{fig:cf-interpolation:cell}}~~~
    \subfloat[]{\epsfig{file=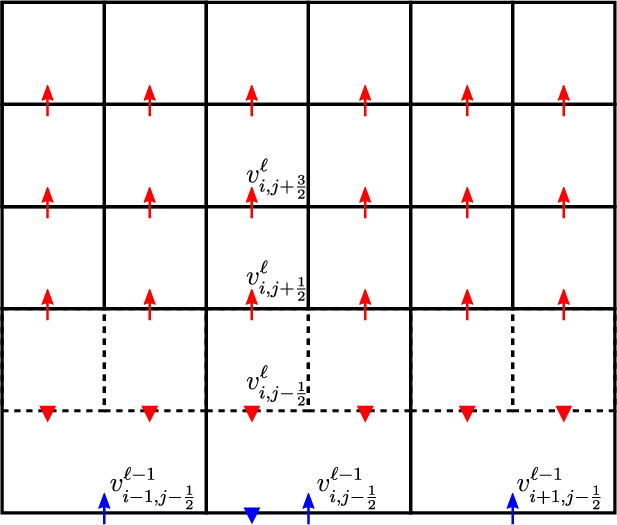, width=0.46\textwidth, height=0.38\textwidth}\label{fig:cf-interpolation:side}}
    \end{center}
    \caption{The degree of freedom and coarse-fine interface for cell and side centered grids with a refinement ration of $r = 2$. Ghost cells are depicted with dashed lines, and values in ghost cells are depicted by red triangles. To compute an approximation to a ghost cell, $p_{i,j}^\ell$ in panel (a) or $v_{i,j-\frac{1}{2}}^\ell$ in panel (b), we first compute a quadratic approximation to the value located at the blue triangle. In this figure the nodes depicted by the blue triangle corresponds to $r = 2$ and $\alpha = 1$. We then compute a quadratic approximation of the ghost cell value using values in the normal direction in the fine patch.}
    \label{fig:cf-interpolation}
\end{figure}

To fill a ghost cell $c_{i,j}^\ell$ such as the one depicted in \Cref{fig:cf-interpolation:cell}, for a cell centered quantity $p\parens{\xx,t}$, we first quadratically interpolate the coarse values to find the value in line with $c_{i,j}^\ell$ by
\begin{equation}
    p_{i-\frac{1}{2}+\frac{\alpha}{r+2},j}^{\ell-1} = \frac{3(r+2)^2 + 4\alpha(r+2)-4\alpha^2}{4(r+2)^2} p_{i,j}^{\ell-1} + \frac{3(r+2)^2-8\alpha(r+2)+4\alpha^2}{8(r+2)^2} p^{\ell-1}_{i-1,j} - \frac{(r+2)^2-4\alpha^2}{8(r+2)^2} p^{\ell-1}_{i+1,j},
\end{equation} 
in which $\alpha\in\{1,\ldots,r+1\}$ refers to the location of the fine cell inside the coarse cell; see \Cref{fig:cf-interpolation}. We then use that value along with $p_{i,j+1}^\ell$ and $p_{i,j+2}^\ell$ to obtain
\begin{equation}
    p_{i,j}^\ell = p_{i,j+1}^\ell \frac{2r-2}{r+1} + p_{i,j+2}^\ell \frac{1-r}{3+r} + p_{i-\frac{1}{2}+\frac{\alpha}{r+2},j}^{\ell-1}\frac{8}{3+4r+r^2}.
\end{equation}
For side centered quantities, the component tangential to the patch's coarse-fine boundary is computed using the same interpolation scheme as centered quantities. For the component normal to the patch's coarse-fine boundary, we perform a similar procedure. To compute an approximation to $v_{i,j-\frac{1}{2}}^\ell$ depicted in \Cref{fig:cf-interpolation:side}, we first quadratically interpolate the coarse values to find an approximation to $v_{i-\frac{1}{2}+\frac{\alpha}{r+2},j-\frac{1}{2}}^{\ell-1}$ by
\begin{equation}
    v_{i-\frac{1}{2}+\frac{\alpha}{r+2},j-\frac{1}{2}}^{\ell-1} = \frac{3(r+2)^2-8\alpha(r+2)+4\alpha^2}{8(r+2)^2}v_{i-1,j-\frac{1}{2}}^{\ell-1} + \frac{3(r+2)^2 + 4\alpha(r+2)-4\alpha^2}{4(r+2)^2}v_{i,j-\frac{1}{2}}^{\ell-1} - \frac{(r+2)^2-4\alpha^2}{8(r+2)^2}v_{i+1,j-\frac{1}{2}}^{\ell-1}.
\end{equation}
We then compute a quadratic approximation to $v_{i,j-\frac{1}{2}}^\ell$ by
%\begin{equation}
%    v_{i,j-\frac{1}{2}}^\ell = v_{i,j+\frac{1}{2}}^\ell - \frac{1}{3}v_{i,j+\frac{3}{2}}^\ell - \frac{1}{3}v_{i-\frac{1}{2}+\frac{\alpha}{r+1},j-\frac{1}{2}}^{\ell-1}.
%\end{equation}
\begin{equation}
    v_{i,j-\frac{1}{2}}^\ell = \frac{r}{2r+2} v_{i,j+\frac{1}{2}}^\ell +\frac{r}{2r-2} v_{i,j+\frac{3}{2}}^\ell - \frac{1}{r^2-1} v_{i-\frac{1}{2}+\frac{\alpha}{r+2},j-\frac{1}{2}}^{\ell-1}.
\end{equation}
In locations that do not have enough data to perform these interpolations, for example, at the corners of the coarse-fine interface, we use linear interpolation from coarse values to fill ghost cells. We note that this leads to first order accuracy in these cells. This reduction in order is localized to corners in the coarse fine interface, which is a set of codimension 2, and, as we will demonstrate, second order accuracy is retained sufficiently far away from the interface.

\subsection{Finite Difference Operators}
We require finite difference operators for the discrete gradient, discrete divergence, and variable coefficient viscous operator. The cell centered discrete gradient $\Gctof$ acts on cell centered quantities and returns side centered quantities. The discrete divergence $\Dftoc$ acts on side centered quantities and returns cell centered quantities. The variable coefficient viscous operator $\Lapftof[\theta] = [{\Lapftof}_0[\theta], {\Lapftof}_1[\theta]]^\text{T}$ acts on side quantities and returns side centered quantities. We define the action of the operators for a patch level $\ell$ and cell $c_{i,j}^\ell$ by
\begin{align}
    \parens{\Gctof}^\ell_{i+\frac{1}{2},j} &= \frac{p^\ell_{i+1,j} - p^\ell_{i,j}}{\Delta x^\ell}, \\[1em]
    \parens{\Dftoc}^\ell_{i,j} &= \frac{u^\ell_{i+\frac{1}{2},j} - u^\ell_{i-\frac{1}{2},j}}{\Delta x^\ell} + \frac{v^\ell_{i,j+\frac{1}{2}} - v^\ell_{i,j-\frac{1}{2}}}{\Delta y^\ell},\\[1em]
    \parens{{\Lapftof}_0[\theta]}^\ell_{i + \frac{1}{2}, j} &= \frac{1}{\Delta x}\parens{\theta_{i+1,j}\frac{u_{i+\frac{3}{2},j} - u_{i+\frac{1}{2},j}}{\Delta x}-\theta_{i,j}\frac{u_{i+\frac{1}{2},j} - u_{i-\frac{1}{2},j}}{\Delta x}} \nonumber\\
							             & +  \frac{1}{\Delta y}\parens{\theta_{i+\frac{1}{2},j+\frac{1}{2}}\frac{u_{i+\frac{1}{2},j+1} - u_{i+\frac{1}{2},j}}{\Delta y}-\theta_{i+\frac{1}{2},j-\frac{1}{2}}\frac{u_{i+\frac{1}{2},j} - u_{i+\frac{1}{2},j-1}}{\Delta y}} \nonumber\\
						                    & + \frac{1}{\Delta y}\parens{\theta_{i+\frac{1}{2},j+\frac{1}{2}}\frac{v_{i+1,j+\frac{1}{2}} - v_{i,j+\frac{1}{2}}}{\Delta x}-\theta_{i+\frac{1}{2},j-\frac{1}{2}}\frac{v_{i+1,j-\frac{1}{2}} - v_{i,j-\frac{1}{2}}}{\Delta x}} \nonumber\\
								   & -\frac{1}{\Delta x}\parens{\theta_{i+1,j}\frac{v_{i+1,j+\frac{1}{2}} - v_{i+1,j-\frac{1}{2}}}{\Delta y}-\theta_{i,j}\frac{v_{i,j+\frac{1}{2}} - v_{i,j-\frac{1}{2}}}{\Delta y}},
\end{align}

\begin{align}
    \parens{{\Lapftof}_1[\theta]}^\ell_{i, j + \frac{1}{2}} &=  \frac{1}{\Delta y}\parens{\theta_{i,j+1}\frac{v_{i,j+\frac{3}{2}} - v_{i,j+\frac{1}{2}}}{\Delta y}-\theta_{i,j}\frac{v_{i,j+\frac{1}{2}} - v_{i,j-\frac{1}{2}}}{\Delta y}} \nonumber \\
							             & +  \frac{1}{\Delta x}\parens{\theta_{i+\frac{1}{2},j+\frac{1}{2}}\frac{v_{i+1,j+\frac{1}{2}} - v_{i,j+\frac{1}{2}})}{\Delta x}-\theta_{i-\frac{1}{2},j+\frac{1}{2}}\frac{(v_{i,j+\frac{1}{2}} - v_{i-1,j+\frac{1}{2}}}{\Delta x}} \nonumber \\
						                    & + \frac{1}{\Delta x}\parens{\theta_{i+\frac{1}{2},j+\frac{1}{2}}\frac{u_{i+\frac{1}{2},j+1} - u_{i+\frac{1}{2},j})}{\Delta y}-\theta_{i-\frac{1}{2},j+\frac{1}{2}}\frac{(u_{i-\frac{1}{2},j+1} - u_{i-\frac{1}{2},j}}{\Delta y}} \nonumber \\		
								   & -\frac{1}{\Delta y}\parens{\theta_{i,j+1}\frac{u_{i+\frac{1}{2},j+1} - u_{i-\frac{1}{2},j+1})}{\Delta x}-\theta_{i,j}\frac{(u_{i+\frac{1}{2},j} - u_{i-\frac{1}{2},j}}{\Delta x}},
\end{align}
in which $\theta_{i-\frac{1}{2},j-\frac{1}{2}}$ is the volume fraction interpolated to a cell corner. We perform this interpolation by taking the average of the surrounding cell values, $\theta_{i-\frac{1}{2},j-\frac{1}{2}} = \frac{\theta_{i-1,j-1} + \theta_{i,j-1} + \theta_{i-1,j} + \theta_{i,j}}{4}$. We additionally define an interpolation operator $\interpctof$ that interpolates cell centered quantities to cell sides
\begin{equation}
    \interpctof \left[\theta_{i,j}\right] = \left[\theta_{i+\frac{1}{2},j}, \theta_{i,j+\frac{1}{2}}\right]^\text{T} = \left[\frac{\theta_{i,j} + \theta_{i+1,j}}{2}, \frac{\theta_{i,j} + \theta_{i,j+1}}{2}\right]^\text{T}.
\end{equation}
This operator acts on scalar quantities and returns staggered vector quantities.

For viscoelastic flows, we require a discrete divergence of the stress tensor $D_h^\text{stress}$. This operator is defined by
\begin{align}
    D_h^\text{stress}{\CC}_{i,j} &= \left[\begin{array}{c}D_h^\text{stress,0} \\ D_h^\text{stress,1}
    \end{array}\right]{\CC}_{i,j}, \\
    D_h^\text{stress,0}{\CC}_{i,j} &= \frac{\sigma_{{xx}_{i,j}} - \sigma_{{xx}_{i-1,j}}}{\Delta x} + \frac{\sigma_{{xy}_{i+1,j+1}}+\sigma_{{xy}_{i-1,j+1}}-\sigma_{{xy}_{i+1,j-1}}-\sigma_{{xy}_{i-1,j-1}}}{4\Delta y}, \\
    D_h^\text{stress,1}{\CC}_{i,j} &= \frac{\sigma_{{xy}_{i+1,j+1}}+\sigma_{{xy}_{i+1,j-1}}-\sigma_{{xy}_{i-1,j+1}}-\sigma_{{xy}_{i-1,j-1}}}{4\Delta x} + \frac{C_{{xx}_{i,j}} - C_{{xx}_{i,j-1}}}{\Delta y},
\end{align}
in which the components of the symmetric tensor $\CC$ are written as $\CC = \left[\begin{array}{cc} \sigma_{xx} & \sigma_{xy} \\ \sigma_{xy} & \sigma_{yy}\end{array}\right]$.

With these operators defined, the momentum and mass \cref{eq:forcebalance_net,eq:forcebalance_sol,eq:incompressibility_a} are discretized in space, resulting in the differential-algebraic equations
\begin{subequations}\label{eq:mom_mass_spatial_system}
\begin{align}
    \rho\frac{\text{d} \interpctof\parens{\thn}\vun}{\text{d} t} &= -\interpctof\parens{\thn}\Gctof p + \mun\Lapftof[\thn]\vun + D_h^\text{stress}\btaun - \xi\interpctof\parens{\thn\ths}\parens{\vun-\vus} \\
    \rho\frac{\text{d} \interpctof\parens{\ths}\vus}{\text{d} t} &= - \interpctof\parens{\ths}\Gctof p + \mus\Lapftof[\ths]\vus - \xi\interpctof\parens{\thn\ths}\parens{\vus-\vun}, \\
    0 &= \Dftoc\parens{\interpctof\parens{\thn} \vun} + \Dftoc\parens{\interpctof\parens{\ths}\vus}.
\end{align}
\end{subequations}
In addition, the transport equation for the network volume fraction given by \cref{eq:net_vol_frac} is discretized by
\begin{align}\label{eq:vol_frac_disc}
    \frac{\text{d}\thn}{\text{d} t} &= -\NN\left[\vun,\thn\right],
\end{align}
in which $\NN\left[\vun,\thn\right]$ is an approximation to $\grad\cdot\parens{\vun\thn}$ using a cubic upwinded interpolation (CUI) scheme \cite{roe1982}. The solvent volume fraction is computed as $\ths = 1 - \thn$.

For the viscoelastic stress tensor equation, we require stencils for the velocity gradient. These are computed by
\begin{equation}
    \grad_h\uu = \left[ \begin{array}{cc} \frac{u_{i+\frac{1}{2},j} - u_{i-\frac{1}{2},j}}{\Delta x} & \frac{u_{i+\frac{1}{2},j} + u_{i-\frac{1}{2},j} - u_{i+\frac{1}{2},j-1}-u_{i-\frac{1}{2},j-1}}{4\Delta y} \\
    \frac{v_{i+1,j+\frac{1}{2}} + v_{i+1,j-\frac{1}{2}} - v_{i-1,j+\frac{1}{2}} - v_{i-1,j-\frac{1}{2}}}{4\Delta x} & \frac{v_{i,j+\frac{1}{2}} - v_{i,j-\frac{1}{2}}}{\Delta y}\end{array}\right]
\end{equation}
in which the components of the velocity are denoted as $\uu = \left[u, v\right]^\text{T}$. The stress tensor and elastic modulus equations given by \cref{eq:StressTensor,eq:ElasticModulus} are therefore discretized as
\begin{align}\label{eq:discrete_ConformTensor}
    \frac{\text{d}\CC}{\text{d}t} &= \mathbb{F}\parens{\vun,\CC,z} = -\NN\left[\vun,\CC\right] + \CC\parens{\grad_h\vun}^\text{T} + \grad_h\vun\CC - \frac{1}{\lambda}\parens{\CC - z\II},\\
    \frac{\text{d}z}{\text{d}t} &= -\NN\left[\vun,z\right].
\end{align}

We note that the advective discretizations and interpolation procedures we utilize herein are conservative, both on uniform and adaptively refined grids. As such, the total mass of each phase, defined as $\int_\Omega\rho\theta_\text{i}\parens{\xx,t}\text{d}\xx$ for $\text{i}=\text{n},\text{s}$, is conserved to machine precision in all the results that follow.

\subsection{Temporal Discretization}
We discretize \cref{eq:mom_mass_spatial_system} using an implicit trapezoidal rule for the viscous stress and drag and a second order Adams-Bashforth scheme for the transport \cref{eq:vol_frac_disc,eq:discrete_ConformTensor}. To advance the solution from time $t^k$ to time $t^{k+1} = t^k + \Delta t$, we first compute $\thn^{k+1}$, $\ths^{k+1}$, $\CC^{k+1}$, and $z^{k+1}$ by evaluating
\begin{align}
    \thn^{k+1} &= \thn^{k} + \Delta t\parens{\frac{3}{2} \NN\left[{\vun^k},{\thn^k}\right] - \frac{1}{2} \NN\left[{\vun^{k-1}},{\thn^{k-1}}\right]}, \\
    \ths^{k+1} &= 1 - \thn^{k+1}, \\
    \CC^{k+1} &= \CC^k + \Delta t\parens{\frac{3}{2} \mathbb{F}\parens{\vun^k,\CC^k,z^k} - \frac{1}{2}\mathbb{F}\parens{\vun^{k-1},\CC^{k-1},z^{k-1}}}, \\
    z^{k+1} &= z^k - \Delta t\parens{\frac{3}{2}\NN\left[\vun^k,z^k\right] - \frac{1}{2}\NN\left[\vun^{k-1},z^{k-1}\right]}.
\end{align}
Then we compute the velocities $\vun^{k+1}$ and $\vus^{k+1}$ and the pressure $p^{k+1}$ by solving the saddle point system
\begin{equation}\label{eq:saddle_point}
    \mathcal{A}\left[\begin{array}{c}
         \vun^{k+1} \\
         \vus^{k+1} \\
         p^{k+1}
    \end{array}\right] = \left[\begin{array}{c}
        \frac{1}{\Delta t}\interpctof\thn^k\vun^k + \frac{\mun}{2}\Lapftof[\thn^k]\vun + D_h^\text{stress}\frac{\btaun^{k+1} + \btaun^k}{2} - \frac{\xi}{2}\interpctof\thn^k\ths^k\parens{\vun^k-\vus^k}  \\
         \frac{1}{\Delta t}\interpctof\ths^k\vus^k + \frac{\mus}{2}\Lapftof[\ths^k]\vus - \frac{\xi}{2}\interpctof\thn^k\ths^k\parens{\vus^k-\vun^k} \\
        0
    \end{array}\right],
\end{equation}
in which $\mathcal{A}$ is the linear operator defined by
\begin{equation}\label{eq:full_operator}
    \mathcal{A} = \left[ \begin{array}{cc}
        \mathcal{L} & \mathcal{G} \\
        \mathcal{D} & 0
    \end{array}\right],
\end{equation}
with block operators
\begin{align}
    \mathcal{L} &= \left[ \begin{array}{cc}
        \frac{1}{\Delta t}\interpctof\thn^{k+1} - \frac{\mun}{2}\Lapftof[\thn^{k+1}] - \frac{\xi}{2}\interpctof\thn^{k+1}\ths^{k+1} & \frac{\xi}{2}\interpctof\thn^{k+1}\ths^{k+1} \\
         \frac{\xi}{2}\interpctof\thn^{k+1}\ths^{k+1} & \frac{1}{\Delta t}\interpctof\ths^{k+1} - \frac{\mus}{2}\Lapftof[\ths^{k+1}] - \frac{\xi}{2}\interpctof\thn^{k+1}\ths^{k+1} 
    \end{array}\right], \\
    \mathcal{G} &= \left[ \begin{array}{c}
         \interpctof\thn^{k+1}\Gctof \\
         \interpctof\ths^{k+1}\Gctof
    \end{array}\right], \\
    \mathcal{D} &= \left[ \begin{array}{cc}
         -\Dftoc\interpctof\thn^{k+1}& -\Dftoc\interpctof\ths^{k+1}  
    \end{array}\right].
\end{align}
This extended saddle point system in \cref{eq:saddle_point} is solved using GMRES. To obtain an efficient linear solver algorithm, we precondition the system using a geometric multigrid method as described next.

\subsection{Multigrid Preconditioner}
We precondition the saddle point system in \cref{eq:saddle_point} with an FAC multigrid preconditioner \cite{briggs2000} based on previous work by Wright et al.\ \cite{wright2008}. We construct additional coarser levels that cover the same physical space as $\Omega^0$. We denote these coarser levels by $\Omega^\ell$ in which we now let $\ell$ be negative. These coarser patch levels are only used during the preconditioning step to smooth the solution. We define the prolongation operator $\mathcal{I}_{\ell\rightarrow\ell+1}$ as the operator that prolongs the solution from level $\Omega^\ell$ to level $\Omega^{\ell+1}$. We also define the restriction operator $\mathcal{I}_{\ell\leftarrow\ell+1}$ as the operator that restricts the residual from level $\Omega^{\ell+1}$ to level $\Omega^\ell$. Both of these operators are constructed using a conservative linear interpolation procedure. We further denote explicitly the spatial dependence of solution operator $\mathcal{A}_\ell$ on a given patch level $\Omega^\ell$, in which $\mathcal{A}_\ell$ is a direct rediscretization of the multiphase equations. A single application of the preconditioner consists of one standard V-cycle \cite{briggs2000}.

During the pre- and post-smoothing steps, because of the presence of a zero block in the operator $\mathcal{A}_\ell$ in \cref{eq:full_operator}, standard relaxation schemes can not be applied directly. Therefore, we use the box relaxation smoother developed by Wright et al.\ \cite{wright2008}. Box relaxation is a generalization of point smoothers like Jacobi and Gauss-Seidel to simultaneously relax all velocity and pressure degrees of freedom within rectangular patches of grid cells. For all the simulations shown, we use either three or five pre- and post-smoothing operations. The number of smoothing operations will affect both the strength and cost of the preconditioner, with additional smoothing iterations tending to decrease the overall Krylov iteration count but increasing the runtime cost of the preconditioner. For the results considered herein, three to five smoothing operations were deemed to give good performance. On the coarsest level, which consists of either $N = 4, 8, \text{ or } 16$ grid points in each direction, we perform ten smoothing iterations, which in our numerical experiments, was sufficient to obtain reasonable convergence for our preconditioner. During each application of the smoother, we solve the following nine-by-nine system of equations in each grid cell
\begin{equation} \label{box_relax_linear_system}
    \begin{bmatrix}
	\mun\mathcal{L}^h[\thn]-\xi\mathcal{C}^h & \xi\mathcal{C}^h & \mathcal{G}^h[\thn] \\
	\xi\mathcal{C}^h        &      \mus\mathcal{L}^h[\ths]-\xi\mathcal{C}^h & \mathcal{G}^h[\ths] \\
	(\mathcal{G}^h[\ths])^\text{T}  & (\mathcal{G}^h[\ths])^\text{T} & 0 \\
	\end{bmatrix}
	\begin{bmatrix}
	({\vun}_{i,j})^{m+1} \\
	({\vus}_{i,j})^{m+1} \\
	(p_{i,j})^{m+1} \\
	\end{bmatrix}
	= \begin{bmatrix}
	F_\text{n} + \mun b[\thn,\vun^{m+1}] \\
	F_\text{s} + \mus b[\ths,\vus^{m+1}]  \\
	0 \\
	\end{bmatrix},
\end{equation}
in which components of $\vun$ and $\vus$ consist of velocity values stored at each grid cell face as follows
	\begin{equation*}
	({\vun}_{i,j})^{m+1} = 
	\begin{bmatrix}
	(u^\text{n}_{i-1/2,j})^{m+1}  \\
	(u^\text{n}_{i+1/2,j})^{m+1}  \\
	(v^\text{n}_{i,j-1/2})^{m+1}   \\
	(v^\text{n}_{i,j+1/2})^{m+1}  \\
	\end{bmatrix}
\qquad \text{and} \qquad
	({\vus}_{i,j})^{k+1} = 
	\begin{bmatrix}
	(u^\text{s}_{i-1/2,j})^{m+1}  \\
	(u^\text{s}_{i+1/2,j})^{m+1}  \\
	(v^\text{s}_{i,j-1/2})^{m+1}   \\
	(v^\text{s}_{i,j+1/2})^{m+1}  \\
	\end{bmatrix},
	\end{equation*}
and the operators $\mathcal{L}^h[\theta]$, $\mathcal{C}^h$, and $\mathcal{G}^h[\theta]$ are defined in the Appendix.

We process cells in a red-black ordering. We combine the smoother with under-relaxation to achieve better smoothing properties:
\begin{equation}
    \begin{bmatrix}
        ({\vun}_{i,j})^{m+1} \\
        ({\vus}_{i,j})^{m+1} \\
        (p_{i,j})^{m+1}
    \end{bmatrix} 
    = (1-\omega)
    \begin{bmatrix}
        ({\vun}_{i,j})^{m} \\
        ({\vus}_{i,j})^{m} \\
        (p_{i,j})^{m}
    \end{bmatrix}
     + \omega     \begin{bmatrix}
	\mun\mathcal{L}^h[\thn]-\xi\mathcal{C}^h & \xi\mathcal{C}^h & \mathcal{G}^h[\thn] \\
	\xi\mathcal{C}^h        &      \mus\mathcal{L}^h[\ths]-\xi\mathcal{C}^h & \mathcal{G}^h[\ths] \\
	(\mathcal{G}^h[\ths])^\text{T}  & (\mathcal{G}^h[\ths])^\text{T} & 0 \\
	\end{bmatrix}^{-1}
 \begin{bmatrix}
        (F_\text{n} + \mun b[\thn,\vun^{m+1}] \\
        (F_\text{s} + \mus b[\ths,\vus^{m+1}] \\
        0
    \end{bmatrix}, 
\end{equation}
in which $\omega$ is the under-relaxation parameter. For all numerical experiments in the following section, we set $\omega = 0.75$ , which was chosen through numerical experimentation. When applying the smoother on patch levels that do not cover the entire computational domain, we use the same ghost cell filling operation as described in \Cref{sec:cf_ghost_cells} to fill ghost cells at coarse-fine interfaces. We note that in this case, the solution on the coarse grid is not changed by the application of the smoother.

\subsection{Regridding Criteria}\label{sec:regridding}
The locally refined meshes can be static, that is, they occupy a fixed region in the domain $\Omega^\ell$, or they can be adaptive, that is, certain cells undergo dynamic refinement based on a regridding criteria. 

For a dynamic grid, after a specified number of time steps, we perform a regridding operation in which cells are tagged for refinement. In this work, cells are tagged for refinement based on the magnitude of the gradient of the volume fraction. Specifically, for each patch level $\ell$, a cell $c^\ell_{i,j}$ is tagged for refinement if $\size{\grad_h {\thn}^\ell_{i,j}} > \varepsilon^\ell$, in which
\begin{equation}\label{eq:regrid_criteria}
    \grad_h {\thn}^\ell_{i,j} = \left[ \frac{{\thn}^\ell_{i+1,j} - {\thn}^\ell_{i-1,j}}{2\Delta x^\ell}, \frac{{\thn}^\ell_{i,j+1}-{\thn}^\ell_{i,j-1}}{2\Delta y^\ell}\right]^\text{T},
\end{equation}
is the discrete gradient of the network volume fraction, and $\varepsilon^\ell$ is the gradient tolerance threshold. These values will depend on the characteristics of the solution that should be refined. In practice, we obtain these values by numerical experimentation: first running a coarse simulation and determining features in the solution which should be resolved. Note that we are not restricted to using a particular criteria and more complex regridding criteria may be used for future applications. In newly refined regions, the velocities, $\vun\parens{\xx,t}$ and $\vus\parens{\xx,t}$, and pressure, $p\parens{\xx,t}$ are interpolated from the old coarse grid using a conservative linear interpolation scheme \cite{griffith2007}. In newly coarsened regions, all of the quantities are updated using conservative averages of the old fine grid data.

\section{Results}
We present convergence results for the numerical method for several test problems using manufactured solutions. The test problems include both prescribed volume fractions as well as evolved volume fractions for both uniform and statically refined grids. We also present results in an idealized four roll mill geometry for both a viscous and viscoelastic network. We show that the iteration count of the linear solver on the adaptive mesh is independent of grid spacing. We show that the numerical method offers significant speed and efficiency improvements for problems that require a high degree of resolution to capture interesting fluid flow phenomena, such as formation of singular structures in the network stress field, in dynamically-changing regions.

We define the $L^1$, $L^2$, and $L^\infty$ discrete norms by
\begin{align}
    \size{p}_1 &= \sum_{\ell = 0}^{\ell_\text{max}}\sum_{\text{valid} (i,j)\in\Omega^\ell}|p_{i,j}|\Delta x^\ell\Delta y^\ell, \\
    \size{p}_2^2 &= \sum_{\ell = 0}^{\ell_\text{max}}\sum_{\text{valid} (i,j)\in\Omega^\ell}p_{i,j}^2\Delta x^\ell\Delta y^\ell, \\
    \size{p}_\infty &= \max_{\ell=0,\ldots,\ell_\text{max}}\max_{\text{valid}(i,j)\in\Omega^\ell}|p_{i,j}|,
\end{align}

in which a cell is valid if it is not covered by a refined cell on a finer level of the grid hierarchy. The discretization is run on a sequence of grids with the coarsest grid having $N = 8$ points in each dimension up to $N = 128$ points. For cases in which we use a statically refined region, the refinement ratio is $r = 4$. The time step size, $\rm{\Delta t}$, is chosen to satisfy the advective CFL condition $\Delta t = \CFL \Delta x^{\ell_\text{max}}/u_{max}$, in which the advective CFL number $\CFL$ is fixed to enforce numerical stability and $\Delta x^{\ell_\text{max}}$ is the grid spacing on the most refined level. For all the simulations reported herein, we use periodic boundary conditions.%(chosen arbitrarily to be a value between 0 and 1 based on heuristic tests),

For simulations that use a statically refined region, we use an L-shaped refined region, as shown in \Cref{fig:errors_at_cf}b. The re-entrant corner along the coarse-fine interface allows us to test the impact of this corner on overall accuracy. We note that the discretization utilized here does not conserve stresses along re-entrant corners, which may limit the effectiveness of the discretization at zero Reynolds numbers \cite{gerya2013}. However, as we demonstrate, the discretization is effective at the non-zero Reynolds numbers tested.

As noted earlier, the advective discretizations and interpolation procedures we utilize herein are conservative, both on uniform and adaptively refined grids. As such, the total mass of each phase, defined as $\int_\Omega\rho\theta_\text{i}\parens{\xx,t}\text{d}\xx$ for $\text{i}=\text{n},\text{s}$, is conserved to machine precision in all the results that follow.

\subsection{Multiphase flow with prescribed, variable volume fraction}
This section analyzes the convergence behavior of the linear solver when simulating multiphase flow with a prescribed, variable network volume fraction. The governing equations for this flow are given by \cref{eq:incompressibility_a,eq:forcebalance_net,eq:forcebalance_sol}. To analyze convergence of the linear solver, a manufactured test problem is set-up using forcing functions so that the solution components are given by
\begin{subequations}\label{eq:prescribed_fcns}
\begin{align}
    \vun\parens{\xx,t} &=  
    \begin{pmatrix}
        \cos(2\pi(x-t))\sin(2\pi(y-t)) \\
        \sin(2\pi(x-t))\cos(2\pi(y-t))  \\
    \end{pmatrix}, \label{eq:prescribed_fcns:un}\\
    \vus\parens{\xx,t} &=  
    \begin{pmatrix}
        -\cos(2\pi(x-t))\sin(2\pi(y-t)) \\
        -\sin(2\pi(x-t))\cos(2\pi(y-t))  \\
    \end{pmatrix}, \label{eq:prescribed_fcns:us}\\
    p\parens{\xx,t} &= 0, \label{eq:prescribed_fcns:p}\\
    \thn\parens{\xx,t} &=  \frac{1}{4}\sin(2\pi(x-t))\sin(2\pi(y-t)) + \frac{1}{2}. \label{eq:prescribed_fcns:thn}
    %\ffp\parens{\xx,t} &= \frac{\pi}{2}\bigg(\cos(4\pi(x-t)-\cos(4\pi(x-y))+\cos(4\pi(y-t))-\cos(4\pi(-2t+x+y))\bigg), \\
\end{align}
\end{subequations}

We also use these functions as the initial conditions. The system parameters are defined in \Cref{tab:variable_thn_params}.

\begin{table}[tb]
    \caption{Parameter values for multiphase flow with a prescribed, variable volume fraction.}
    \centering
    \label{tab:variable_thn_params}%
    \begin{tabular}{|c c c|}     
    \hline
         \textbf{Parameter} & \textbf{Symbol}& \textbf{Value}\\ \hline
         Network viscosity      &    $\mun$         &    4.0          \\ 
         Solvent viscosity      &    $\mus$         &    0.004        \\ 
         Drag coefficient       &    $\xi$           &    250$\mus$  \\
         Density                &    $\rho$          &    1.0          \\
         Maximum CFL number     &    $\CFL$          &    0.5         \\
    \hline
    \end{tabular}
\end{table}

\begin{figure}[h]
    \subfloat[Log-log plot of error in $\vun\parens{\xx,t}$]{\epsfig{file=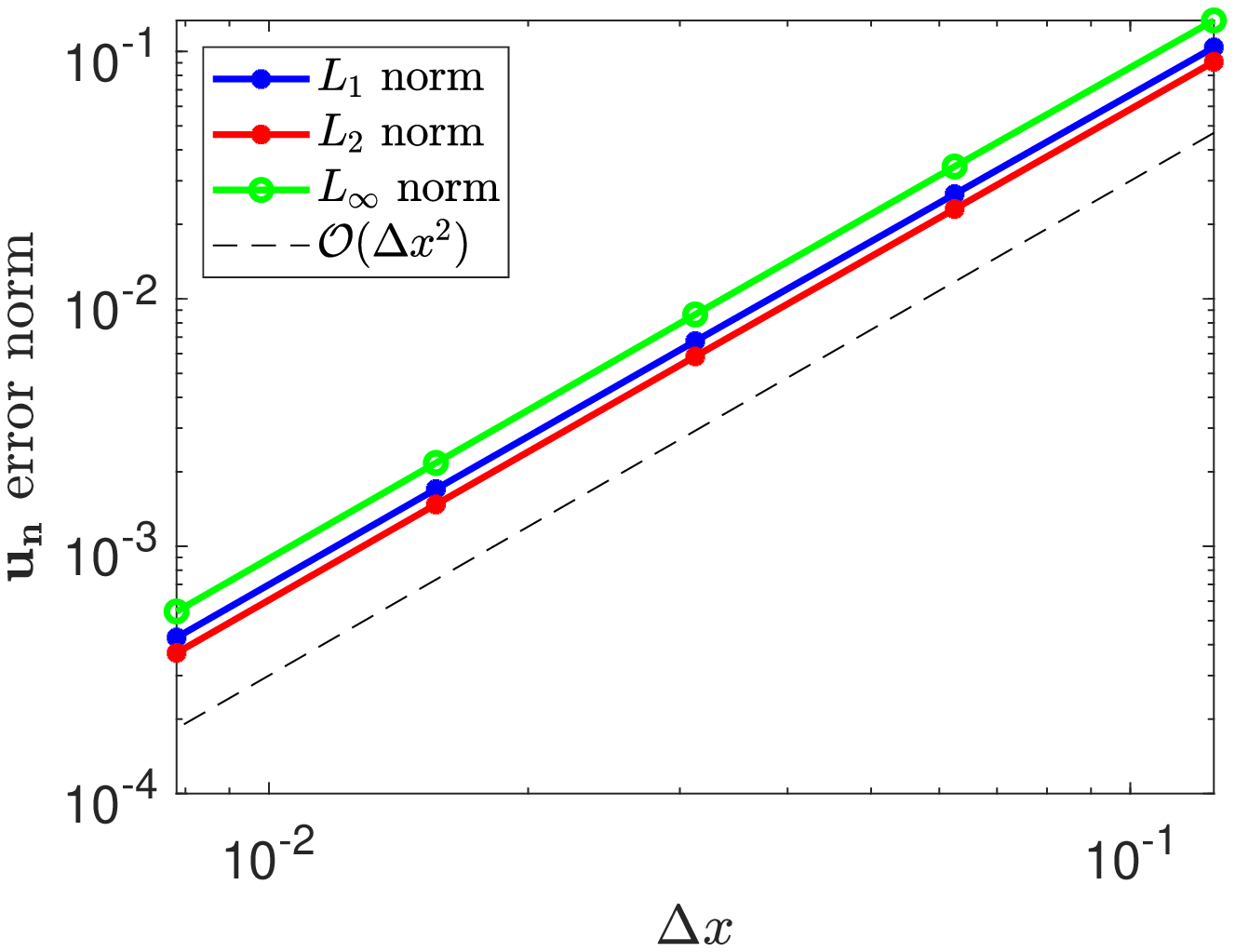, width=0.32\textwidth}}~
    \subfloat[Log-log plot of error in $\vus\parens{\xx,t}$]{\epsfig{file=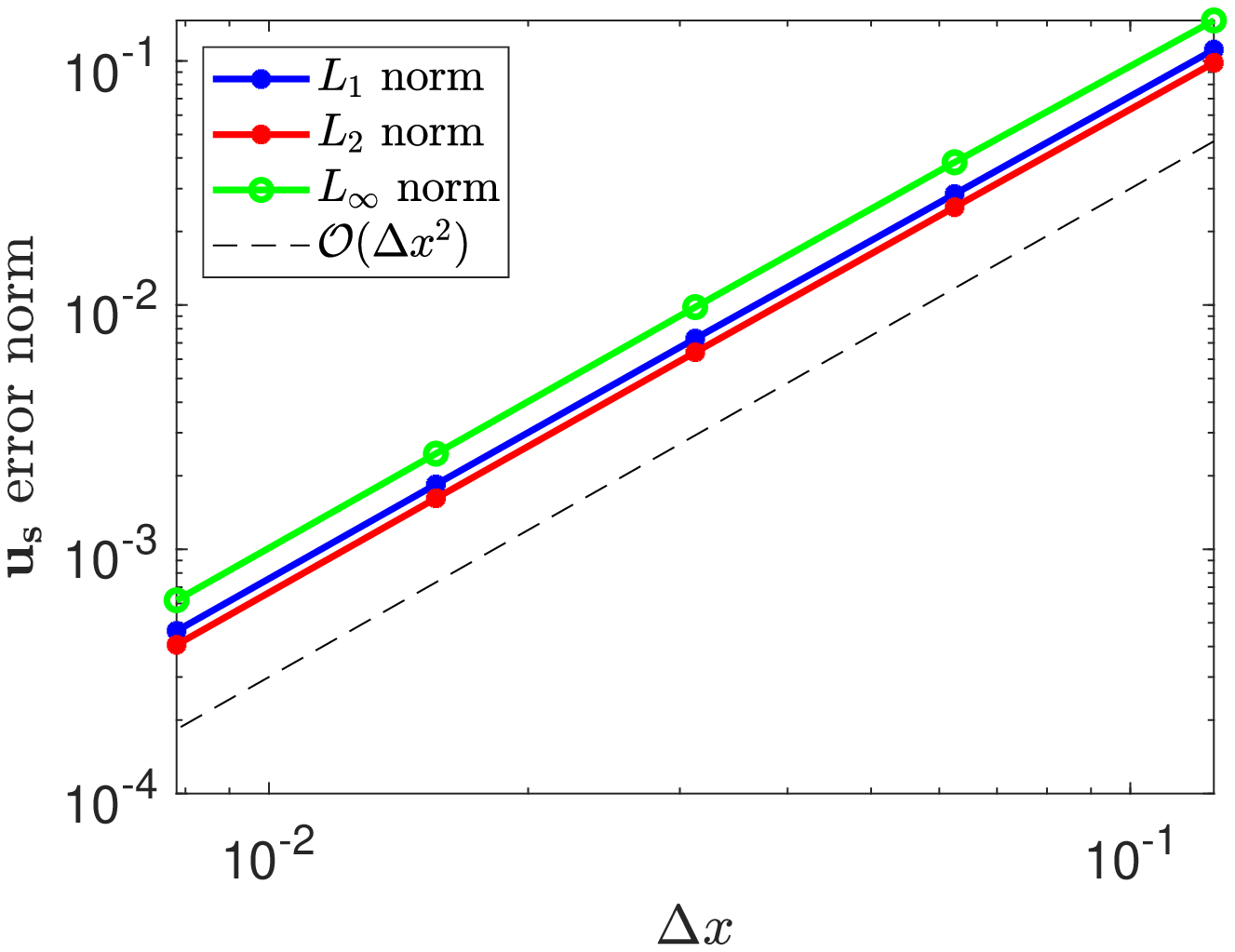, width=0.32\textwidth}}~
    \subfloat[Log-log plot of error in $p\parens{\xx,t}$]{\epsfig{file=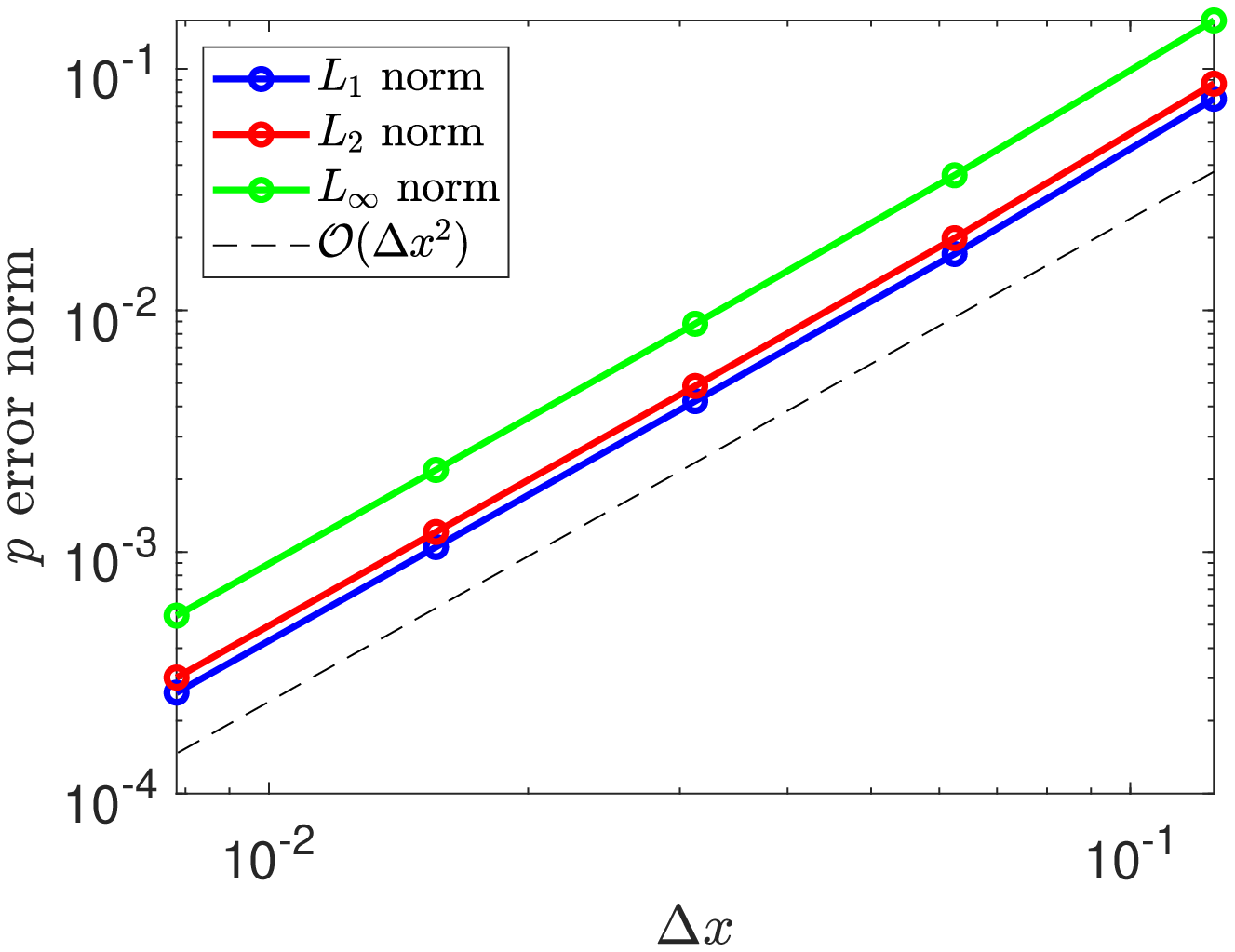, width=0.32\textwidth}}
    \caption{Log-log plots indicating convergence behavior of the momentum linear solver for a prescribed, variable network volume fraction $\thn\parens{\xx,t}$ on a uniform grid for each of the solution components: (a) $\vun\parens{\xx,t}$, (b) $\vus\parens{\xx,t}$, (c) $p\parens{\xx,t}$. We obtain second order convergence for all three solution components in the $L^1$, $L^2$ and $L^{\infty}$ norms.}
    \label{fig:varThn_uniGrid_error}
\end{figure}

\begin{figure}[h]
    \subfloat[Log-log plot of error in $\vun\parens{\xx,t}$]{\epsfig{file=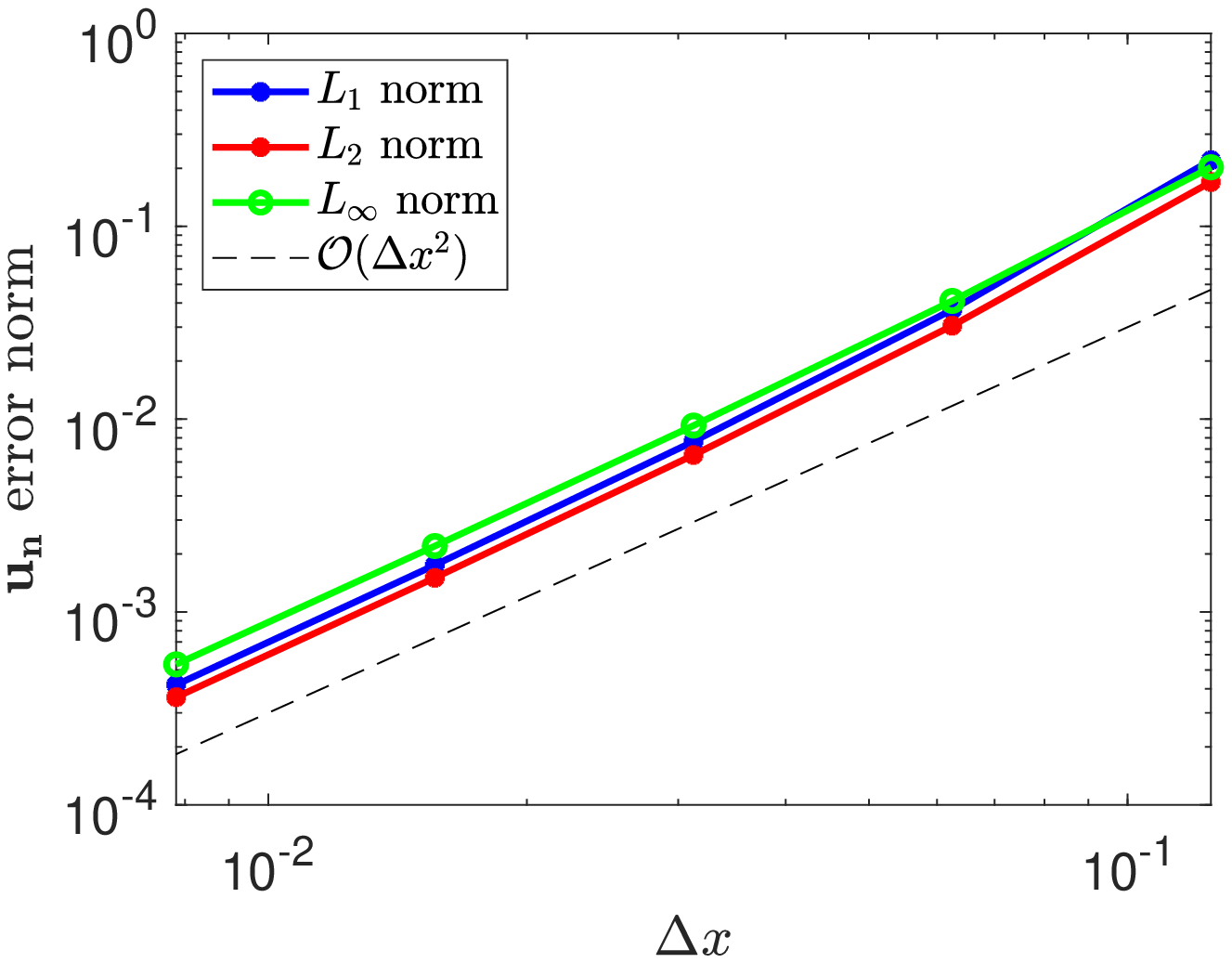, width=0.32\textwidth}}~
    \subfloat[Log-log plot of error in $\vus\parens{\xx,t}$]{\epsfig{file=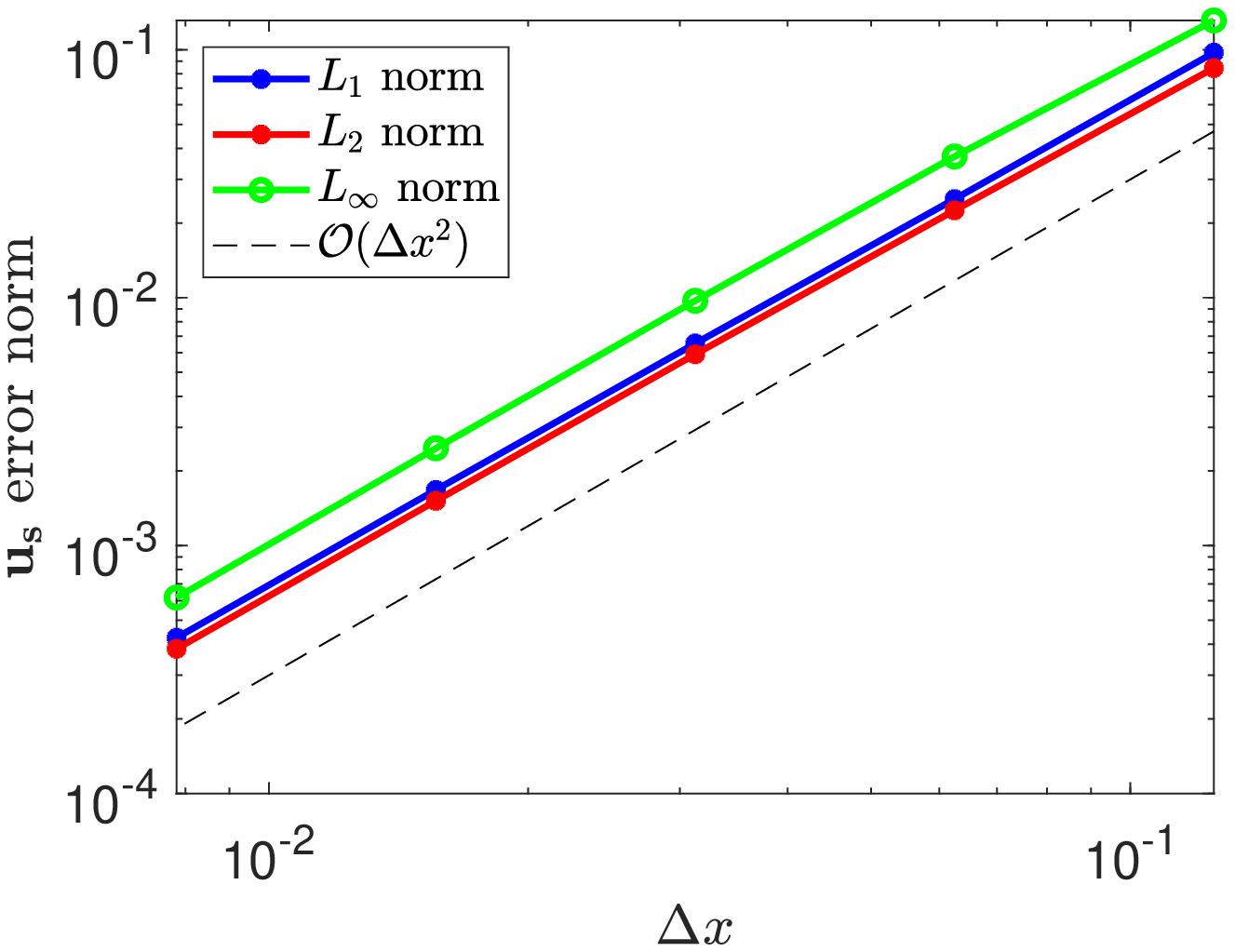, width=0.32\textwidth}}~
    \subfloat[Log-log plot of error in $p\parens{\xx,t}$]{\epsfig{file=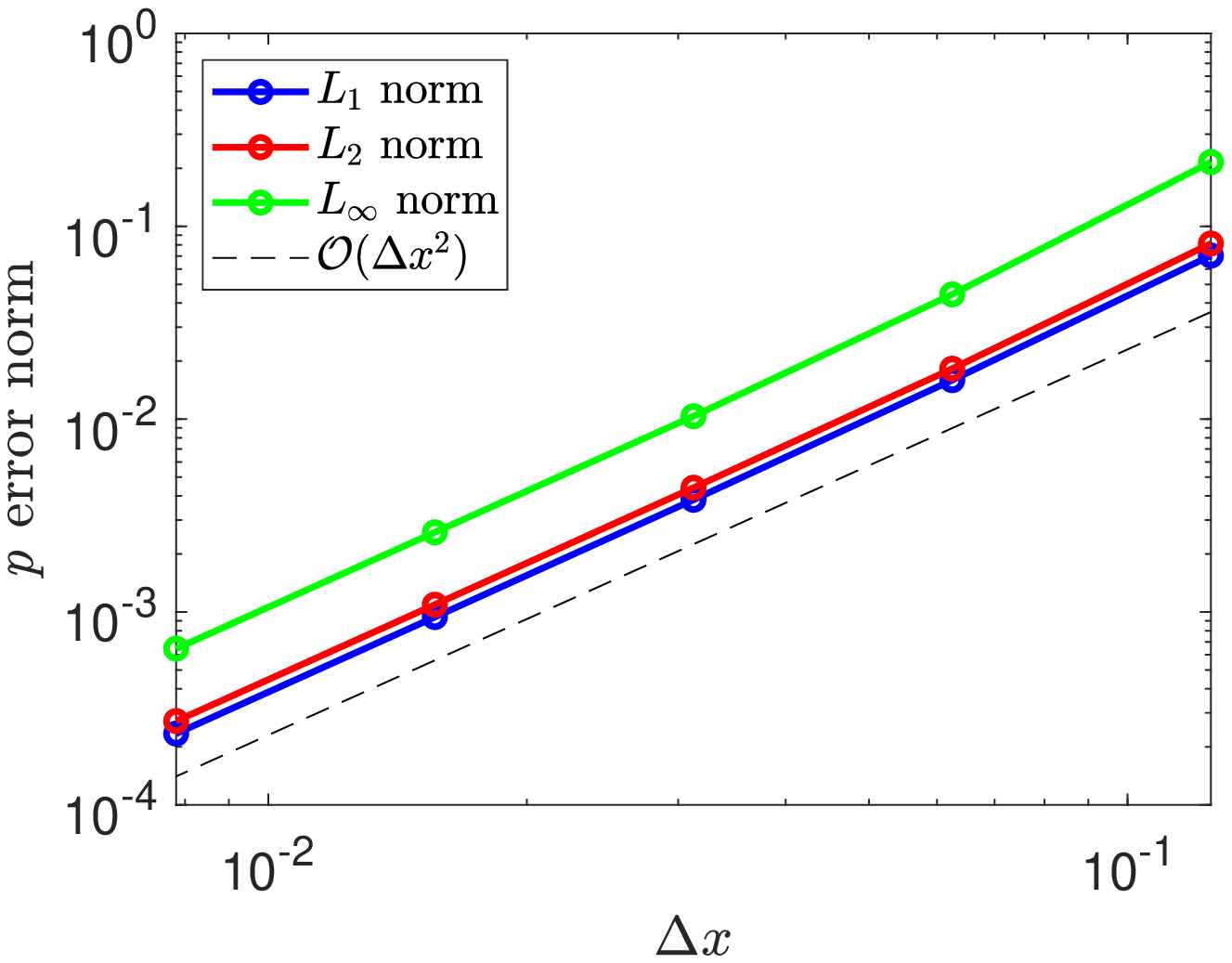, width=0.32\textwidth}}
    \caption{Log-log plots indicating convergence behavior of the momentum linear solver for a prescribed, variable network volume fraction $\thn\parens{\xx,t}$ on a static locally-refined grid for each of the solution components: (a) $\vun\parens{\xx,t}$, (b) $\vus\parens{\xx,t}$, (c) $p\parens{\xx,t}$. We obtain second order convergence for all three solution components in the $L^1$, $L^2$ and $L^{\infty}$ norms.}
    \label{fig:varThn_refGrid_error}
\end{figure}

The error norms between the numerical and analytical solution for the solution variables, $\vun\parens{\xx,t}, \vus\parens{\xx,t}$ and $p\parens{\xx,t}$, are calculated on both uniform grids and grids with locally refined, static L-shaped regions. The error norms vs. the grid size are plotted in \Cref{fig:varThn_uniGrid_error,fig:varThn_refGrid_error} for the simulations performed on uniform grids and static locally-refined grids, respectively. The plots show that adaptive discretization achieves second order accuracy for all solution variables across all error norms for the solutions obtained using a uniform grid and those obtained using a static locally-refined grid.

\subsection{Multiphase flow with volume fraction advection}
We allow the network and solvent volume fractions to advect with the fluid flow by solving an advection equation for $\thn\parens{\xx,t}$ and set $\ths\parens{\xx,t} = 1 - \thn\parens{\xx,t}$, in addition to solving the momentum equations. That is, we solve \cref{eq:net_vol_frac,eq:incompressibility_a,eq:forcebalance_net,eq:forcebalance_sol} to simulate multiphase Newtonian fluid flow for the case where the volume fractions $\thn\parens{\xx,t}$ and $\ths\parens{\xx,t}$ undergo advection with their respective velocity field. This means that $\thn\parens{\xx,t}$ is an additional solution component whose error norm with respect to grid size needs to be analyzed to determine the convergence behavior of the advection discretization. To analyze convergence of the discretization, a manufactured test problem is set-up using forcing functions so that the solution components are again given by \cref{eq:prescribed_fcns}.

The remaining system parameters are defined in \Cref{tab:adv_thn_params}. Because we are solving the full non-linear problem, additional time step size constraints must be followed, as shown by  Wright et al.\ \cite{wright2008}. Here, we use a smaller maximum advective CFL number $\CFL = 0.1$ to achieve a stable simulation. We have found that this advective CFL number constraint was sufficient to maintain stability in these simulations.

\begin{table}[bt]
\centering
    \caption{Parameter values for multiphase flow with advecting volume fraction.}
    \label{tab:adv_thn_params}%
    \begin{tabular}{|c c c|} 
    \hline
         \textbf{Parameter} & \textbf{Symbol}& \textbf{Value}\\ \hline
         Network Viscosity      &    $\mun$         &    4.0          \\ 
         Solvent viscosity      &    $\mus$         &    0.4        \\ 
         Drag coefficient       &    $\xi$           &    250$\mus$  \\
         Density                &    $\rho$          &    1.0          \\
         Maximum CFL number     &    $\CFL$          &    0.1         \\
    \hline
    \end{tabular}
\end{table}

\begin{figure}
\begin{center}
    \subfloat[Error in pressure]
    {\epsfig{file=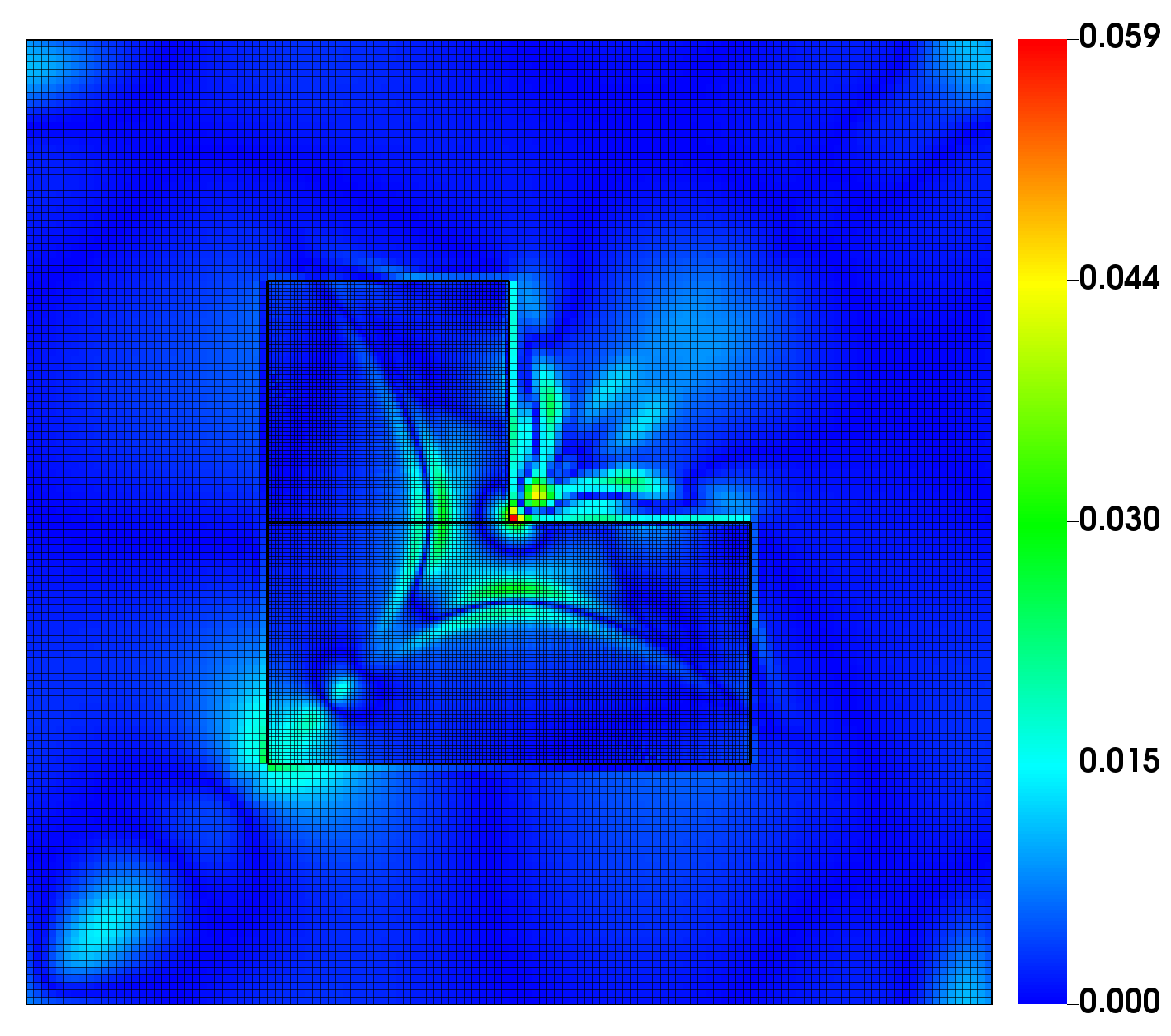, width=0.45\textwidth }} \qquad
    \subfloat[Statically refined grid]
    {
    \begin{tikzpicture}
    \draw[step=0.5cm,gray,very thin] (-1,-1) grid (5,5);
    \draw[step=0.125cm,gray,very thin] (0.5,0.5) grid (3.5,1.5);
    \draw[step=0.125cm,gray,very thin] (0.5,1.5) grid (1.5,3.5);
    \draw[very thick,-] (0.5,0.5) -- (3.5,0.5);
    \draw[very thick,-] (3.5,0.5) -- (3.5,1.5);
    \draw[very thick,-] (0.5,0.5) -- (0.5,3.5);
    \draw[very thick,-] (0.5,3.5) -- (1.5,3.5);
    \draw[very thick,-] (1.5,3.5) -- (1.5,1.5);
    \draw[very thick,-] (1.5,1.5) -- (3.5,1.5);
    \end{tikzpicture}}
\end{center}
    \caption{(a) The absolute error in pressure at $t=0.4$ for the fluid simulation with advection of network volume fraction $\thn\parens{\xx,t}$ on a static locally-refined grid. The greatest error occurs near the re-entrant corner of the refined region and along the lengths of the coarse-fine interfaces adjacent to this corner. This is consistent with the first-order discretization implemented at corners of the coarse-fine interface. The statically refined grid in (b) has an L-shaped refined region where the ratio of refinement between the coarse level and the refined level is $r=4$}
    \label{fig:errors_at_cf}
\end{figure}

\begin{figure}[h]
\begin{center}
    \subfloat[Log-log plot of error in $\vun\parens{\xx,t}$]{\epsfig{file=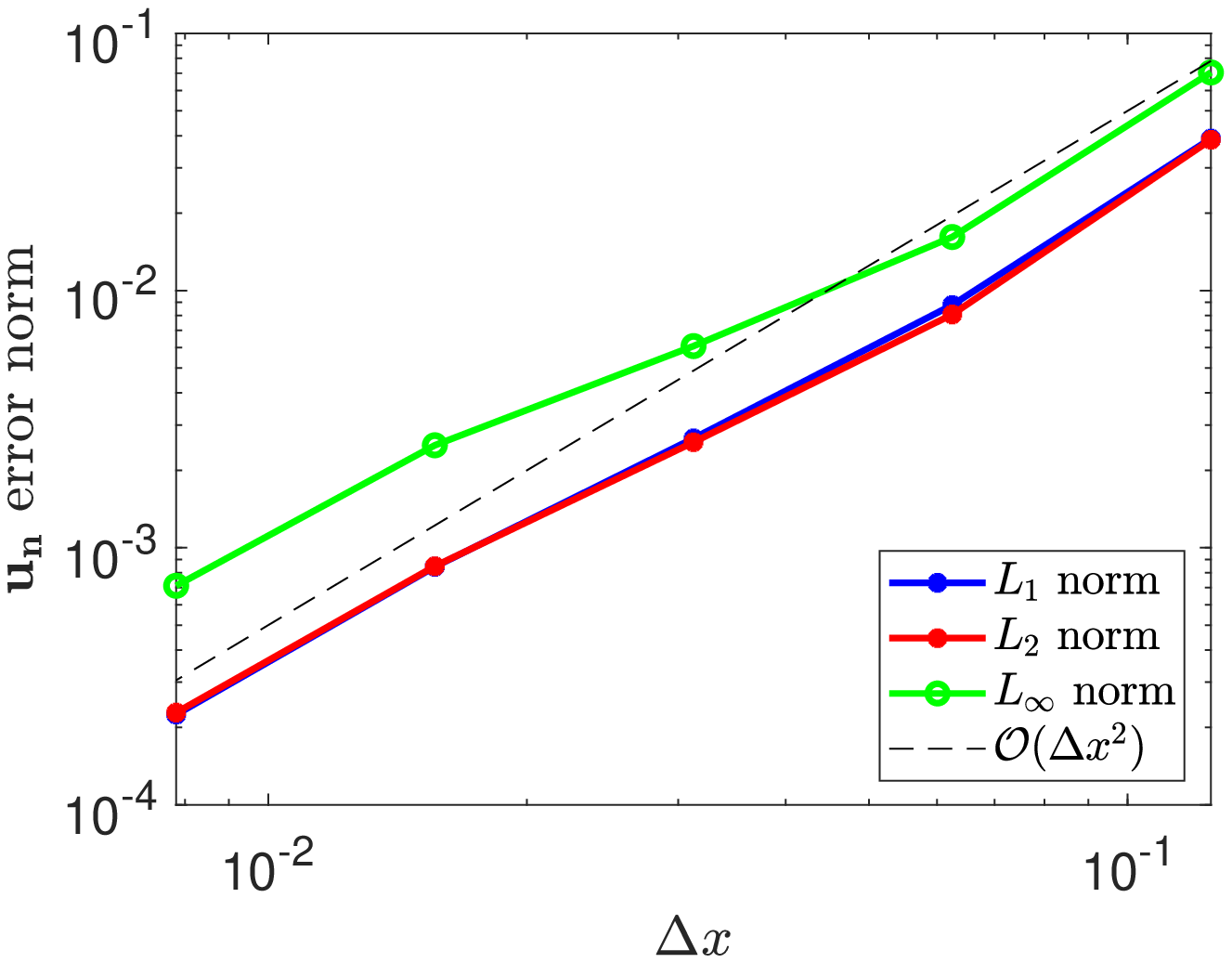, width=0.4\textwidth}}~ \qquad
    \subfloat[Log-log plot of error in $\vus\parens{\xx,t}$]{\epsfig{file=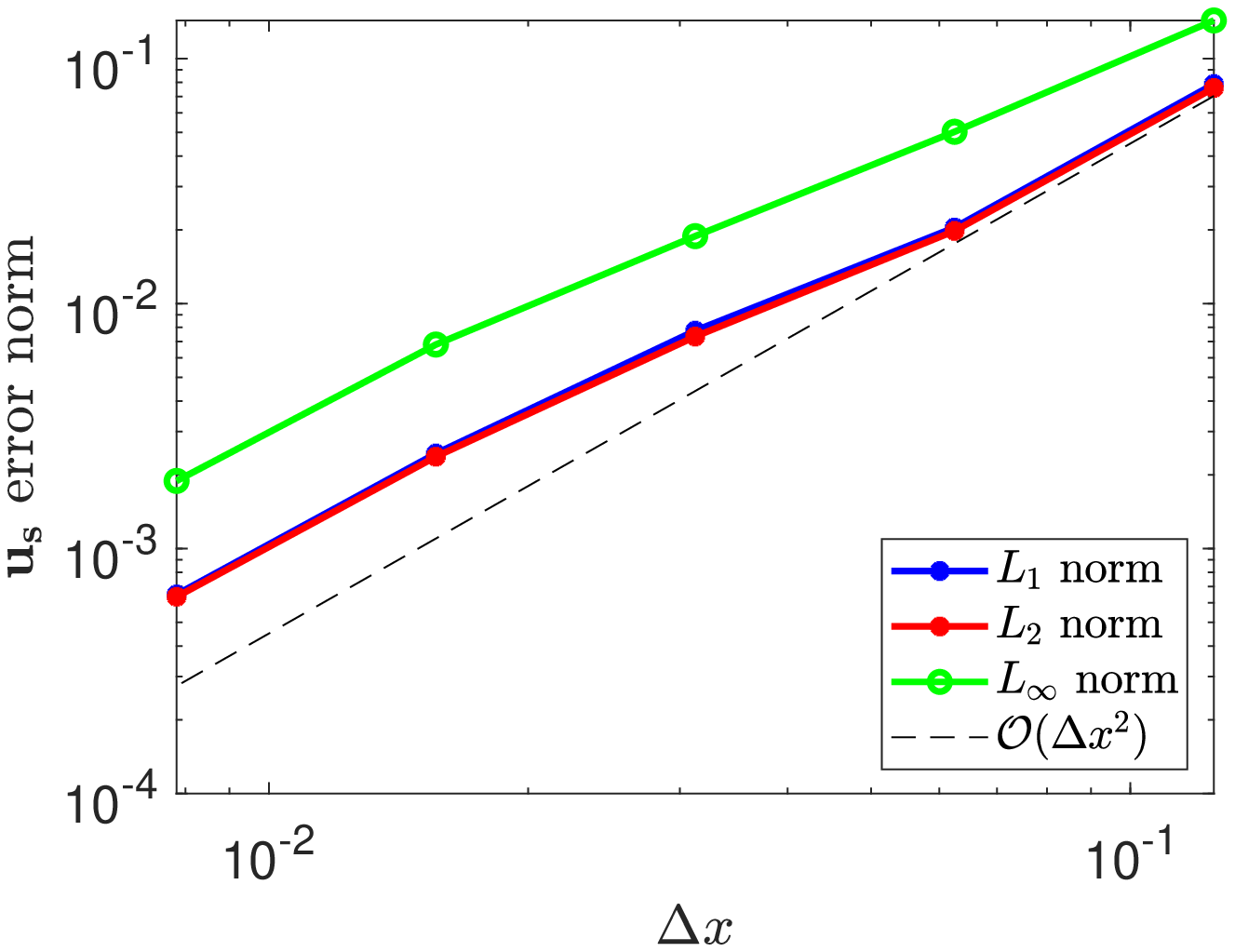, width=0.4\textwidth}}  \\
    \subfloat[Log-log plot of error in $p\parens{\xx,t}$]{\epsfig{file=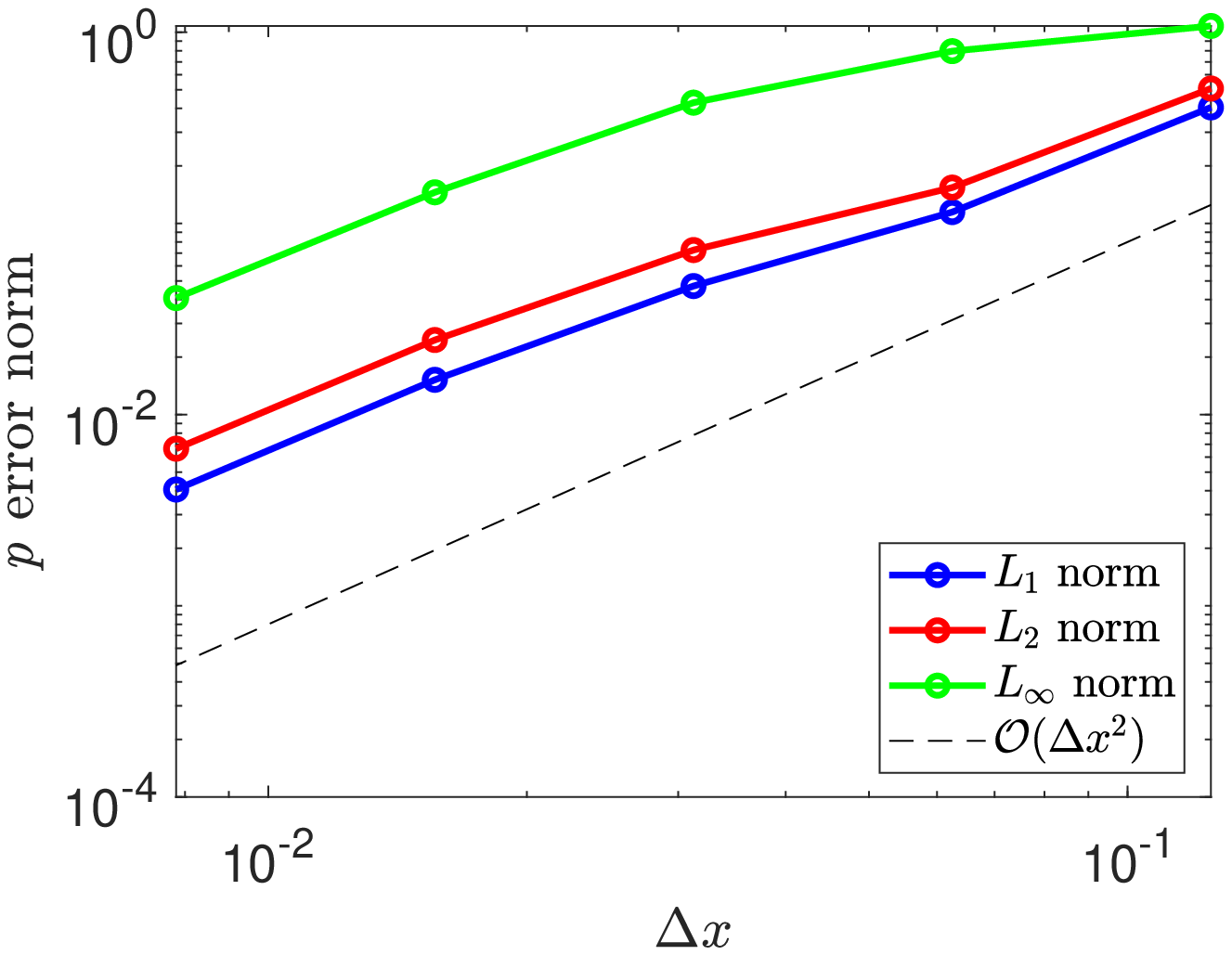, width=0.4\textwidth}}~ \qquad
    \subfloat[Log-log plot of error in $\thn\parens{\xx,t}$]{\epsfig{file=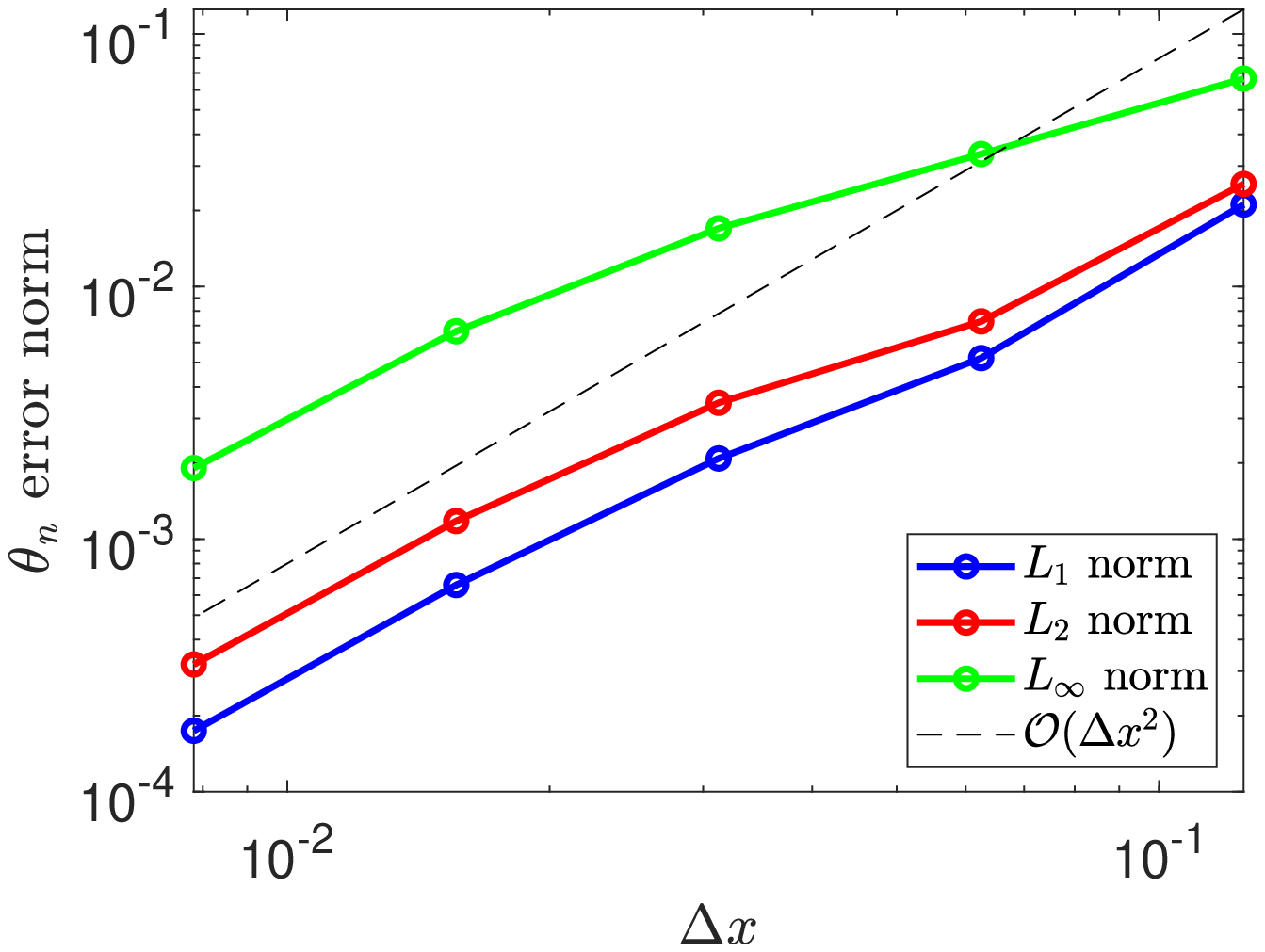, width=0.4\textwidth}}
\end{center}
    \caption{Log-log plots indicating convergence behavior of the fluid discretization with advection of network volume fraction $\thn\parens{\xx,t}$ on a uniform grid for each of the solution components:(a) $\vun\parens{\xx,t}$, (b) $\vus\parens{\xx,t}$, (c) p and (d) $\thn\parens{\xx,t}$. We obtain second order convergence for all four solution components in the $L^1$, $L^2$ and $L^{\infty}$ norms.}
    \label{fig:advect_uniGrid_error}
    \end{figure}

\begin{figure}[h]
\begin{center}
    \subfloat[Log-log plot of error in $\vun\parens{\xx,t}$]{\epsfig{file=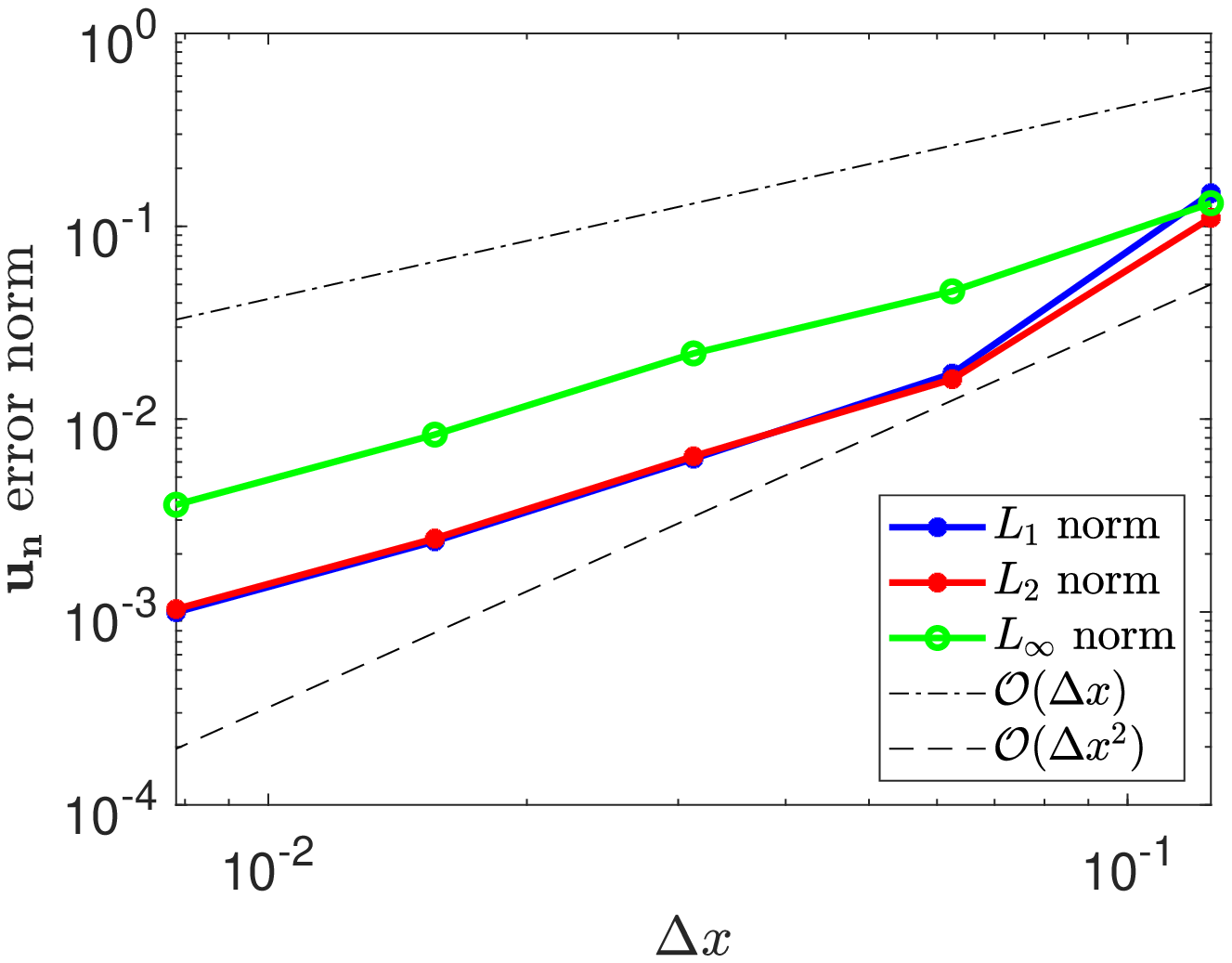, width=0.4\textwidth}}~ \qquad
    \subfloat[Log-log plot of error in $\vus\parens{\xx,t}$]{\epsfig{file=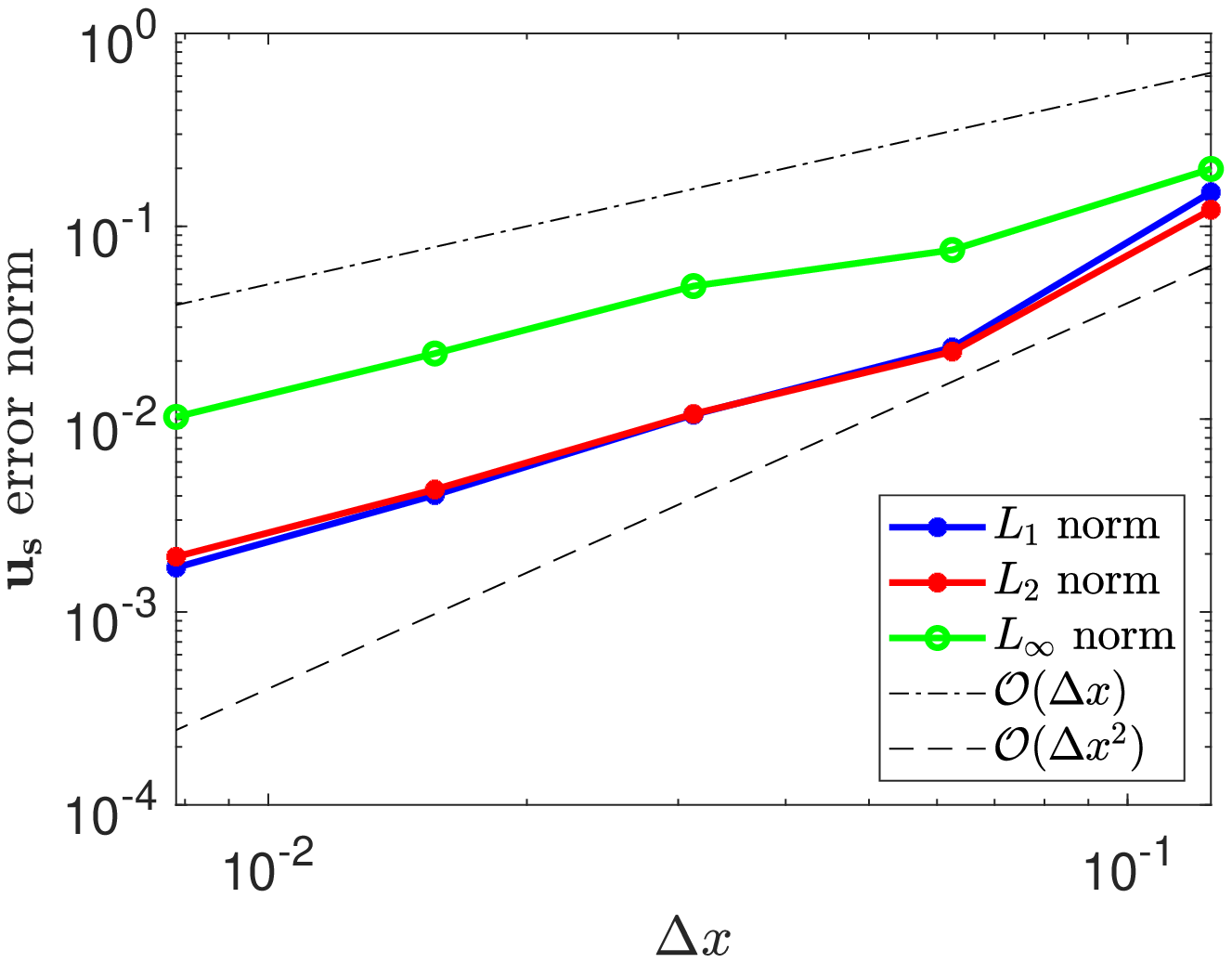, width=0.4\textwidth}}\\
    \subfloat[Log-log plot of error in $p\parens{\xx,t}$]{\epsfig{file=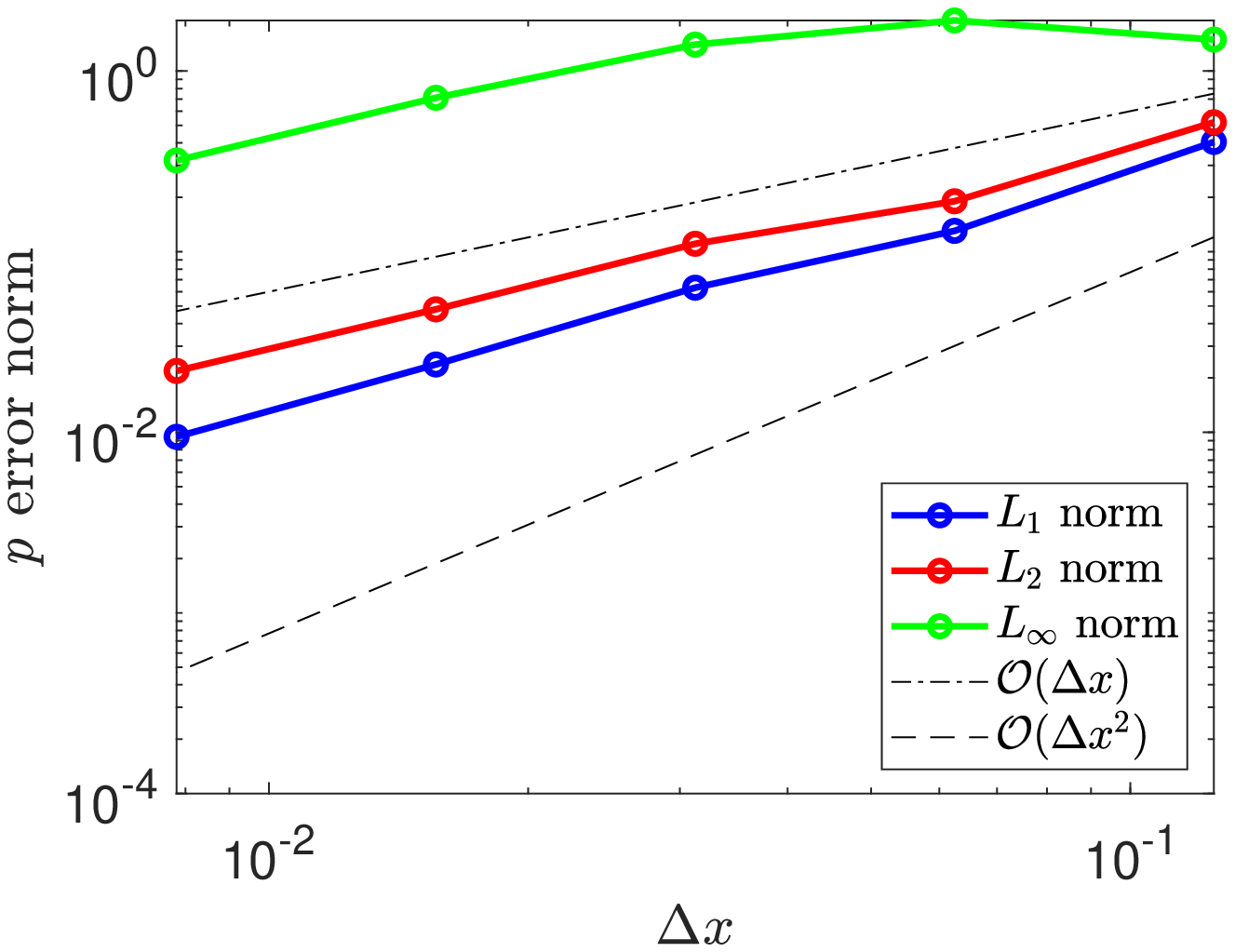, width=0.4\textwidth}}~ \qquad
    \subfloat[Log-log plot of error in $\thn\parens{\xx,t}$]{\epsfig{file=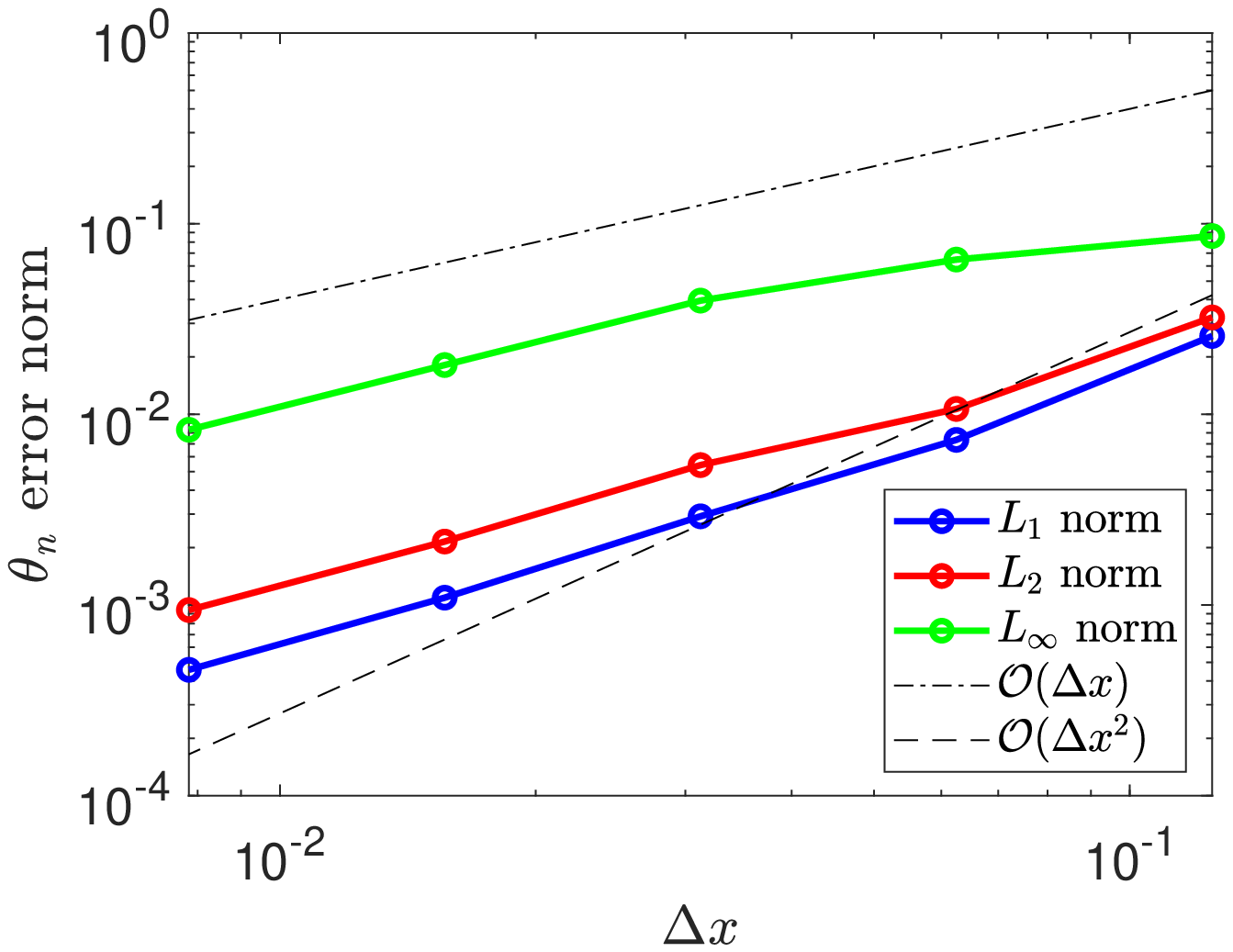, width=0.4\textwidth}}
    \end{center}
    \caption{Log-log plots indicating convergence behavior of the fluid discretization with advection of network volume fraction $\thn\parens{\xx,t}$ on a static locally-refined grid for each of the solution components:(a) $\vun\parens{\xx,t}$, (b) $\vus\parens{\xx,t}$, (c) $p\parens{\xx,t}$ and (d) $\thn\parens{\xx,t}$. The statically refined grid has an L-shaped refined region where the ratio of refinement between the coarse level and the refined level is $r=2$. We obtain convergence that is asymptotically approaching second order accuracy for all four solution components in the $L^1$, $L^2$ and $L^{\infty}$ norms.}
    \label{fig:advect_refGrid_error}
\end{figure}

As before, the error norms between the numerical and analytical solution for the solution variables, $\vun\parens{\xx,t}, \vus\parens{\xx,t}$, $p\parens{\xx,t}$, and $\thn\parens{\xx,t}$ are calculated on both uniform grids and grids with locally refined, static L-shaped regions. The error norms as a function of the grid size are plotted in \Cref{fig:advect_uniGrid_error,fig:advect_refGrid_error} for the simulations performed on uniform grids and static locally-refined grids, respectively. For the simulations performed using a uniform grid, the plots indicate that the discretization achieves second order accuracy for all solution variables across all error norms. For the simulations performed using a static locally-refined L-shaped grid, the discretization achieves between first and second order accuracy for all solution variables for each error norm. Note that this reduction in order of accuracy at coarse-fine interfaces is consistent with the first order discretization that is utilized at corners of the coarse-fine interface. This is illustrated in Figure \ref{fig:errors_at_cf} which shows the greatest error in the pressure at $t=0.4$ is occurring near the re-entrant corner of the refined region and along the lengths of the coarse-fine interfaces adjacent to this corner.

\subsection{Idealized Four Roll Mill}
This experiment demonstrates the ability of the AMR discretization to adaptively refine the mesh in critical regions of interest when simulating a two-phase flow model with a variable network volume fraction, $\thn\parens{\xx,t}$. We select the four roll mill first introduced by Taylor \cite{taylor1934} as our test problem for illustrating the AMR capabilities on an adaptive grid following the regridding criteria defined in \Cref{sec:regridding}. Here, the flow of the gel is driven by body forces, $\ffn\parens{\xx,t}$ and $\ffs\parens{\xx,t}$ acting on the network and solvent, respectively, resulting in the formation of vortices in each of the quadrants of the bi-periodic computational domain, $\Omega =[-0.5, 0.5] \times [-0.5, 0.5]$. The four roll mill has been used in other numerical studies of single-phase \cite{thomases2007, thomases2009} and two-phase viscous fluids \cite{wright2008}. The forces generate both rotational and extensional flow, with a stagnation point occuring at the center of the computational domain. For the viscoelastic fluid model, the extensional points generate large elastic stresses, which in single phase fluids with an Oldroyd-B model, can result in unbounded growth in the viscoelastic stress tensor for large enough Weissenberg numbers \cite{thomases2007}.

The initial network volume fraction is set to be a blob concentrated in a circular region of radius 0.175 centered at the origin as displayed in \Cref{fig:frm_AMR}. The exact form of the initial condition is

\begin{equation} \label{eq:blob_inital}
\thn\parens{\xx,0} = \frac{1}{4} + \frac{1}{4}\begin{cases}
      \cfrac{1989}{896\pi}\Bigg(1-\Big(\dfrac{x^2+y^2}{\delta^2}\Big)^4\Bigg)^4 \Bigg(4 \Big(\dfrac{x^2+y^2}{\delta^2}\Big)^4 + 1\Bigg)^4,       & \text{if} \quad x^2 + y^2 < \delta^2, \\
      0, & \text{otherwise,}
\end{cases}  
\end{equation}
in which $\delta = 0.175$.

The forcing functions $\ffn\parens{\xx,t}$ and $\ffs\parens{\xx,t}$ acting on the network and solvent, respectively, are defined as 
\begin{align} \label{FnAMR}
	\ffs\parens{\xx,t} = \ffn\parens{\xx,t} = 
\begin{bmatrix}
	\parens{2\pi \sin(2\pi x)\cos(2\pi x) + 8 \pi^2 \sin(2\pi x) \cos(2 \pi y)}  \\
	\parens{2\pi \sin(2\pi y)\cos(2\pi y) - 8 \pi^2 \sin(2\pi y) \cos(2 \pi x)}
\end{bmatrix}.
%\label{FsAMR}
\end{align}

We run two simulations, one with a Newtonian network and one with a viscoelastic network. The shear viscosity values ($\mun, \mus$) are selected such that the network is 100 times more viscous than the solvent. 

\begin{table}[hb] 
\centering
    \caption{Parameter values for viscous two-phase fluid simulation under four roll mill forcing.}
    \label{tab:frm_params}
    \begin{tabular}{|c c c|} 
        \hline
        \textbf{Parameter} & \textbf{Symbol}& \textbf{Value}\\ \hline
         Network Viscosity      &    $\mun$              &    4.0       \\ 
         Solvent viscosity      &    $\mus$              &    0.04       \\ 
         Drag coefficient       &    $\xi$                &    250 $\mus$   \\ 
         Density                &    $\rho$               &    1.0       \\
         Maximum CFL number     &    CFL             &    0.1         \\
          Gradient Threshold     &    $\varepsilon_\ell$  & 0.75, 0.75 \\
        \hline
    \end{tabular}
\end{table}

\subsubsection{Convergence of Newtonian Network}\label{sec:Newtonian_Network}

\begin{figure}[tb]
    \centering
    % NOTE: trim={<left> <lower> <right> <upper>}
    \subfloat[$t=0$.]{\epsfig{file=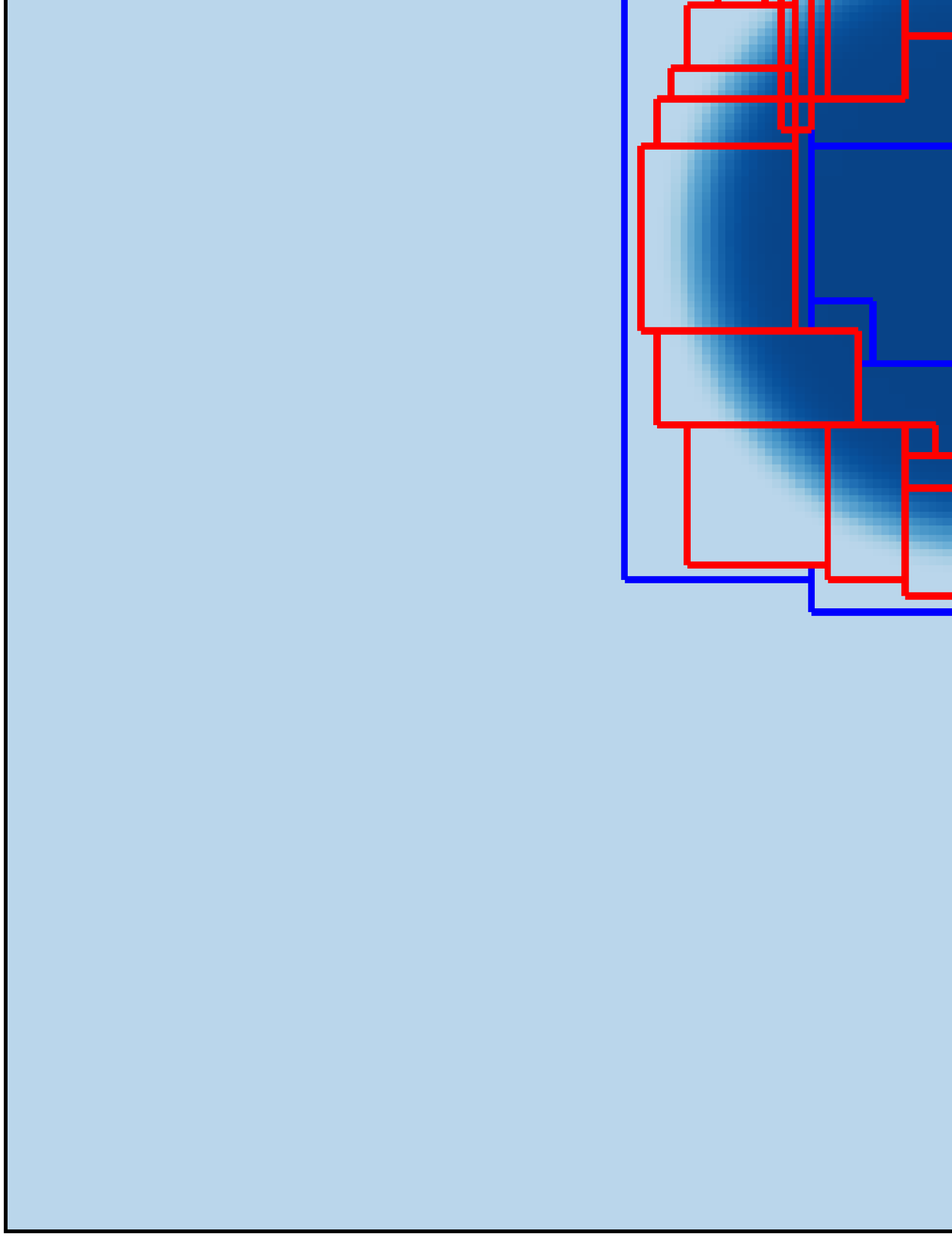, width=0.33\textwidth}}~
    \subfloat[$t=0.18$, $\size{\vun}^{\rm{max}} = 0.46$.]{\epsfig{file=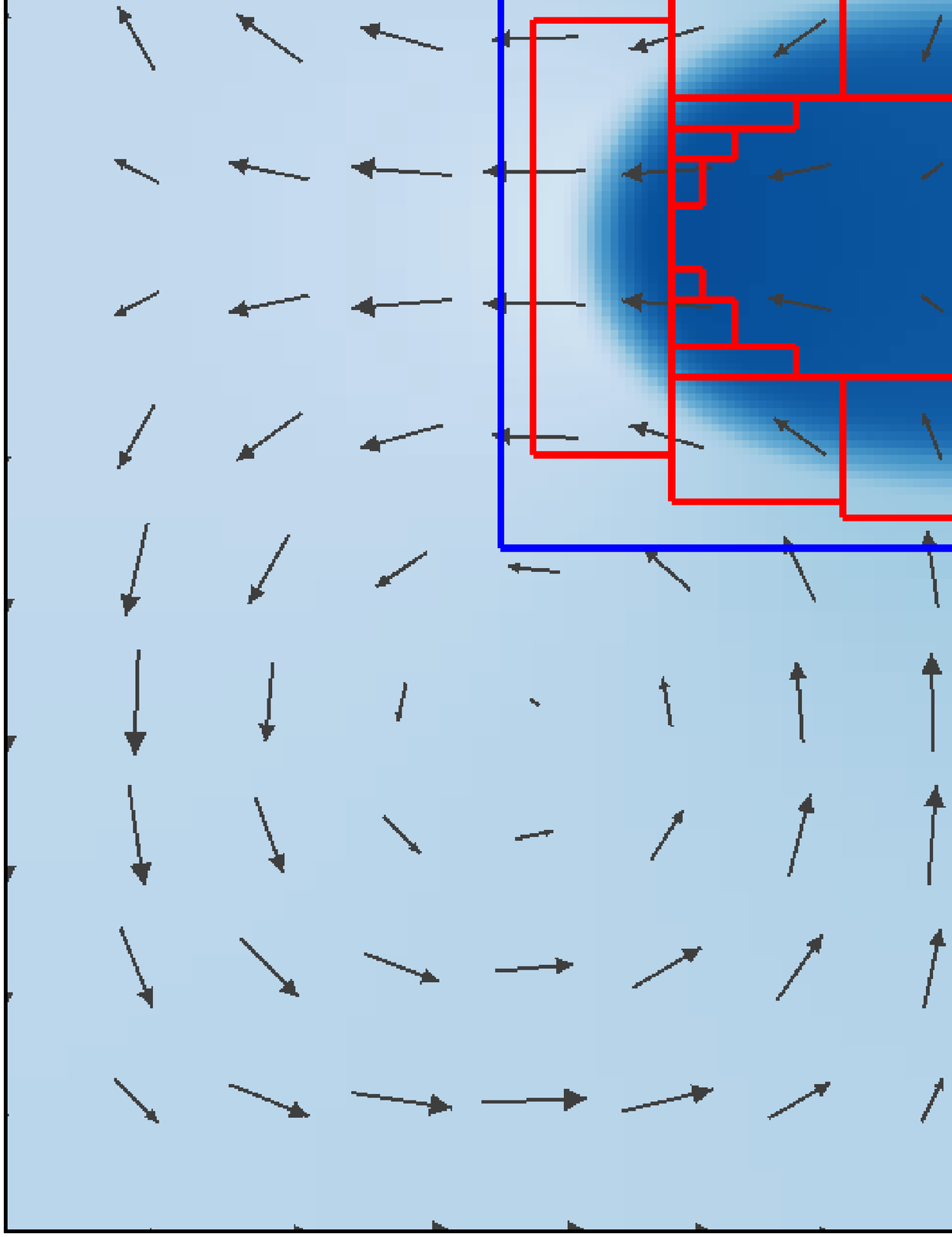, width=0.33\textwidth}}~
    \subfloat[$t=0.375$, $\size{\vun}^{\rm{max}} = 0.60$.]{\epsfig{file=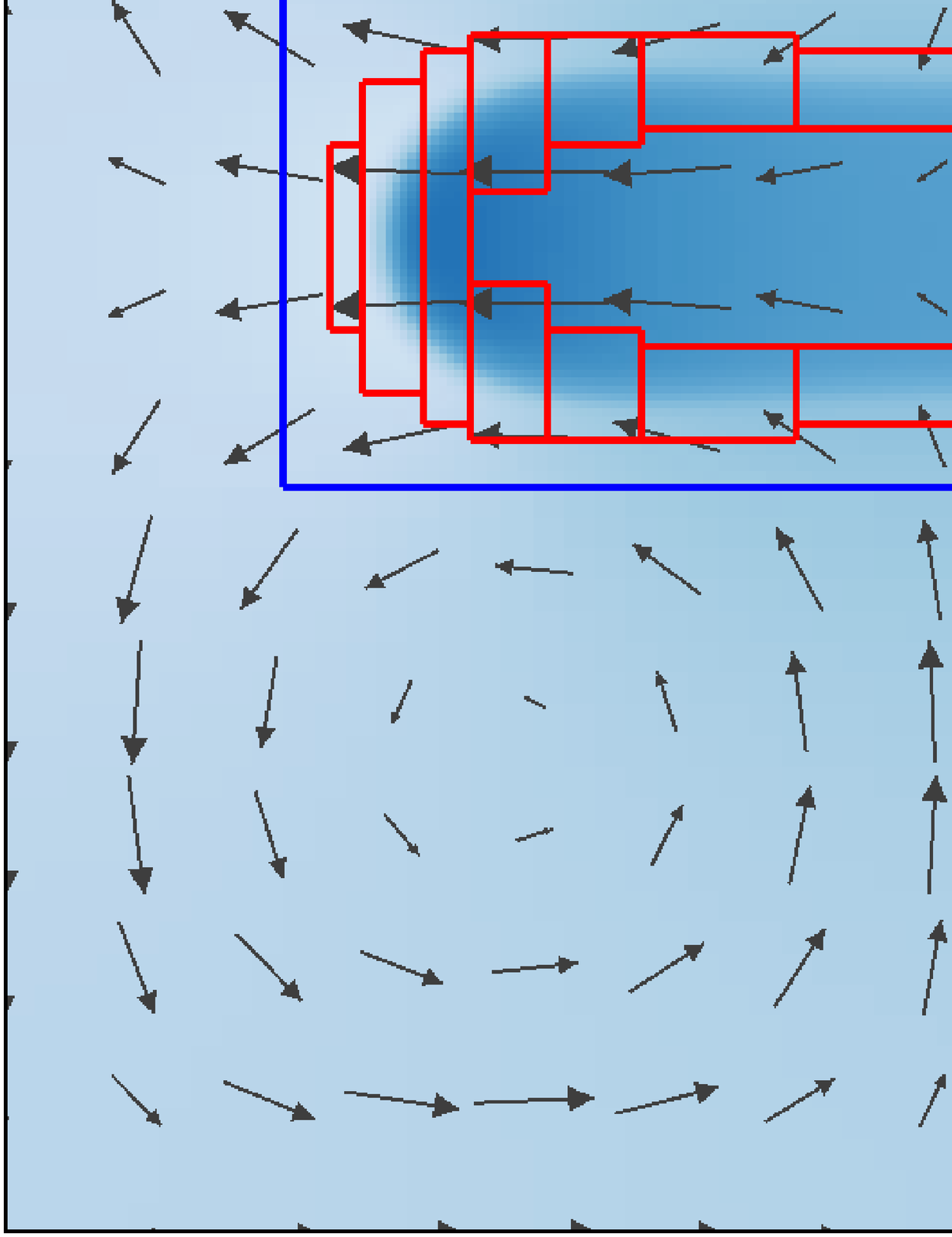, width=0.33\textwidth}}\\
    \subfloat[$t=0.52$, $\size{\vun}^{\rm{max}} = 0.61$.]{\epsfig{file=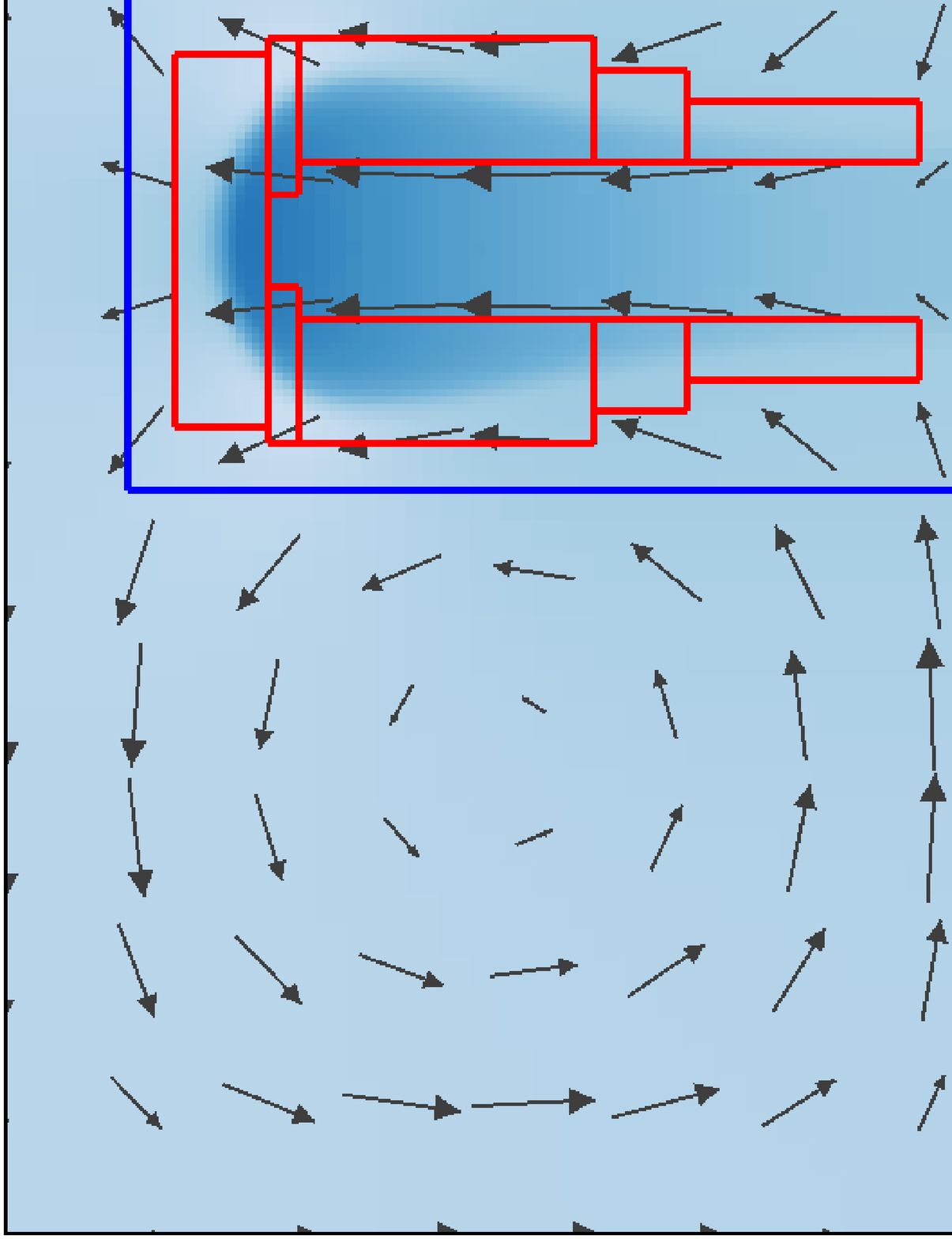, width=0.33\textwidth}}~
    \subfloat[$t=0.68$, $\size{\vun}^{\rm{max}} = 0.56$.]{\epsfig{file=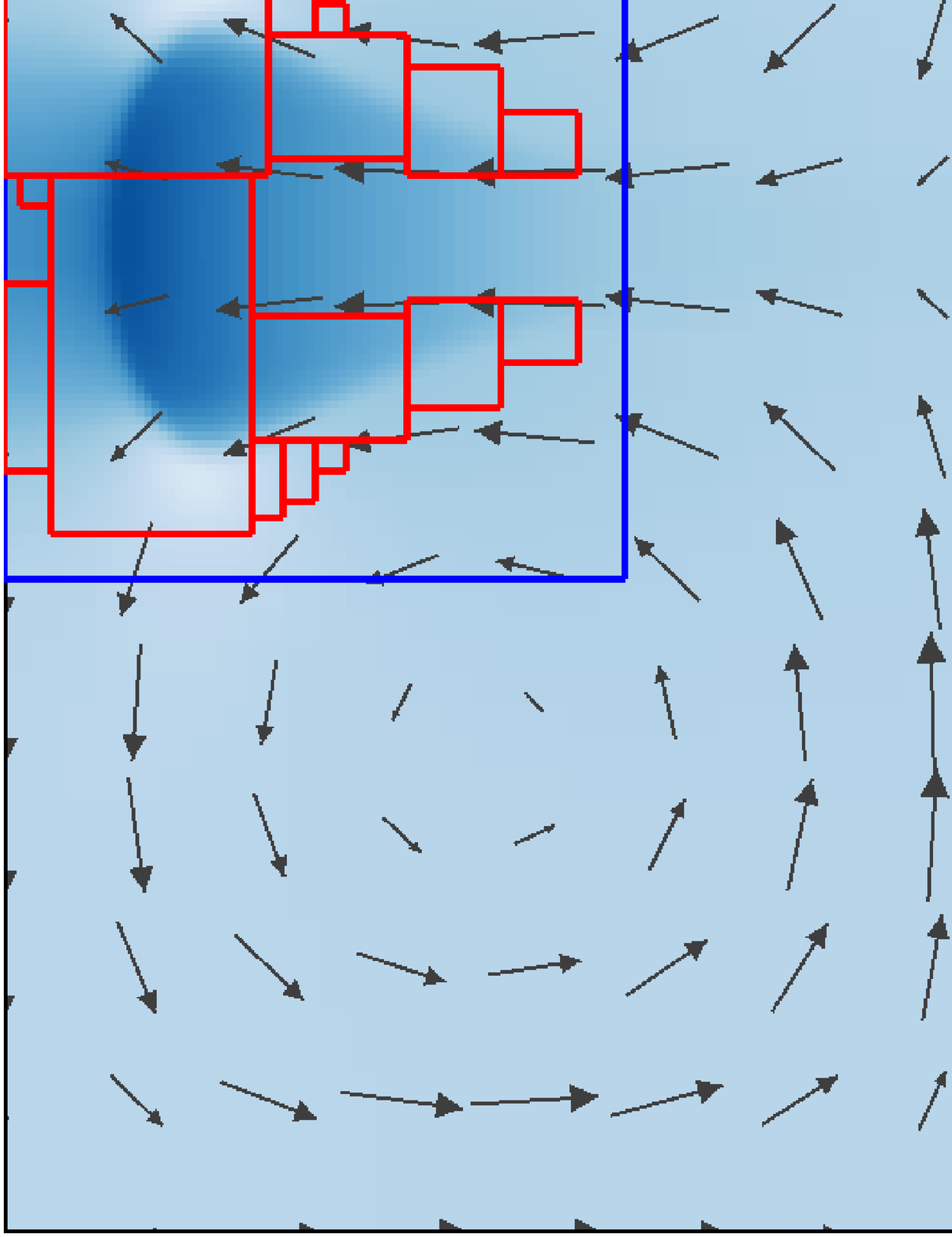, width=0.33\textwidth}}~
    \subfloat[$t=1.0$, $\size{\vun}^{\rm{max}} = 0.57$.]{\epsfig{file=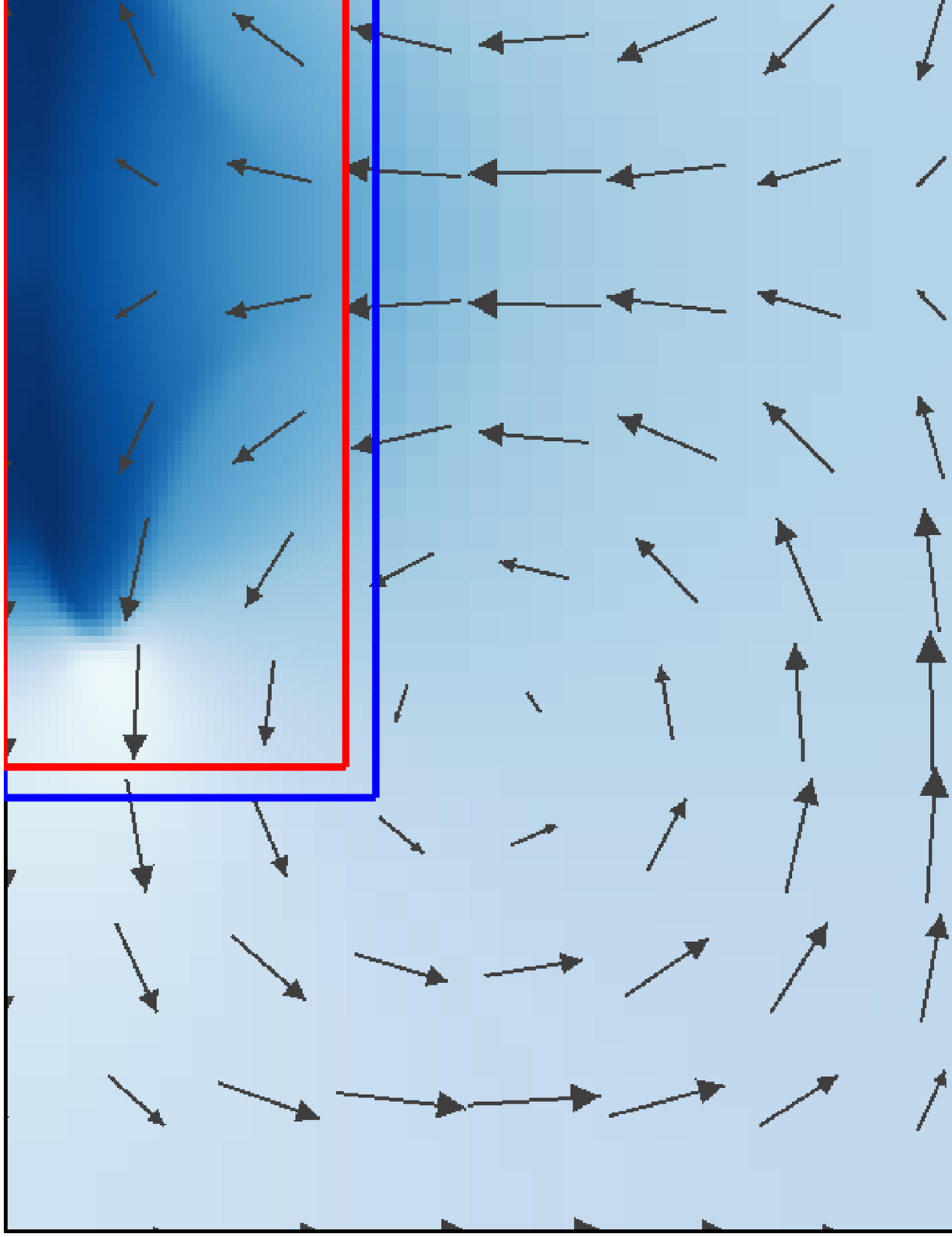, width=0.33\textwidth}}
    \caption{Simulation results for a blob of viscous network immersed in a viscous solvent bath, with the mixture driven by four roll mill forcing. The distribution of $\thn\parens{\xx,t}$ and $\vun\parens{\xx,t}$ at various times on an AMR grid are plotted, whereby the coarsest level grid size is $64 \times 64$. The grid is adaptively refined with up to 2 additional levels of refinement, as indicated by the blue and red levels, if the absolute gradient of $\thn\parens{\xx,t}$ exceeds 0.75 for both $\Omega^0$ and $\Omega^1$. Vectors at different times have the same scale.}
    \label{fig:frm_AMR}
\end{figure}

We simulate a two-phase viscous mixture consisting of a Newtonian network immersed in a solvent bath subject to four roll mill forcing with fluid parameters as outlined in \Cref{tab:frm_params}. The simulation employs an AMR grid and runs up to a final time of $T=1.0$. The plots of the distribution of $\thn\parens{\xx,t}$ and $\vun\parens{\xx,t}$ at different times are shown in \Cref{fig:frm_AMR}. The coarsest level, $\Omega^0$, of the AMR grid is of size 64-by-64 and covers the entire computational domain. The discretization uses two additional levels of refinement, $\Omega^1$ and $\Omega^2$, with coarse-to-fine refinement ratios of two between each level. These refined levels appear in the plot with blue and red outlines, respectively. The plots indicate that the fluid discretization is able to adaptively refine regions of interest, which in this case are areas where the absolute gradient of $\thn\parens{\xx,t}$ exceeds a threshold of 0.75 for both levels, $\Omega^0$ and $\Omega^1$. Initially, the refined levels cover the entire blob of network immersed in the solvent bath. As time goes on, the four roll mill forcing results in extension of the blob of network. As the blob extends horizontally, the network volume fraction accumulates at the ends of the elongated blob. The refined regions follow the edge of the blob, as it fully separates into two blobs. % Timing tests between an AMR grid and a uniform grid with a grid size equal to the highest refinement level in the AMR grid show that the AMR solver is 8.6x faster.

Since we do not know the analytical solution to two-phase viscous flow under four roll mill forcing, we approximate the order of convergence $k$ using Richardson extrapolation:
\begin{equation}\label{eq:richardson}
     k = \log_2\parens{\frac{\|f_{4h} - f_{2h}\|_p}{\|f_{2h}-f_{h}\|_p}},
\end{equation}
in which $f_{4h}$, $f_{2h}$, $f_h$ are the solutions on grids where the size of the coarsest level is increased, but the number of patch levels and the refinement ratios are held fixed.

Since we are dealing with adaptively refined grids, $f_h$ is the solution obtained on a computational domain whose coarsest level has a grid spacing of $h$. The norm of difference, $\size{f_h - f_{2h}}_p$ is calculated by interpolating the fine solution, $f_{h}$, onto the coarser AMR patch hierarchy, $f_{2h}$ whose coarsest level has a grid spacing of $2h$.

\begin{table}[tb]
\centering
\caption{Order of accuracy, k, at $t = 0.15$ using three solutions with coarsest level grid sizes (a) N = 16, N = 32, N = 64, and (b)  N = 32, N = 64, N = 128, corresponding to solutions $f_{4h}, f_{2h}, f_h$, respectively. The discretization is asymptotically approaching second order convergence in the $L^1$, $L^2$ and $L^\infty$ norm as the grids are refined.}
\setlength{\tabcolsep}{10pt}
\renewcommand{\arraystretch}{1.0}
\begin{tabular}{!{\vrule width 2pt} c !{\vrule width 2pt} c | c !{\vrule width 2pt} c | c !{\vrule width 2pt} c | c !{\vrule width 2pt}}
\hlineB{5}
\multicolumn{1}{!{\vrule width 2pt}c !{\vrule width 2pt}}{Component} & \multicolumn{6}{c !{\vrule width 2pt} }{\textbf{Order of accuracy, $k$}}                              \\ \noalign{\hrule height 2pt}
 \multicolumn{1}{!{\vrule width 2pt} c !{\vrule width 2pt}}{}   &  \multicolumn{2}{c !{\vrule width 2pt}}{\hspace*{2.5mm} \textbf{$L^1$ norm}  \hspace*{2.5mm} } & \multicolumn{2}{c !{\vrule width 2pt}}{\hspace*{2.5mm} \textbf{$L^2$ norm}  \hspace*{2.5mm}} & \multicolumn{2}{c !{\vrule width 2pt}}{\hspace*{2.5mm} \textbf{$L^\infty$ norm}  \hspace*{2.5mm} }\\ \hline
        & \textbf{(a)}  &  \textbf{(b)}  & \textbf{(a)}  & \textbf{(b)}  & \textbf{(a)}  & \textbf{(b)} \\ \hline   
$\vun\parens{\xx,t}$   &  1.713   &  1.796  &   1.738  &   1.912  &  1.162  &  1.943 \\ \hline   
$\vus\parens{\xx,t}$   &  1.773   &  2.220  &   1.784  &   2.293  &  1.384  &  2.012 \\ \hline 
$\thn\parens{\xx,t}$   &  0.967   &  1.652  &   1.375  &   1.966  &  1.128  &  2.000 \\ \hline
$p\parens{\xx,t}$     &   1.489   &  1.601  &   1.032  &   1.643  &  0.770  &  1.457 \\ \hlineB{5} 
\end{tabular}
\label{table:frm_accuracy_t022}
\end{table}

\begin{table}[tb]
\centering
\caption{Order of accuracy, k, at $t = 0.18$ using three solutions with coarsest level grid sizes (a) N = 16, N = 32, N = 64, and (b)  N = 32, N = 64, N = 128, corresponding to solutions $f_{4h}, f_{2h}, f_h$, respectively. The discretization is asymptotically approaching second order convergence in the $L^1$, $L^2$ and $L^\infty$ norm as the grids are refined.}
\setlength{\tabcolsep}{10pt}
\renewcommand{\arraystretch}{1.0}
\begin{tabular}{!{\vrule width 2pt} c !{\vrule width 2pt} c | c !{\vrule width 2pt} c | c !{\vrule width 2pt} c | c !{\vrule width 2pt}}
\hlineB{5}
\multicolumn{1}{!{\vrule width 2pt}c !{\vrule width 2pt}}{Component} & \multicolumn{6}{c !{\vrule width 2pt} }{\textbf{Order of accuracy, $k$}}                              \\ \noalign{\hrule height 2pt}
 \multicolumn{1}{!{\vrule width 2pt} c !{\vrule width 2pt}}{}   &  \multicolumn{2}{c !{\vrule width 2pt}}{\hspace*{2.5mm} \textbf{$L^1$ norm}  \hspace*{2.5mm} } & \multicolumn{2}{c !{\vrule width 2pt}}{\hspace*{2.5mm} \textbf{$L^2$ norm}  \hspace*{2.5mm}} & \multicolumn{2}{c !{\vrule width 2pt}}{\hspace*{2.5mm} \textbf{$L^\infty$ norm}  \hspace*{2.5mm} }\\ \hline
        & \textbf{(a)}  &  \textbf{(b)}  & \textbf{(a)}  & \textbf{(b)}  & \textbf{(a)}  & \textbf{(b)} \\ \hline
$\vun\parens{\xx,t}$ & 2.054 &  1.714   & 2.117 &  1.791   & 2.683 &  1.024  \\ \hline   
$\vus\parens{\xx,t}$ & 2.201 &  2.131   & 2.244 &  2.144   & 1.955 &  1.621  \\ \hline 
$\thn\parens{\xx,t}$ & 0.983 &  1.620   & 1.184 &  1.931   & 0.680 &  2.015  \\ \hline
$p\parens{\xx,t}$    & 1.232 &  1.680   & 0.857 &  1.685   & 0.254 &  1.682  \\ \hlineB{5}
\end{tabular}
\label{table:frm_accuracy_t025}
\end{table}

\begin{figure}[ht!]
    \begin{center}
    % NOTE: trim={<left> <lower> <right> <upper>}
    \subfloat[Pressure at t = 0.2]{\epsfig{file=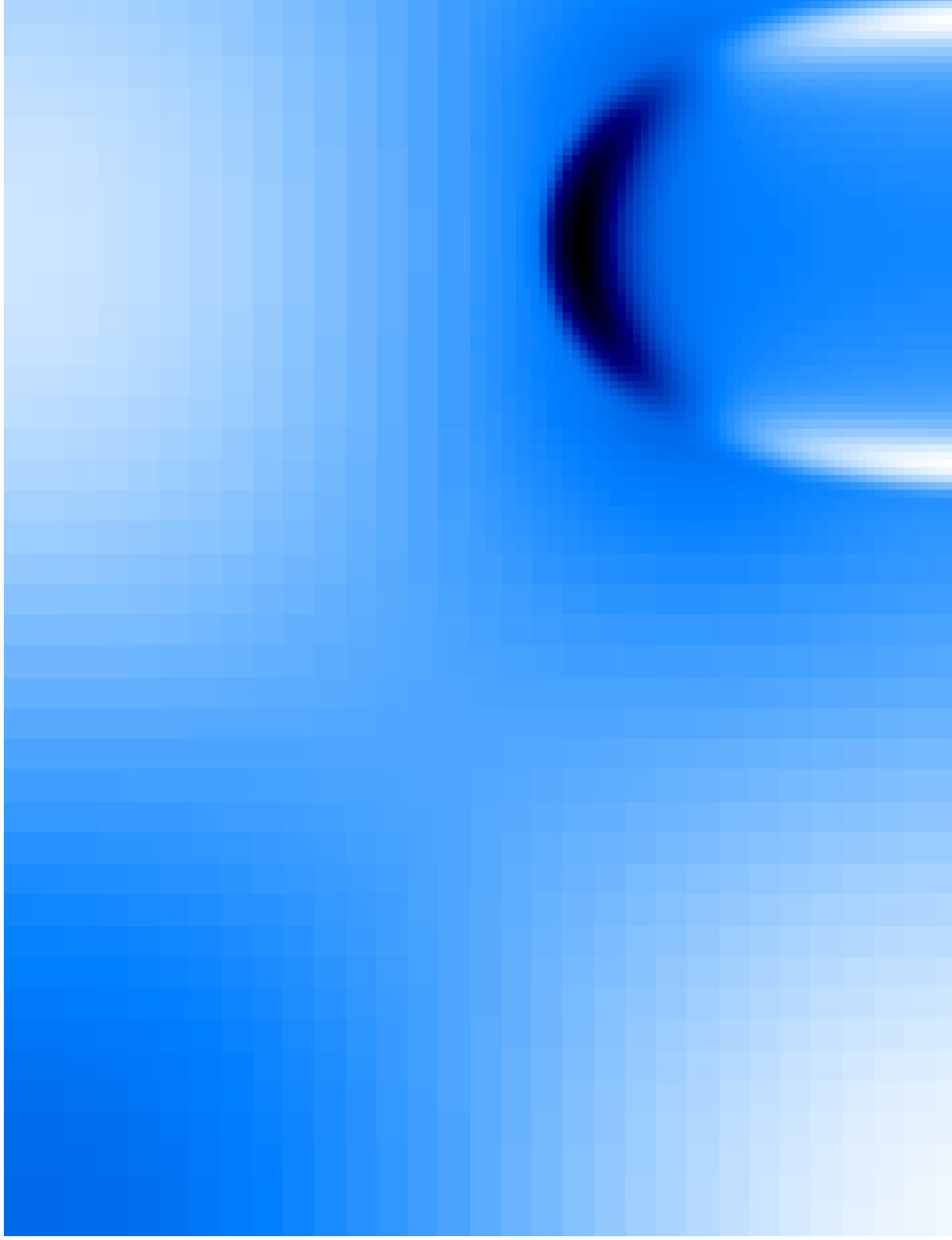, width=0.4\textwidth}}~
    \subfloat[Pressure along the x-axis at t = 0.2]{\epsfig{file=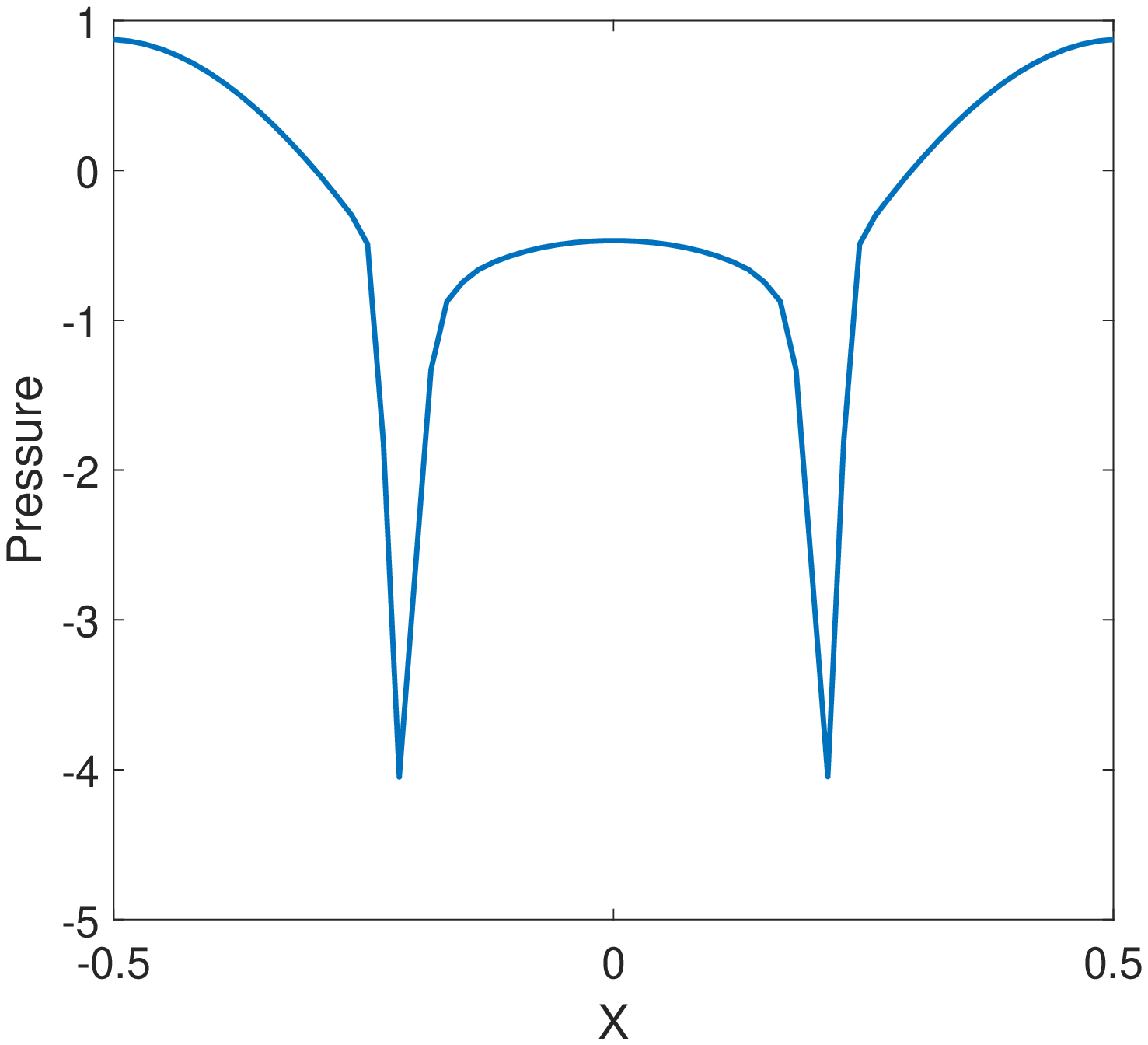, width=0.4\textwidth}}\\
    \subfloat[Pressure at t = 0.4]{\epsfig{file=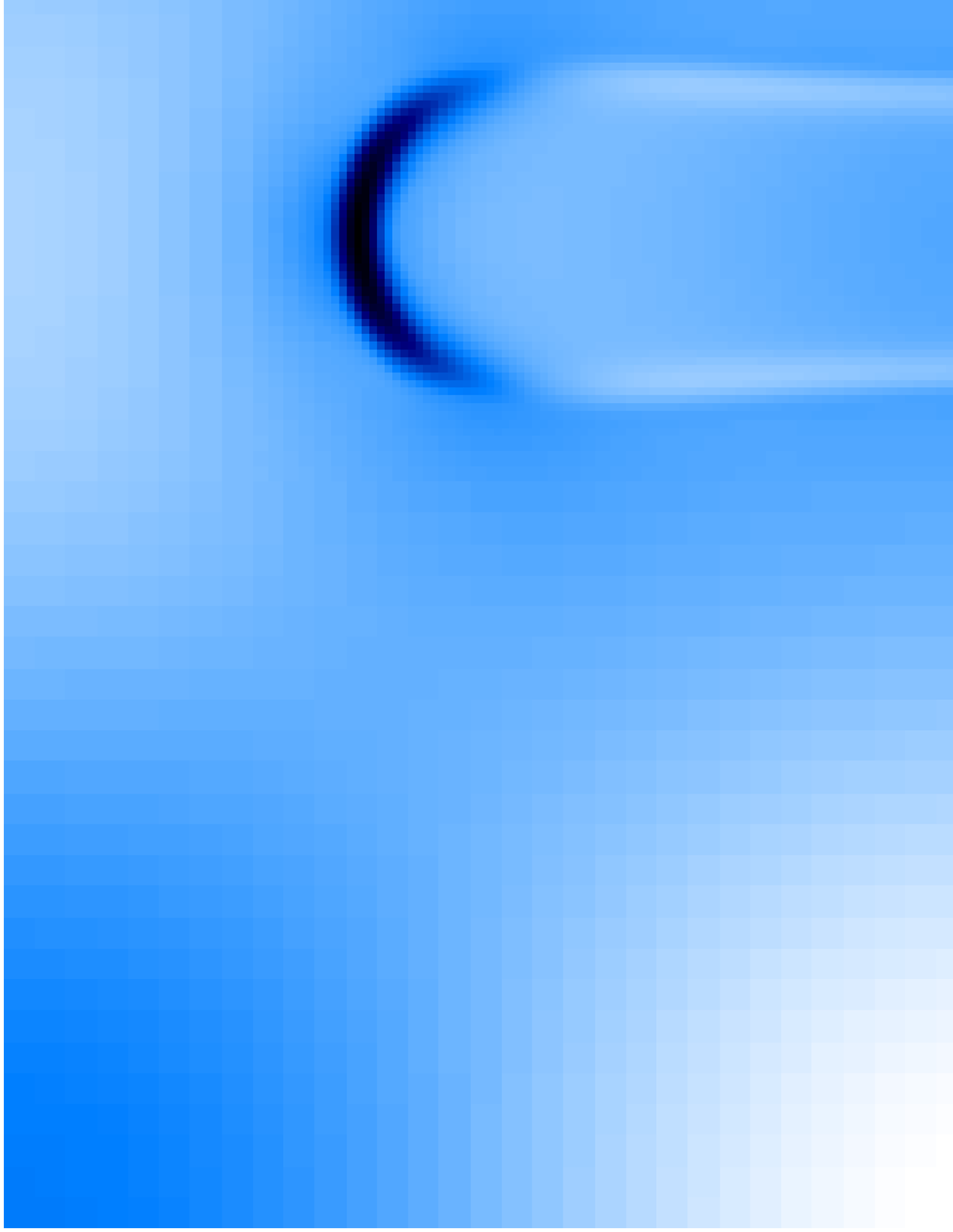, width=0.4\textwidth}}~
    \subfloat[Pressure along the x-axis at t = 0.4]{\epsfig{file=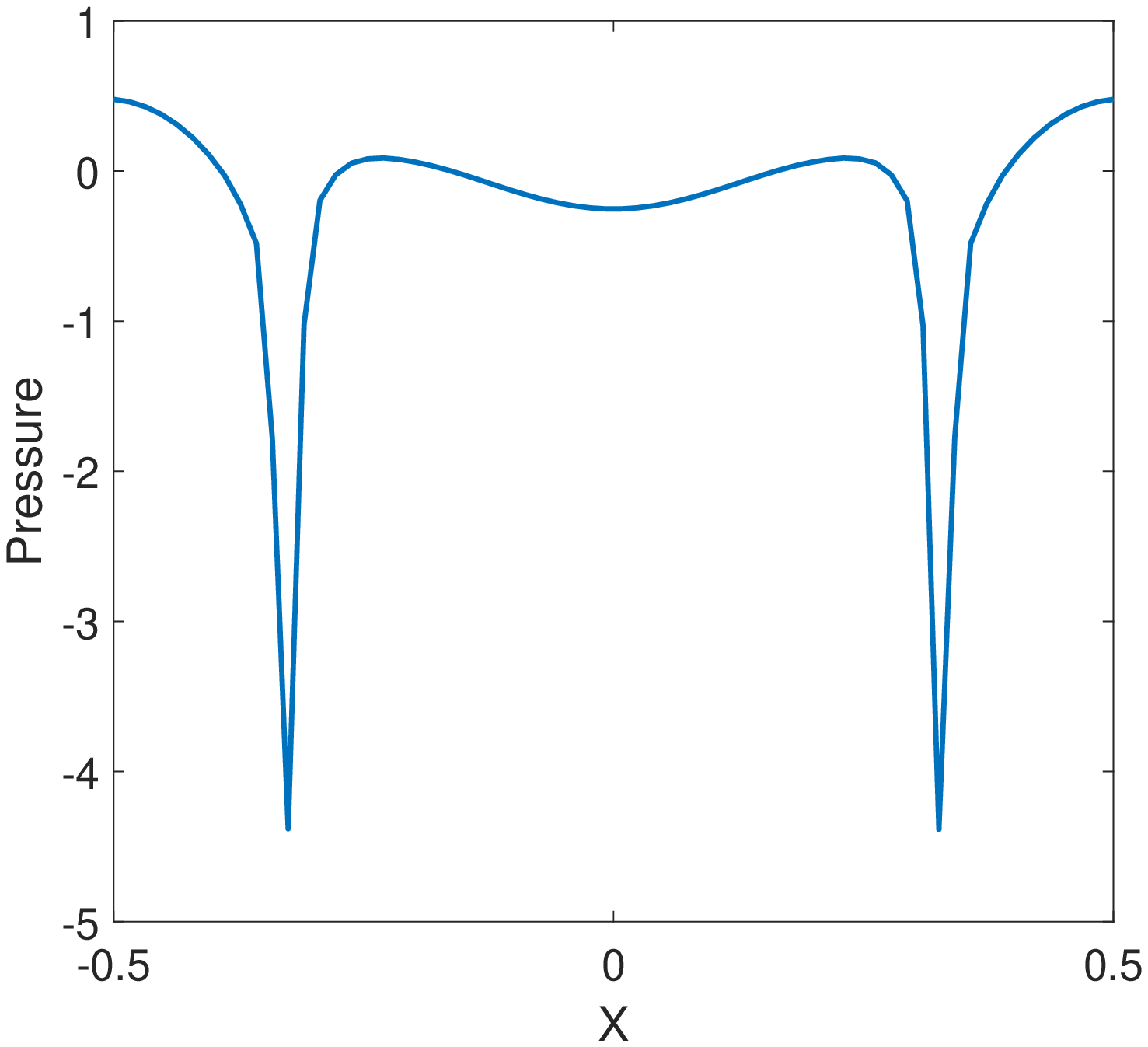, width=0.4\textwidth}}
    \end{center}
    \caption{The (a,c) pressure distribution over the whole domain and (b,d) pressure distribution along the x-axis at two different times for the Newtonian network under four roll mill forcing simulated on an AMR grid. The sharp pressure gradients can be observed in the 1D pressure distribution plots in (b,d).}
    \label{fig:sharp_pressure}
\end{figure}

The order of accuracy values at $t=0.15$ and $t=0.18$ are presented in \Cref{table:frm_accuracy_t022,table:frm_accuracy_t025}. These times are selected for computing convergence behavior because these are times before the blob begins to separate. Beyond these times, pressures with sharp gradients form as the blob fully separates, as shown in Figure \ref{fig:sharp_pressure}, and these limit the convergence of the method. We expect that with sufficient resolution, second order accuracy will be restored. The data in these tables shows that the discretization achieves convergence for all components across all norms, whereby the discretization is asymptotically approaching second order convergence in the $L^1$, $L^2$ and $L^\infty$ norm as the grids are refined.

\begin{figure}[h!]
    \centering
    \epsfig{file=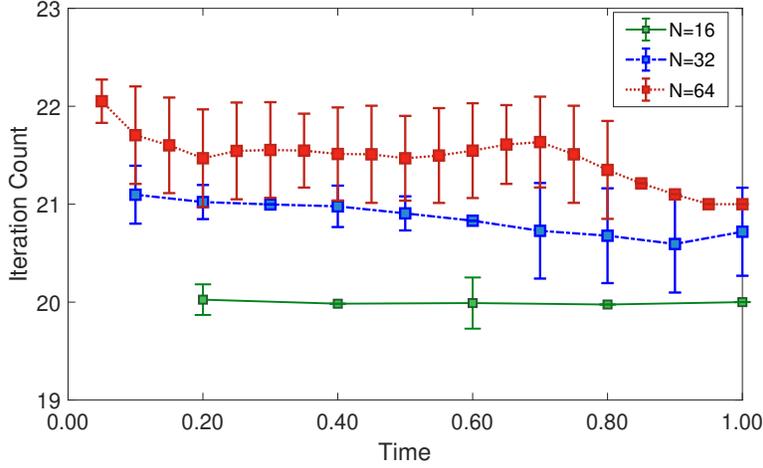, width=0.7\textwidth}
    \caption{A rolling average (computed every $640\Delta t$) of the number of iterations against the time to converge to a relative residual tolerance of $10^{-14}$ for the blob of viscous network immersed in a viscous solvent bath under four roll mill forcing simulated using different AMR grid sizes. $N$ refers to the size of the coarsest level grid on the patch hierarchy.  Since the iteration count is confined to a narrow range of 19 to 23, there is clear evidence that the number of iterations required for the linear solver to converge at each time step is independent of the grid spacing.}
    \label{fig:num_iter_frm}
\end{figure}

To illustrate that the number of iterations required for the linear solver to converge at each time step is independent of grid spacing, we plot the number of iterations it takes for the linear solver to achieve a relative residual tolerance of $10^{-14}$ against the time step number. We calculate a rolling average by computing the mean number of iterations needed to converge every 640 time steps. \Cref{fig:num_iter_frm} shows that the linear solver remains relatively consistent throughout the simulation time for a given AMR grid, and the number of iterations taken to achieve convergence at each time step is very similar across the different grids, with the value always falling between 19 to 23 iterations. This indicates that the multigrid solver serves as an effective preconditioner on adaptive grids.

\subsubsection{Convergence of Viscoelastic Network}
\begin{figure}[h!]
\centering
    % NOTE: trim={<left> <lower> <right> <upper>}
\begin{center}
    \subfloat[$\text{tr}(\CC)\parens{\xx,t}$ at $t=0.5$, $\size{\vun}^{\rm{max}}= 0.52$.]{\epsfig{file=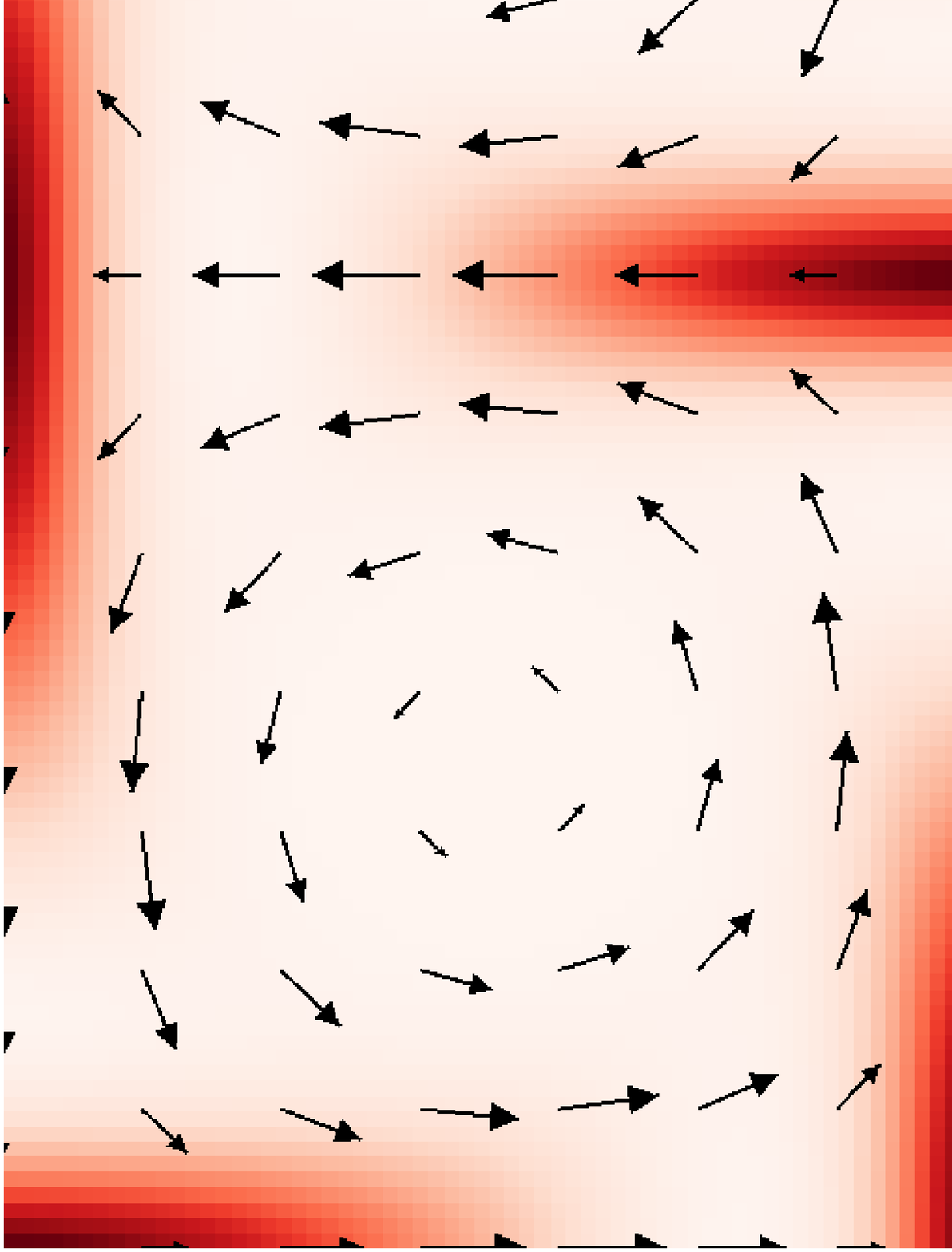, width=0.4\textwidth}}
    ~
    \subfloat[$\thn\parens{\xx,t}$ at $t=0.5$, $\size{\vun}^{\rm{max}}=0.52$.]{\epsfig{file=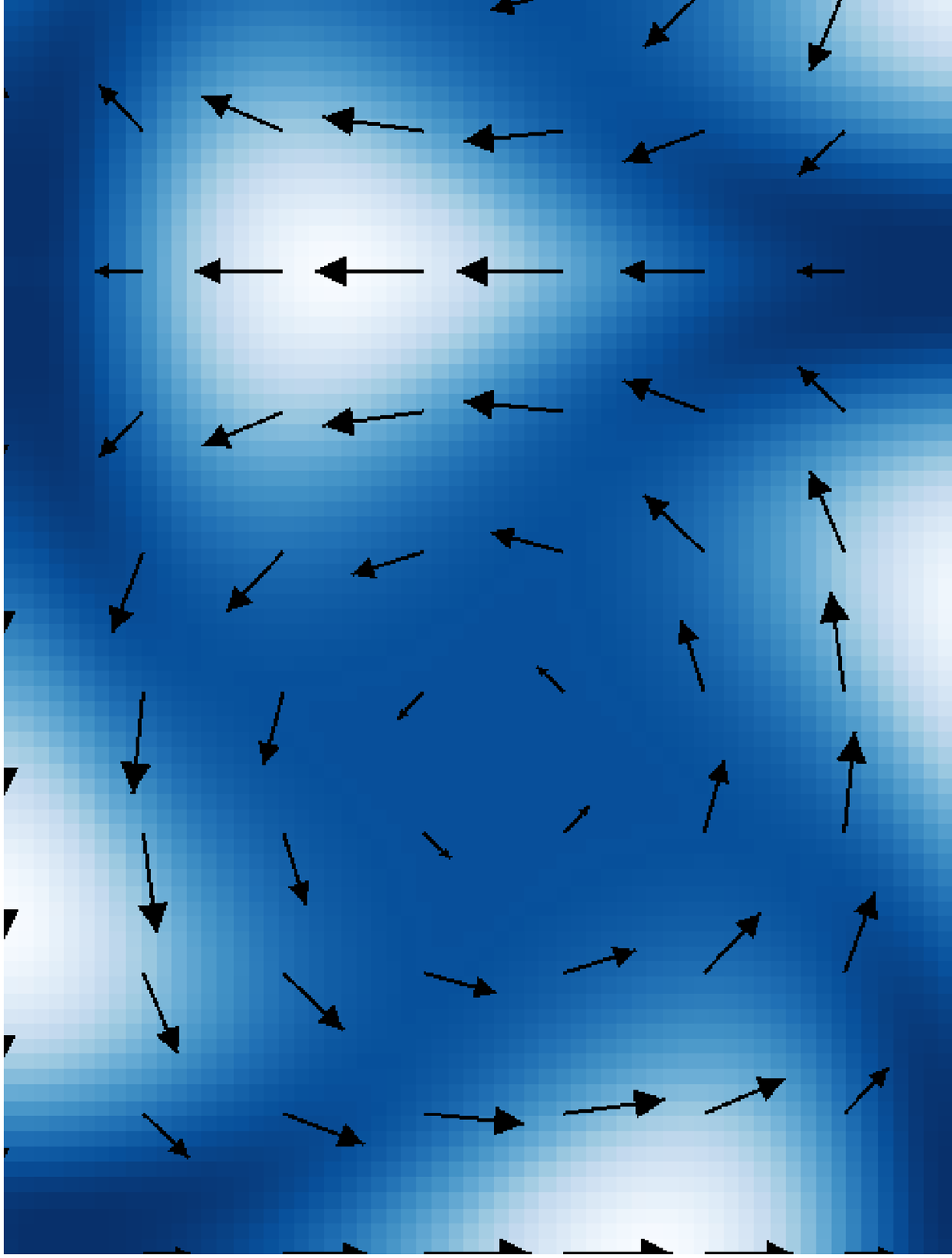, width=0.4\textwidth}} 
    \\
    \subfloat[$\text{tr}(\CC)\parens{\xx,t}$ at $t=1.0$, $\size{\vun}^{\rm{max}}=0.22$.]{\epsfig{file=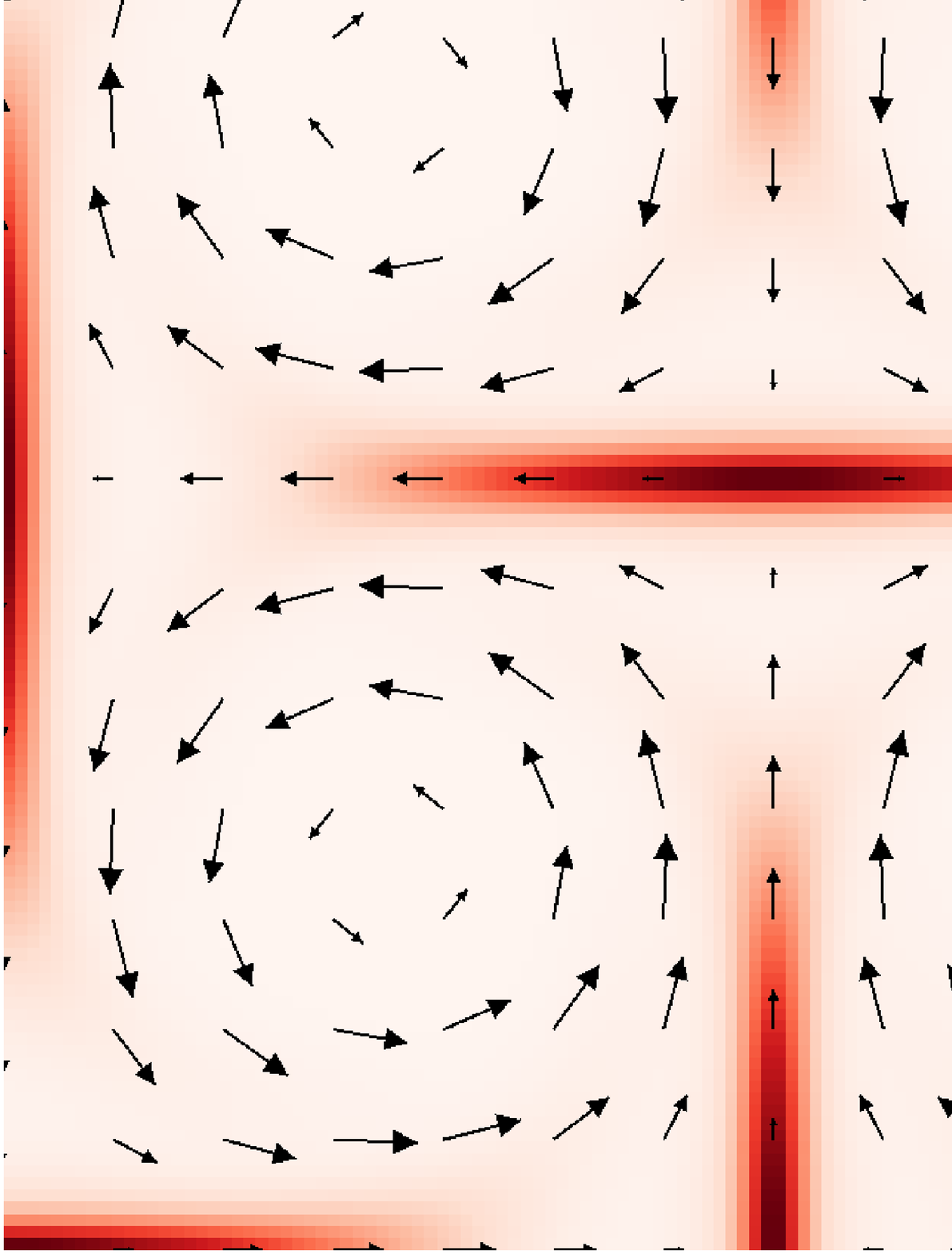, width=0.4\textwidth}}
    ~
    \subfloat[$\thn\parens{\xx,t}$ at $t=1.0$, $\size{\vun}^{\rm{max}}=0.22$.]{\epsfig{file=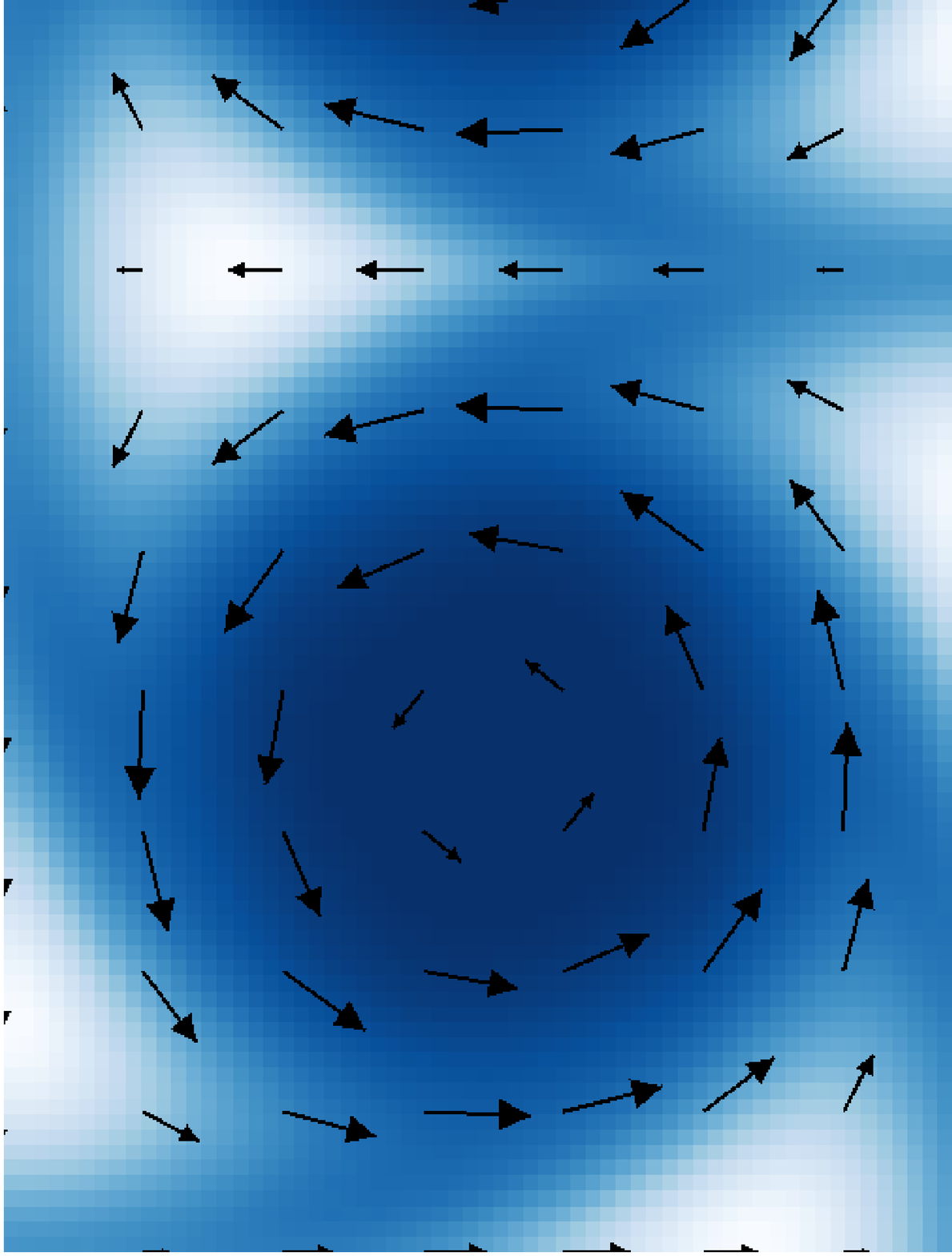, width=0.4\textwidth}} 
\end{center}
    \caption{The distribution of (a,c) the trace of $\CC\parens{\xx,t}$ and (b,d) $\thn\parens{\xx,t}$ at $t=0.5$ and $t=1.0$ along with the network velocity field, $\vun\parens{\xx,t}$, on a uniform grid of size $128 \times 128$. Initially, $\thn\parens{\xx,t}$ is  distributed uniformly across the entire domain and gets advected due to four roll mill forcing, resulting in the distribution shown in (b,d). The background forcing causes the multiphase fluid to elongate along the x-axis and compress along the y-axis. This is apparent from the distribution of $\text{tr}(\CC)\parens{\xx,t}$, which develops cusps (identified by taking a cross-section across the dark shaded regions to reveal a symmetric curve with a singularity point) as the simulation proceeds, and achieves the greatest values at locations with stagnation points. Vectors at different times do not have the same scale.}
    
    \label{fig:cf_frm_uni_tr_thn}
\end{figure}

We simulate a two-phase viscoelastic fluid that is subject to four roll mill forcing using a uniform grid to showcase the convergence behavior of our discretization. The advected quantities, $\thn\parens{\xx,t}$ and $z\parens{\xx,t}$, are initially set to spatially-uniform values of 0.5 and 1.0, respectively, and are advected with the network velocity field.  The stress tensor, $\CC\parens{\xx,t}$, is initially set to the identity tensor. The fluid parameters are outlined in \Cref{tab:cf_frm_uni}. This set-up was selected because, while it is a relatively simple set of initial conditions, the simulation produces non-trivial results as shown in \Cref{fig:cf_frm_uni_tr_thn}. The order of accuracy $k$ at two different times is obtained using Richardson's extrapolation as in \cref{eq:richardson}. Convergence results are presented in \Cref{table: cf_frm_uni_accuracy_t_05,table: cf_frm_uni_accuracy_t_1} and the corresponding distributions of the trace of the stress tensor and the network volume fraction are displayed in \Cref{fig:cf_frm_uni_tr_thn}.

\begin{table}[hb]
\centering
    \caption{Parameter values for the two-phase viscoelastic fluid under four roll mill forcing on a uniform grid.}
    \begin{tabular}{|c c c|} 
    \hline
        \textbf{Parameter} & \textbf{Symbol}& \textbf{Value}\\ \hline
         Network Viscosity      &    $\mun$              &    2.0       \\ 
         Solvent Viscosity      &    $\mus$              &    0.04       \\ 
         Polymeric contribution to Viscosity & $\munp$  &    2.0     \\
         Polymer Relaxation Time &   $\lambda$            &    2.0 \\
         Drag Coefficient       &    $\xi$                &    250 $\mus$   \\ 
         Density                &    $\rho$               &    1.0       \\        
    \hline
    \end{tabular}
    \label{tab:cf_frm_uni}
\end{table}

\begin{table}[tb]
\centering
\caption{Order of accuracy $k$ at $t = 0.5$ using three solutions with coarsest level grid sizes: $N = 64$, $N = 128$, $N = 256$, corresponding to solutions $f_{4h}, f_{2h}, f_h$, respectively. The discretization obtains second order accuracy in the $L^1$ and $L^2$ norms and attains at least 1st order accuracy in the $L^{\infty}$ norm, with most variables approaching second order accuracy in the $L^{\infty}$ norm.}
\renewcommand{\arraystretch}{1.0}
\begin{tabular}{|| c | c |  c |  c ||}
\hline
\multicolumn{1}{||c |}{Component} & \multicolumn{3}{c| |}{\textbf{Order of accuracy, $k$}}                                                   \\ \hline
 \multicolumn{1}{|| c |}{}         &  \multicolumn{1}{c |}{\hspace*{2.5mm} \textbf{$L^1$ norm}  \hspace*{2.5mm} } & \multicolumn{1}{c |}{\hspace*{2.5mm} \textbf{$L^2$ norm}  \hspace*{2.5mm}} & \multicolumn{1}{c ||}{\hspace*{2.5mm} \textbf{$L^\infty$ norm}  \hspace*{2.5mm} }\\ \hline
$\vun\parens{\xx,t}$    &   2.166   &   2.502     &  2.401   \\ \hline   
$\vus\parens{\xx,t}$    &   2.069   &   2.112     &  2.450   \\ \hline 
$\thn\parens{\xx,t}$    &   2.325   &   2.257     &  1.627   \\ \hline
$p\parens{\xx,t}$       &   2.237   &   2.262     &  1.716   \\ \hline 
$\sigma_{xx}\parens{\xx,t}$  &   2.374   &   2.176     &  1.615   \\ \hline
$\sigma_{yy}\parens{\xx,t}$  &   2.374   &   2.176     &  1.615   \\ \hline
$\sigma_{xy}\parens{\xx,t}$  &   1.996   &   1.755     &  1.325   \\ \hline \hline
\end{tabular}
\label{table: cf_frm_uni_accuracy_t_05}
\end{table}

\begin{table}[tb]
\centering
\caption{Order of accuracy $k$ at $t = 1.0$ using three solutions with coarsest level grid sizes: $N = 64$, $N = 128$, $N = 256$, corresponding to solutions $f_{4h}, f_{2h}, f_h$, respectively. The discretization obtains second order accuracy in the $L^1$ and $L^2$ norms and attains at least 1st order accuracy in the $L^{\infty}$ norm, with most variables approaching second order accuracy in the $L^{\infty}$ norm.}
\renewcommand{\arraystretch}{1.0}
\begin{tabular}{|| c | c |  c |  c ||}
\hline
\multicolumn{1}{||c |}{Component} & \multicolumn{3}{c| |}{\textbf{Order of accuracy, $k$}}                                                   \\ \hline
 \multicolumn{1}{|| c |}{}         &  \multicolumn{1}{c |}{\hspace*{2.5mm} \textbf{$L^1$ norm}  \hspace*{2.5mm} } & \multicolumn{1}{c |}{\hspace*{2.5mm} \textbf{$L^2$ norm}  \hspace*{2.5mm}} & \multicolumn{1}{c ||}{\hspace*{2.5mm} \textbf{$L^\infty$ norm}  \hspace*{2.5mm} }\\ \hline
$\vun\parens{\xx,t}$    &   2.253   &   1.980     &  1.709  \\ \hline   
$\vus\parens{\xx,t}$    &   1.876   &   1.917     &  1.809  \\ \hline 
$\thn\parens{\xx,t}$    &   1.982   &   1.861     &  1.605  \\ \hline
$p\parens{\xx,t}$       &   1.841   &   1.822     &  1.800  \\ \hline 
$\sigma_{xx}\parens{\xx,t}$  &   2.178   &   2.161     &  1.810  \\ \hline
$\sigma_{yy}\parens{\xx,t}$  &   2.178   &   2.161     &  1.811  \\ \hline
$\sigma_{xy}\parens{\xx,t}$  &   2.030   &   1.838     &  1.537   \\ \hline \hline
\end{tabular}
\label{table: cf_frm_uni_accuracy_t_1}
\end{table}

From these convergence studies, we see that the discretization obtains second order accuracy for velocities, volume fraction, pressure, and components of the stress tensor in the $L^1$ and $L^2$ norms. We attain at least 1st order accuracy in the $L^{\infty}$ norm, with most variables approaching second order of accuracy in the $L^{\infty}$ norm as well. In particular, the off-diagonal component, $\sigma_{xy}\parens{\xx,t}$, of the stress tensor only achieves an order of accuracy of 1.5 in the $L^{\infty}$ at the final time. Note that the advection scheme utilized here reduces to a first order approximation near regions of large gradients or discontinuities, limiting pointwise accuracy to first order.

Plots of the distribution of the trace of the stress tensor, $\text{tr}(\CC\parens{\xx,t})$, along with the network volume fraction, $\thn\parens{\xx,t}$, are shown in \Cref{fig:cf_frm_uni_tr_thn}. The background forcing due to the four roll mill is apparent by the symmetrical distribution of $\thn\parens{\xx,t}$. The lowest $\thn\parens{\xx,t}$ occurs in regions where the magnitude of $\vun\parens{\xx,t}$ is greatest since it is being advected away by the high network velocity, and the greatest $\thn\parens{\xx,t}$ accumulates in regions where the magnitude of $\vun\parens{\xx,t}$ is the lowest. We observe formation of cusps in $\text{tr}(\CC\parens{\xx,t})$ as the simulation proceeds, and the value of $\text{tr}(\CC\parens{\xx,t})$ is greatest at locations with stagnation points. These also happen to be locations where $\thn\parens{\xx,t}$ is observed to accumulate the most by virtue of being advected towards these stagnation points by the background four roll mill forcing.

\subsubsection{Relaxation of Viscoleastic Network}\label{sec:viscoelastic_Network}
We demonstrate the robustness and efficiency of our computational AMR discretization by simulating the dynamics of a viscoelastic material. Simulations such as the one presented in this section are beneficial in the study of soft matter, such as gels, elastomers and biological tissue, which have numerous applications in nature and engineering due to their high versatility. To that end, we simulate a multiphase viscoelastic fluid model based on those used in clotting models \cite{guy2008,du2018}. We subject this fluid model to four roll mill forcing and simulate it's behaviour using an AMR grid up to time $T=2.0$, with fluid parameters as outlined in \Cref{tab:cf_frm_amr}. The parameters are selected to match the fluid parameters of the four roll mill simulation presented in Section \ref{sec:Newtonian_Network}. In particular, the network viscosity is split equally into a viscous contribution and a polymeric contribution whose sum is equal to the viscosity of the Newtonian network simulated previously. The AMR grid used in the simulation has a total of four refinement levels, $\Omega^0$ to $\Omega^3$, starting with a 32-by-32 grid on the coarsest level which covers the entire computational domain $\Omega =[-0.5, 0.5] \times [-0.5, 0.5]$. The refinement ratios are 4, 2, and 2, starting from the coarsest level to the finest level. The initial network volume fraction is set to be concentrated in a circular region of radius 0.175 centered at the origin, as defined in \cref{eq:blob_inital}. The forcing functions acting on the network and solvent are 
\begin{align}
	\ffs\parens{\xx,t} = \ffn\parens{\xx,t} &=
\begin{cases}
    \begin{bmatrix}
    	\parens{2\pi \sin(2\pi x)\cos(2\pi x) + 8 \pi^2 \sin(2\pi x) \cos(2 \pi y)}  \\
    	\parens{2\pi \sin(2\pi y)\cos(2\pi y) - 8 \pi^2 \sin(2\pi y) \cos(2 \pi x)} \\
    \end{bmatrix}, & \text{if} \quad t < 0.20, \\
    \quad \mathbf{0},  & \text{otherwise,}
\end{cases} \label{Fn_AMR}
\end{align}
The background forces are conditionally set to act on the phases if $t < 0.20 $, after which they are switched off. We define the Weissenberg number as $\Wi = \frac{\lambda }{\tau_\text{f}}$, in which $\tau_\text{f}$ is the characteristic time scale of the fluid. The Weissenberg number measures the relaxation of the fluid to the transient processes in the fluid. For large Weissenberg numbers $\Wi \gg 1.0$, the fluid maintains substantial elastic energies and generates large elastic forces over time. As done by Thomases and Shelley \cite{thomases2007}, we choose the time scale of the fluid to be set by the dimensional scaling of the force $\tau_\text{f} = \frac{\mu_\text{n}^\text{v} + \mu_\text{n}^\text{p}}{L F}$ in which $F = 8\pi^2$ and $L = 1$ is the length of the computational domain. This gives a Weissenberg number of approximately $\Wi \approx 80$.

 \begin{table}[tb]
 \centering
     \caption{Parameter values for the viscoelastic fluid model simulation on an AMR grid subject to four roll forcing.}
     \begin{tabular}{|c c c|} 
     \hline
         \textbf{Parameter} & \textbf{Symbol}& \textbf{Value}\\ \hline
          Network viscosity      &    $\mun$              &    2.0      \\ 
          Solvent viscosity      &    $\mus$              &    0.04     \\ 
          Polymeric contribution to viscosity & $\munp$   &    2.0      \\
          Polymer relaxation time &   $\lambda$           &    4.0      \\
          Drag coefficient        &    $\xi$              &    250 $\mus$   \\ 
          Density                &    $\rho$              &    1.0      \\
          Gradient Threshold     &    $\varepsilon_\ell$  & 0.2, 0.8, 1.75 \\
     \hline
     \end{tabular}
     \label{tab:cf_frm_amr}
 \end{table}

\begin{figure}[htb]
    \centering
    % NOTE: trim={<left> <lower> <right> <upper>}
    \subfloat[$t=0.10$, $\size{\vun}^{\rm{max}} = 0.72$.]{\epsfig{file=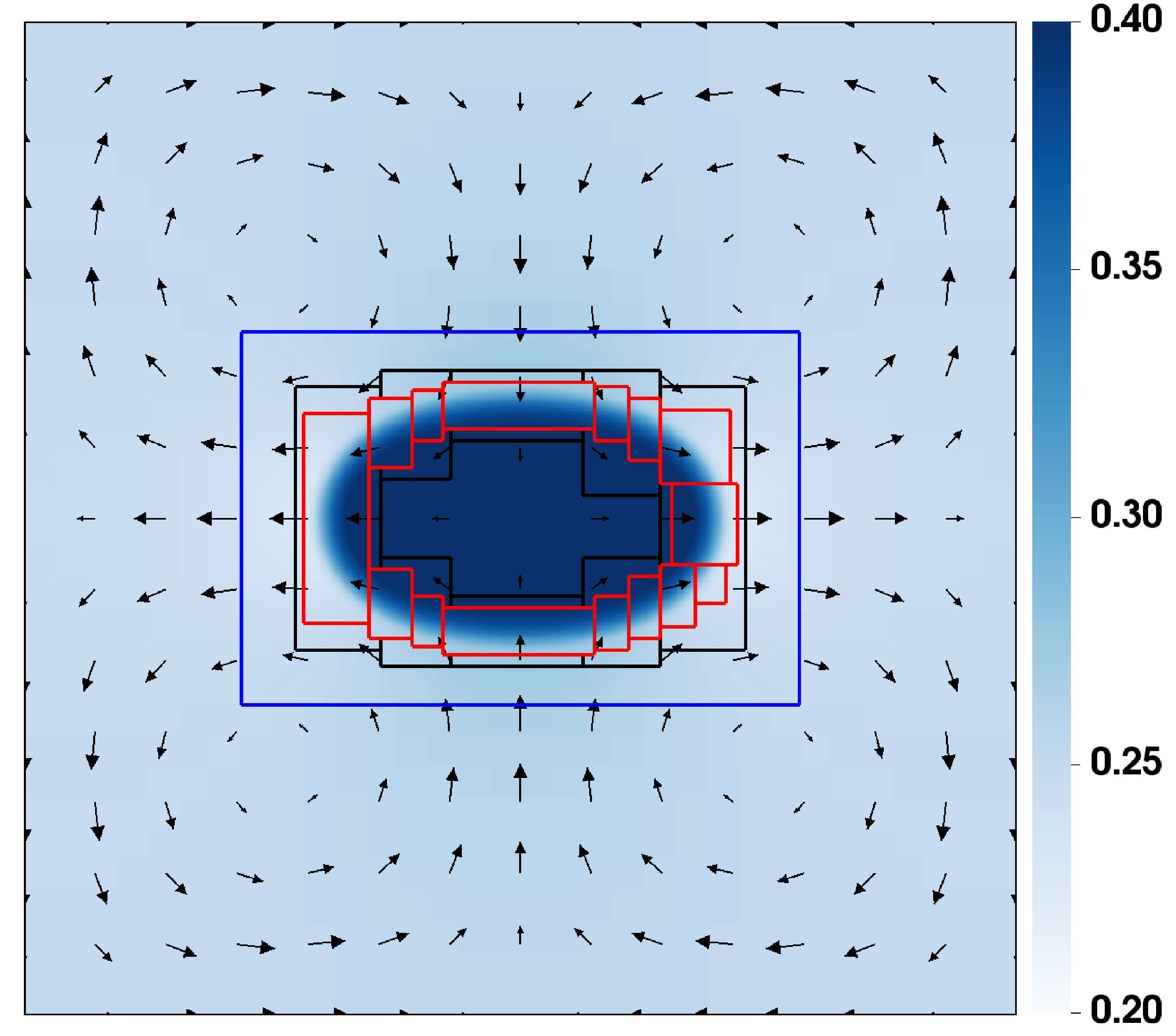, width=0.33\textwidth}}~
    \subfloat[$t=0.40$, $\size{\vun}^{\rm{max}} = 0.04$.]{\epsfig{file=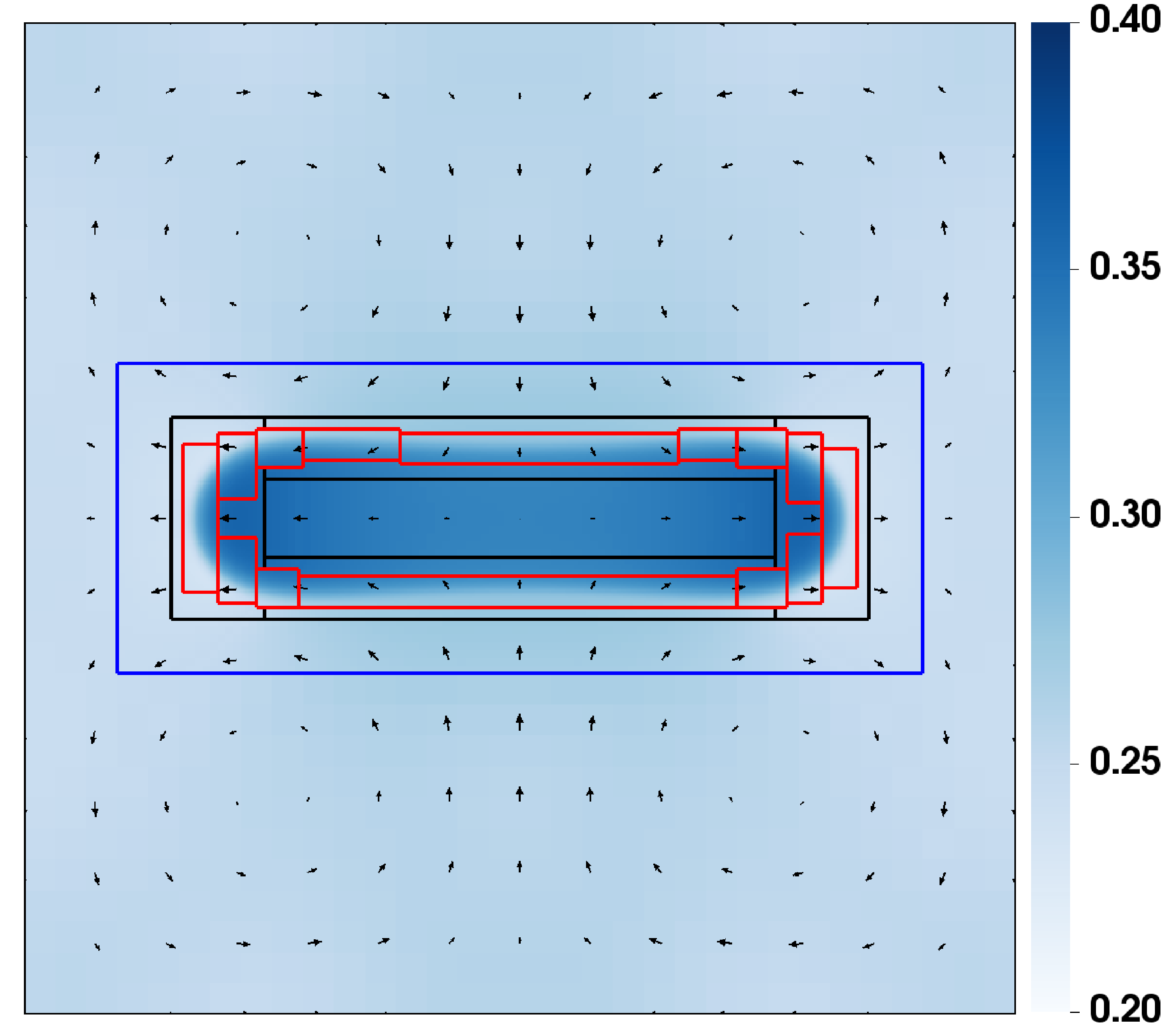, width=0.33\textwidth}}~
    \subfloat[$t=0.80$, $\size{\vun}^{\rm{max}} = 0.08$.]{\epsfig{file=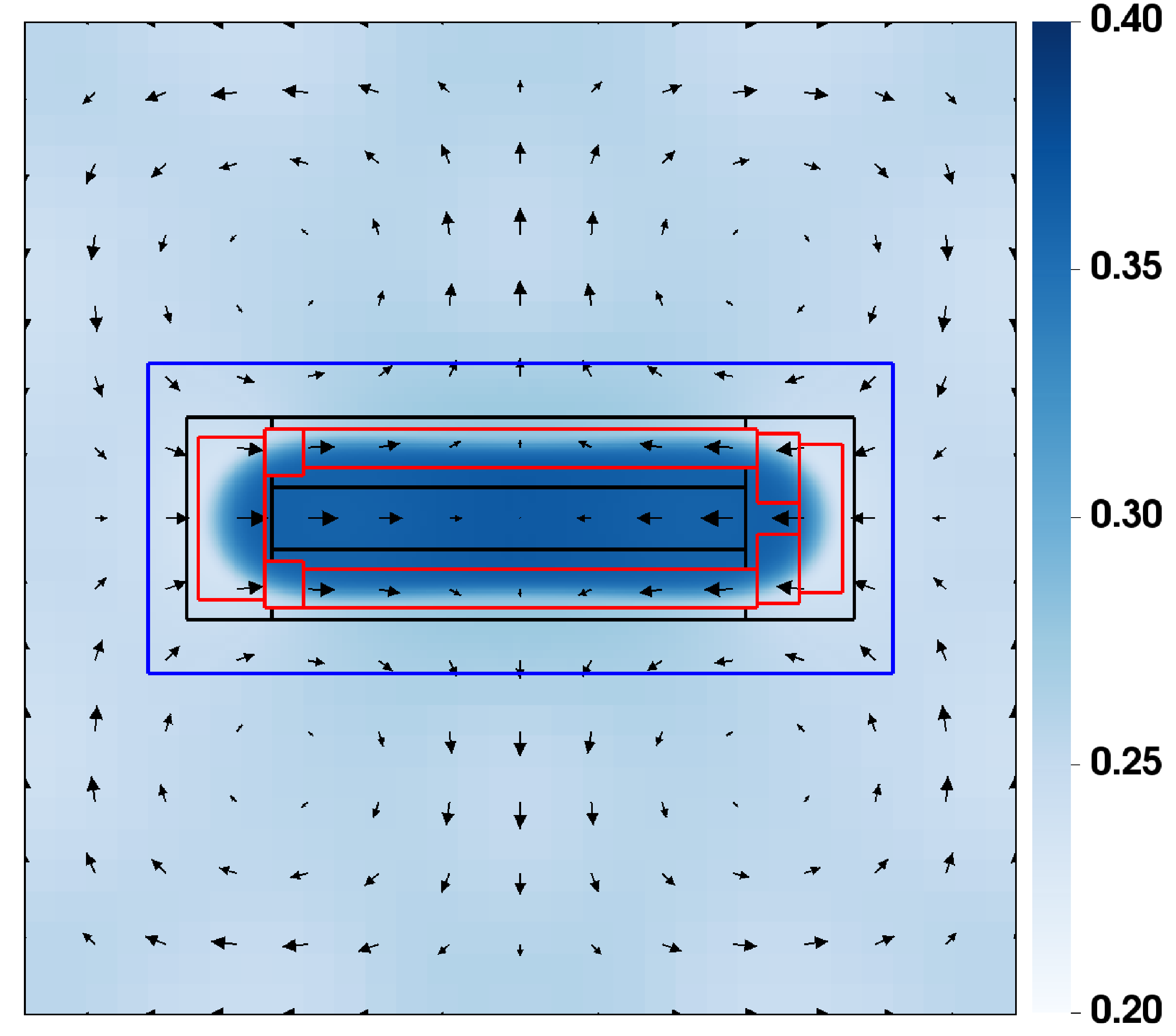, width=0.33\textwidth}}\\
    \subfloat[$t=1.10$, $\size{\vun}^{\rm{max}} = 0.06$.]{\epsfig{file=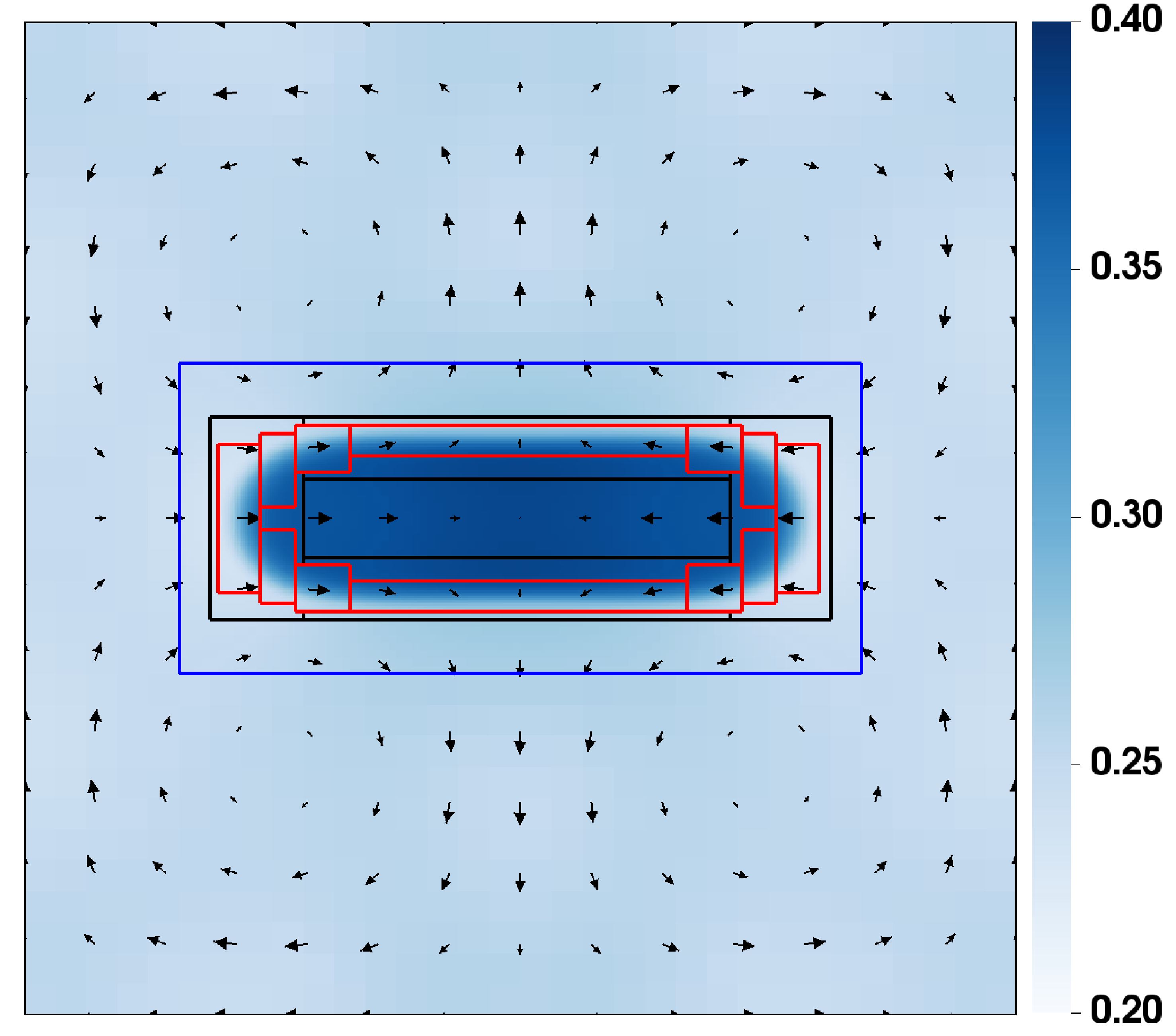, width=0.33\textwidth}} ~
    \subfloat[$t=1.50$, $\size{\vun}^{\rm{max}} = 0.04$.]{\epsfig{file=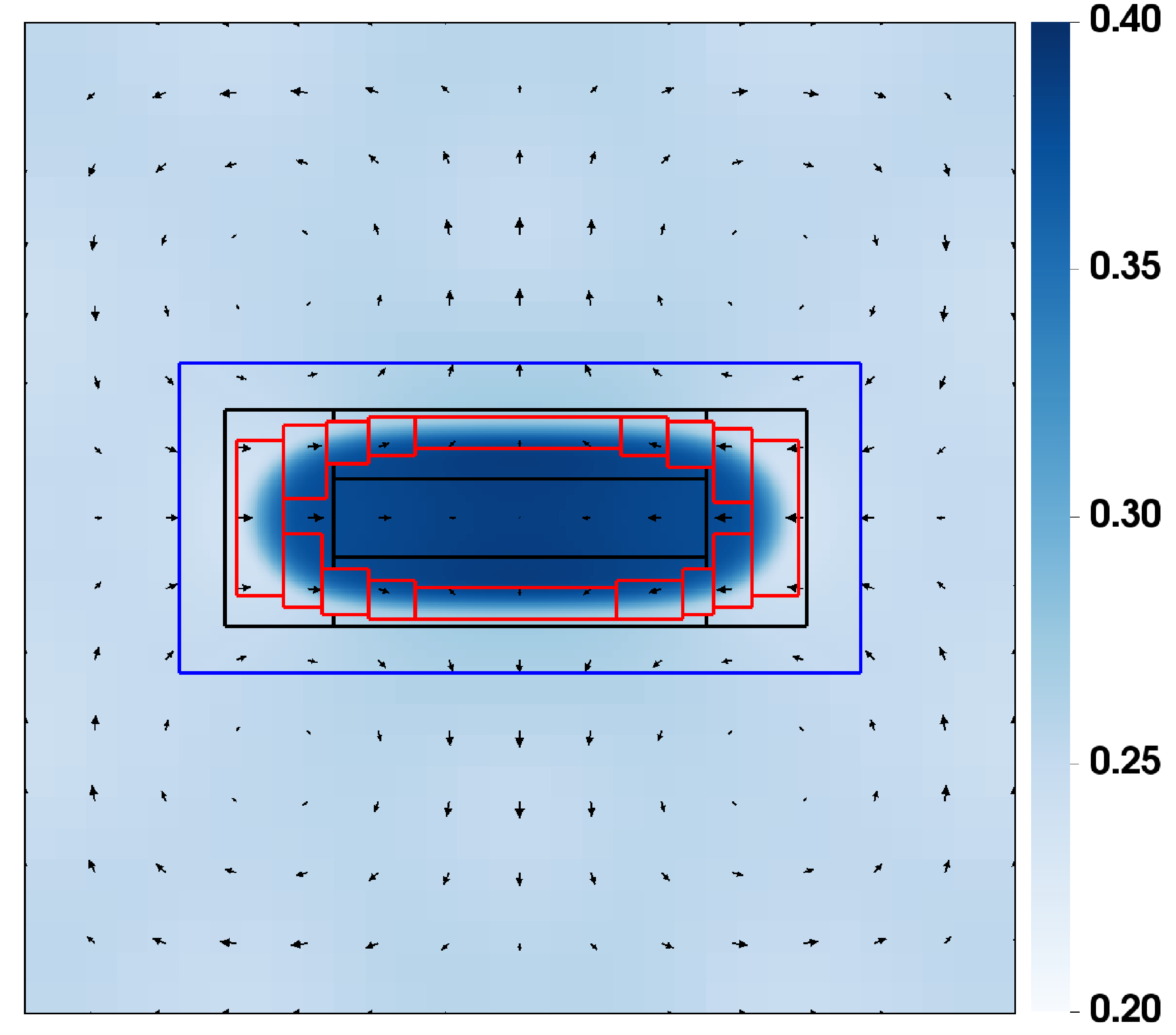, width=0.33\textwidth}}~
    \subfloat[$t=2.00$, $\size{\vun}^{\rm{max}} = 0.03$.]{\epsfig{file=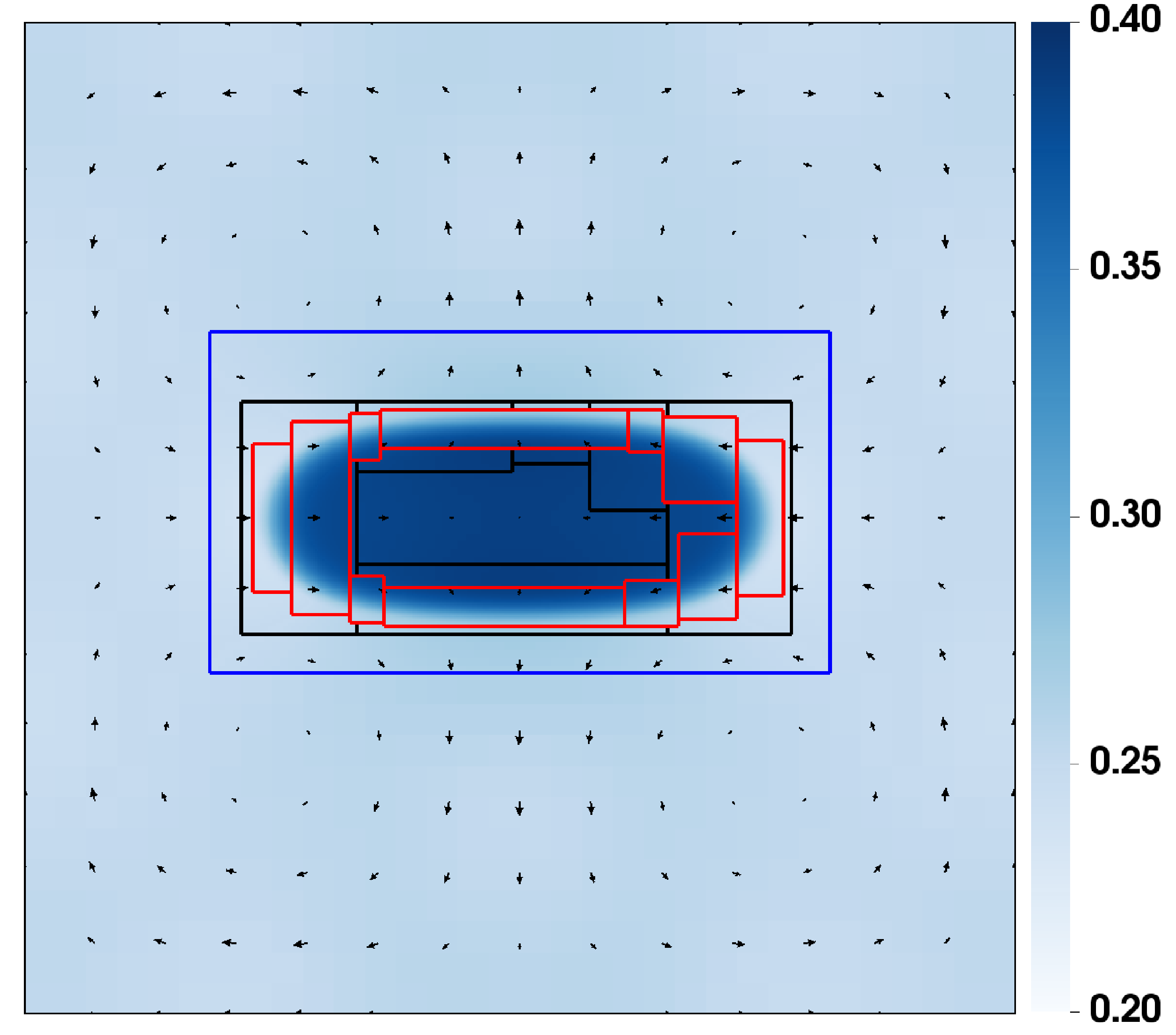, width=0.33\textwidth}} \\
    \caption{Simulation results for a blob of viscoelastic network immersed in a viscous solvent bath, subject to four roll mill forcing. The distribution of $\thn\parens{\xx,t}$ and the network velocity, $\vun\parens{\xx,t}$, are plotted at various times on an AMR grid with a coarsest level of size $32 \times 32$ and 3 additional nested levels of refinement, as indicated by the blue, black and red lines, with refinement ratios of 4, 2, and 2, respectively, between levels. Vectors at different times do not have the same scale.}
    \label{fig:cf_frm_amr_thn}
\end{figure}

The distribution of the network volume fraction $\thn\parens{\xx,t}$ along with the network velocity field $\vun\parens{\xx,t}$ is plotted in \Cref{fig:cf_frm_amr_thn} at several different times. From $t=0$ to $t=0.20$, the network blob experiences forces that tend to pull it apart horizontally, as well as forces that cause compression in the vertical direction. When the background forcing is turned off at $t=0.20$, the network blob continues to stretch horizontally because of the presence of inertia in the system. At around $t = 0.4$, the stretching subsides and the blob begins returning to its initial state because of the elasticity of the fluid.
%The rate of relaxation of the elastic stresses is characterized by the relaxation time, $\lambda$. Because of the elasticity of the fluid, we observe the blob starting to relax towards it's initial state.
The flow dynamics are now dominated by the relaxation of the elastic stresses and the network velocity reverses direction. As the blob of network shrinks, the solvent (not pictured) is correspondingly pushed away from the center of the blob. Each level in the patch hierarchy, from $\Omega^0$ to $\Omega^3$, adaptively refines itself by the prescribed refinement ratio, $r$, whenever the absolute gradient of the network volume fraction exceeds 0.2, 0.8 and 1.75, respectively. The mesh is able to maintain a high resolution near the edge of the blob where the network volume fraction has a steep gradient, while retaining a coarse grid in areas away from the network blob, as indicated by the refinement levels in \Cref{fig:cf_frm_amr_thn}. 

\begin{figure}[ht!]
    \centering
    \epsfig{file=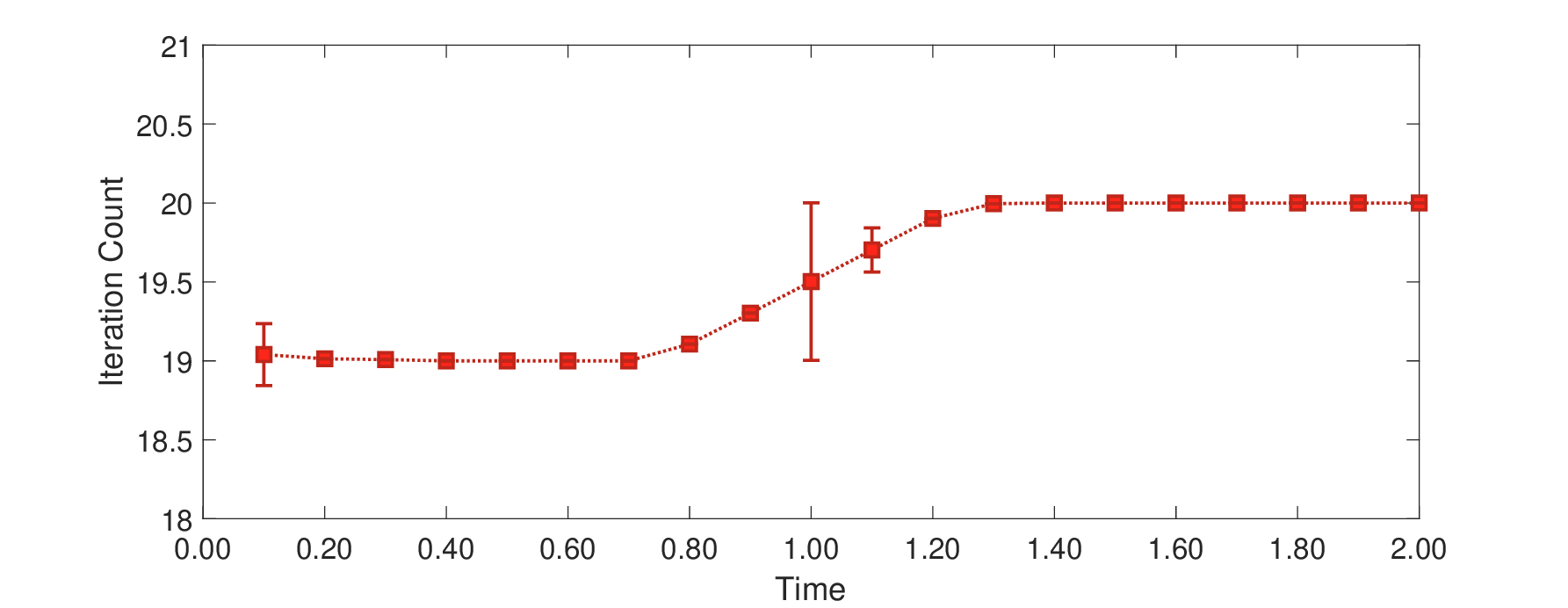, width=0.7\textwidth}
    \caption{A plot of the rolling average (computed every $500\Delta t$) of the number of iterations as a function of time needed to converge to a relative residual tolerance of $10^{-8}$ for the blob of viscoelastic network immersed in a viscous solvent bath under four roll mill forcing simulated using an AMR grid.  Since the iteration count is confined to a narrow range of 19 to 20, there is clear evidence that linear solver convergence is consistent across the whole simulation regardless of the shape and size of the AMR grid.}
    \label{fig:num_iter_cf_frm}
\end{figure}

The number of iterations needed to achieve convergence at each time step were computed and the results are plotted in \Cref{fig:num_iter_cf_frm}. We use a relative residual tolerance of $10^{-8}$ as our convergence criteria and compute a rolling average of the number of iterations needed to converge. The iteration count remains confined between 19 and 20, which indicates that the linear solver is robust and efficient for simulating a complex fluid model with adaptive mesh refinement. The moving average increases slightly after $t = 0.6$ which is consistent with when the blob of network begins to shrink towards its initial state and the network velocity field reverses direction.

\subsection{Comparing Computational Cost between AMR and Uniform Grids}

\begin{figure}[ht!]
\begin{center}
    \subfloat[Newtonian simulation.]{\epsfig{file=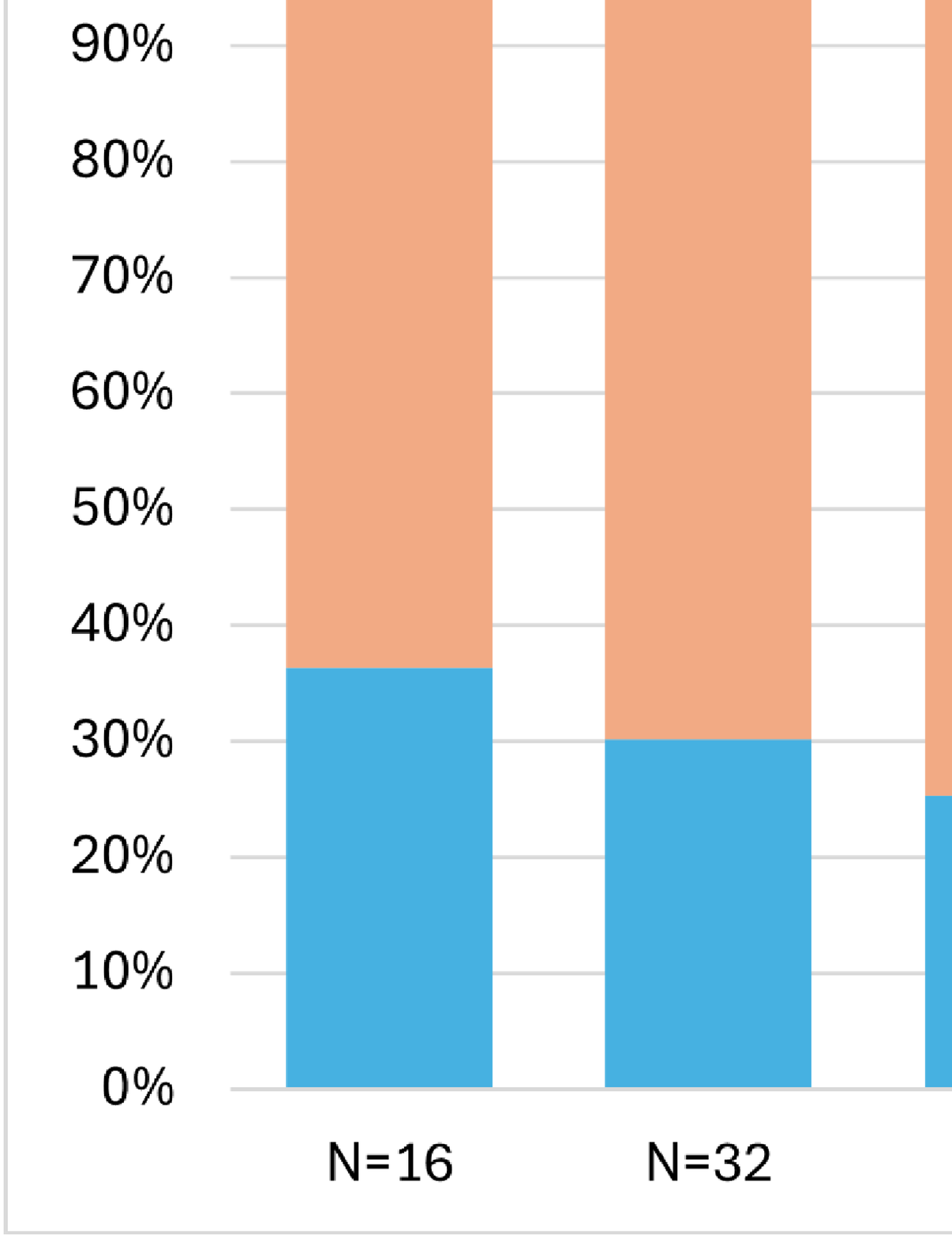, width=0.4\textwidth}} ~
    \subfloat[Viscoelastic simulation.]{\epsfig{file=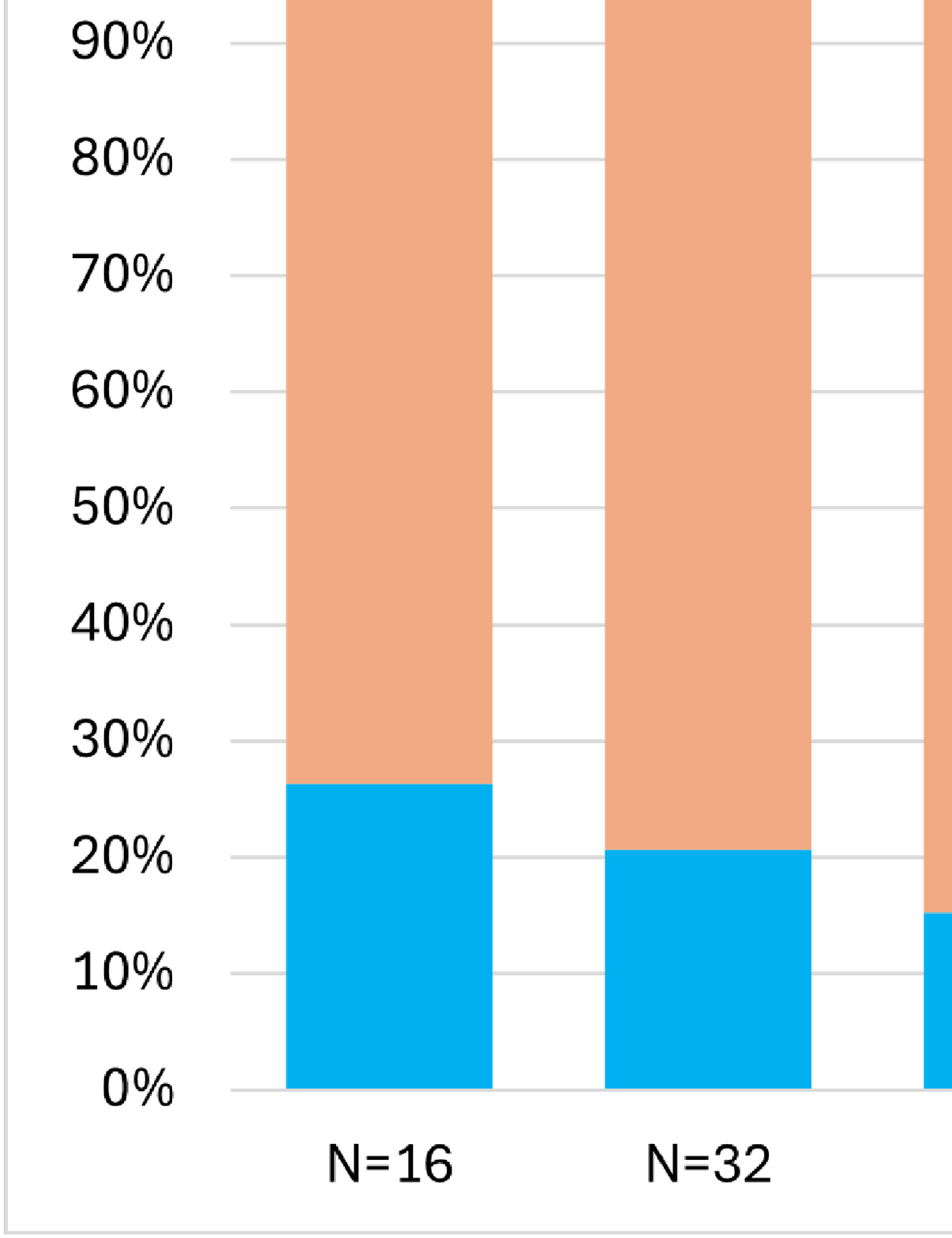, width=0.4\textwidth}} 
\end{center}
    \caption{Plots showing the time taken to complete four roll mill simulations using an AMR grid as a percentage of the time taken using a uniform grid with (a) a Newtonian network and (b) a viscoelastic network. The AMR grid in both (a) and (b) uses a total of 3 adaptive refinement levels. For the Newtonian network on the AMR grid,  a refinement ratio of $r^1=r^2=2$ is used between consecutive levels. For the viscoelastic network on the AMR grid, refinement ratios of $r^1=2$ and $r^2=4$ are used between levels in the AMR hierarchy. The value of $N$ indicates the size of the coarsest level of the AMR grid. The size of the uniform grid is determined using the size of the finest level of the AMR grid. It is observed that AMR offers major cost savings, yielding up to 10x speedup in our experiments, whereby greater speedup is observed as the finest spatial resolution is increased.}
    \label{fig:timing_results}
\end{figure}

To demonstrate the significant savings in computational costs, we perform timing experiments using the method on both uniform and adaptively refined grids. We perform four roll mill simulations for both the Newtonian network and the viscoelastic network from $t=0$ to $t=1$ on different grids whereby the size, $N$, of the coarsest level is increased, but the number of patch levels and refinement ratios are held fixed for each simulation. We use a total of 3 adaptive refinement levels, $\Omega^0$ to $\Omega^2$, for the AMR simulations for both network types. For the Newtonian network on the AMR grid, a refinement ratio of $r^{(1)}=r^{(2)}=2$ is used between consecutive levels. For the viscoelastic network on the AMR grid, refinement ratios of $r^{(1)}=2$ and $r^{(2)}=4$ are used between levels in the AMR hierarchy. We note that this setup is different from that used in Section \ref{sec:viscoelastic_Network} because of the prohibitive expense of running the corresponding uniform grid simulation. Figure \ref{fig:timing_results} shows the time taken using an AMR grid as a percentage of the time taken using a uniform grid to generate four roll mill simulations with (a) a Newtonian network and (b) a viscoelastic network. The size of the uniform grid covering the computational domain is determined using the size of the finest level of the AMR grid. The results indicate that AMR offers a major cost savings benefit, providing up to 10x speed-up in the numerical experiments presented here, with greater speedup observed as the finest spatial resolution is increased. The speed-up is not limited to the results presented here and increased speed-up can be achieved depending on the problem setup.

Table \ref{tab:cost_breakdown} presents a cost breakdown of the total runtime of the viscoelastic four-roll mill simulation over the first 1000 time steps with both a uniform and adaptive grid. Because we use the same routines to fill ghost cells, restrict and prolong, and interpolate during regridding, the cost of all of these operations is presented as ``ghost filling.'' Therefore, the percentages presented in the table are not mutually exclusive and may overlap across different computational components. The majority of the simulation is spent in the preconditioner, specifically in smoothing the error. While the absolute and relative costs of filling ghost cells increases with the use of adaptive grids, those increases are marginal compared to the cost of the preconditioner. The advective discretization for the viscoelastic stress and volume fraction and the grid generation account for less than 1\% of the simulation cost.

\begin{table}
\begin{center}
    \caption{A breakdown of the total runtime of the viscoelastic four-roll mill simulation. An overwhelming majority of the runtime is spent smoothing the error in the preconditioner. Note that the ghost filling operation includes costs of both the ghost filling operations as well as interpolations due to regridding and prolongations and restrictions.} \label{tab:cost_breakdown}
    \begin{tabular}{|c|cc|c|cc|}
        \hline
        \multicolumn{3}{|c|}{Total Run Time} & \multicolumn{3}{|c|}{Krylov Solver} \\
        \hline
        & Uniform & AMR & & Uniform & AMR \\
        \hline
        Ghost Filling & 1.62\% & 11.57\% & GMRES & 1.91\% & 3.85\% \\
        Krylov Solver & 3.63\% & 8.51\% & Krylov Operator & 1.73\% & 4.66\% \\
        Preconditioner & 94.62\% & 89.63\% & Preconditioning: Smoothing & 89.37\% & 72.21\% \\
        & & & Preconditioning: Residual & 2.77\% & 8.00\% \\
        \hline
    \end{tabular}
\end{center}
\end{table}

\section{Conclusion}

Multiphase flows are encountered in a range of diverse settings, ranging from natural phenomena to biomedical applications. In order to study multiphase systems, a widely-used approach is to describe each phase as a continuum in which each point in space can be occupied by a mixture of phases, while additionally specifying the velocity field and stresses of each phase. Depending on the choice of constitutive equation for the Cauchy stress, each phase can be modeled as either a Newtonian or non-Newtonian (complex) fluid. The resulting coupled system of equations pose numerical challenges due to the presence of multiple non-linear terms and a co-incompressibility condition, while the resulting fluid dynamics arising from the model necessitate the development of an adaptive mesh refinement technique to accurately capture regions of interest while keeping computational costs low. 

We have presented an accurate, robust, and efficient computational method for simulating multiphase mixtures on adaptive grids with viscosity ratios in the range of $50-400$. We use second order accurate finite differences to discretize the differential operators, and the resulting saddle point system is solved using GMRES with preconditioning. We adapt a robust geometric multigrid preconditioner with box relaxation smoothing to solve the system on an adaptively refined grid. For single-phase fluids, projection-based methods are simple and effective for solving the momentum equations. However, the co-incompressibility condition leads to additional complications making the details of a projection method unclear. As our numerical studies indicate that the geometric multigrid preconditioner is expensive, future studies will focus on developing efficient projection or block factorization preconditioners, which have yet to be developed for this system. To advance the solution in time, we discretize the viscous stress and drag using an implicit trapezoidal rule and discretize the transport equations using a second order Adams-Bashforth scheme. 

We presented convergence behavior on uniform and refined grids for both prescribed volume fractions and evolved volume fractions. Our results indicate that the discretization achieves second order accuracy in the $L^1$ and $L^2$ norms and approaches second order accuracy in the $L^\infty$ norm. We generated four-roll mill simulations for a Newtonian fluid model and a viscoelastic fluid model that has both viscous and viscoelastic stresses. We demonstrated that the AMR discretization on the Newtonian model asymptotically approaches second order convergence in the $L^1$, $L^2$ and $L^\infty$ norms for all solution variables. For the viscoelastic model subject to four roll mill forcing on a uniform grid, we showed that the discretization obtains second order accuracy for solution components in the $L^1$ and $L^2$ norms and attains at least 1st order accuracy in the $L^\infty$ norm. The final experiment presented a simulation of the viscoelastic fluid model that undergoes deformation up to a certain time after which the forces are turned off, allowing the viscoelastic polymer network to relax to a state of equilibrium. An AMR grid with multiple nested refinement levels was employed for this simulation and we observed that it can sufficiently resolve sharp gradients in the volume fractions. All experiments demonstrated the discretization is stable provided the time step size satisfies the imposed CFL condition. An investigation into the iteration count at every time step showed that the linear iterative solver is robust because its performance remains independent of grid spacing.

\section*{Acknowledgments}
We gratefully acknowledge funding from NIH grants 1R01GM131408, U01HL143336, and R01HL157631 as well as NSF OAC 1931516.
%\bibliographystyle{unsrt}
%\bibliography{references}
\printbibliography

\newpage
\vspace{16 cm}
\section{Appendix}

\subsection{Box relaxation solver}
We define the operators $\mathcal{L}^h[\theta]$, $\mathcal{C}^h$, and $\mathcal{G}^h[\theta]$ that appear in the linear system (see equation \eqref{box_relax_linear_system}) for the box relaxation solver below:

\begin{align}
    \mathcal{L}^h[\theta] =& \\
 &\begin{bmatrix}
	-\frac{\theta_{i,j} + \theta_{i-1,j}}{h^2} -\frac{\theta_{i-\frac{1}{2},j+\frac{1}{2}} + \theta_{i-\frac{1}{2},j-\frac{1}{2}}}{\Delta y^2} & \frac{\theta_{i,j}}{\Delta x^2}  &  -\frac{\theta_{i-\frac{1}{2},j-\frac{1}{2}} -\theta_{i,j}}{\Delta x \Delta y} & \frac{\theta_{i-\frac{1}{2},j+\frac{1}{2}} -\theta_{i,j}}{\Delta x \Delta y} \\ \\
	\frac{\theta_{i,j}}{\Delta x^2} &  -\frac{\theta_{i+1,j} + \theta_{i,j}}{\Delta x^2} - \frac{\theta_{i+\frac{1}{2},j+\frac{1}{2}} + \theta_{i+\frac{1}{2},j-\frac{1}{2}}}{\Delta y^2} &  \frac{\theta_{i+\frac{1}{2},j-\frac{1}{2}} + \theta_{i,j}}{\Delta x \Delta y} &  -\frac{\theta_{i+\frac{1}{2},j+\frac{1}{2}} -\theta_{i,j}}{\Delta x \Delta y} \\ \\
	\frac{\theta_{i-\frac{1}{2},j+\frac{1}{2}}-\theta_{i,j}}{\Delta x \Delta y} &  -\frac{\theta_{i+\frac{1}{2},j+\frac{1}{2}} + \theta_{i,j}}{\Delta x \Delta y} &  \frac{\theta_{i,j}}{\Delta y^2} &  -\frac{\theta_{i,j+1} +\theta_{i,j}}{\Delta y^2} - \frac{\theta_{i+\frac{1}{2},j+\frac{1}{2}} +\theta_{i-\frac{1}{2},j+\frac{1}{2}}}{\Delta x^2} \\	\\
	-\frac{\theta_{i-\frac{1}{2},j-\frac{1}{2}} -\theta_{i,j}}{\Delta x \Delta y}  &  \frac{\theta_{i+\frac{1}{2},j-\frac{1}{2}} -\theta_{i,j}}{\Delta x \Delta y}  &   -\frac{\theta_{i,j} -\theta_{i,j-1}}{\Delta y^2} - \frac{\theta_{i+\frac{1}{2},j-\frac{1}{2}} +\theta_{i-\frac{1}{2},j-\frac{1}{2}}}{\Delta x^2}     &   \frac{\theta_{i,j}}{\Delta y^2}
	\end{bmatrix}, \nonumber \\
 \mathcal{C}^h =& 
 	\begin{bmatrix}
		{\thn}_{i-\frac{1}{2},j}{\ths}_{i-\frac{1}{2},j} & 0 & 0 & 0 \\
		0 & {\thn}_{i+\frac{1}{2},j}{\ths}_{i+\frac{1}{2},j} & 0 & 0 \\
		0 & 0 & {\thn}_{i,j-\frac{1}{2}}{\ths}_{i,jj-\frac{1}{2}} & 0 \\
		0 & 0 & 0 & {\thn}_{i,j+\frac{1}{2}}{\ths}_{i,j+\frac{1}{2}} \\
	\end{bmatrix}, \\
 \mathcal{G}^h[\theta] =& 
 	\begin{bmatrix}
		-\theta_{i-\frac{1}{2},j} & \theta_{i+\frac{1}{2},j} &-\theta_{i,j-\frac{1}{2}} &\theta_{i,j+\frac{1}{2}}
	\end{bmatrix}^\text{T},
\end{align}
and 
\begin{equation}
    b[\theta,\uu] = 
 	\begin{bmatrix}
		-\frac{\theta_{i-1,j} u_{i-\frac{3}{2},j}}{\Delta x^2} - \frac{\theta_{i-\frac{1}{2},j-\frac{1}{2}} u_{i-\frac{1}{2},j+1}}{\Delta y^2} 
		-\frac{\theta_{i-\frac{1}{2},j-\frac{1}{2}} u_{i-\frac{1}{2},j-1}}{\Delta y^2} + v_{i-1,j+\frac{1}{2}}\frac{\theta_{i-\frac{1}{2},j+\frac{1}{2}}-\theta_{i-1,j}}{\Delta x \Delta y} - v_{i-1,j-\frac{1}{2}}\frac{\theta_{i-\frac{1}{2},j-\frac{1}{2}} - \theta_{i-1,j}}{\Delta x \Delta y} - \frac{\theta_{i-\frac{1}{2},j}p_{i-1,j}}{\mu \Delta x}    \\ \\

		-\frac{\theta_{i+1,j} u_{i+\frac{3}{2},j}}{\Delta x^2} - \frac{\theta_{i+\frac{1}{2},j+\frac{1}{2}} u_{i+\frac{1}{2},j+1}}{\Delta y^2} 
		-\frac{\theta_{i+\frac{1}{2},j-\frac{1}{2}} u_{i+\frac{1}{2},j-1}}{\Delta y^2} - v_{i+1,j+\frac{1}{2}}\frac{\theta_{i+\frac{1}{2},j+\frac{1}{2}}-\theta_{i+1,j}}{\Delta x \Delta y} + v_{i+1,j-\frac{1}{2}}\frac{\theta_{i+\frac{1}{2},j-\frac{1}{2}}-\theta_{i+1,j}}{\Delta x \Delta y} +  \frac{\theta_{i+\frac{1}{2},j}p_{i+1,j}}{\mu \Delta x} \\ \\

		-\frac{\theta_{i,j+1} v_{i,j+\frac{3}{2}}}{\Delta y^2} - \frac{\theta_{i+\frac{1}{2},j+\frac{1}{2}} v_{i+1,j+\frac{1}{2}}}{\Delta x^2} 
		-\frac{\theta_{i-\frac{1}{2},j+\frac{1}{2}} v_{i-1,j+\frac{1}{2}}}{\Delta x^2} - u_{i+\frac{1}{2},j+1}\frac{\theta_{i+\frac{1}{2},j+\frac{1}{2}}-\theta_{i,j+1}}{\Delta x \Delta y} + u_{i-\frac{1}{2},j+1}\frac{\theta_{i-\frac{1}{2},j+\frac{1}{2}}-\theta_{i,j+1}}{\Delta x \Delta y} + \frac{\theta_{i,j+\frac{1}{2}}p_{i,j+1}}{\mu \Delta y} \\ \\

		-\frac{\theta_{i,j-1} v_{i,j-\frac{3}{2}}}{\Delta y^2} - \frac{\theta_{i+\frac{1}{2},j-\frac{1}{2}} v_{i+1,j-\frac{1}{2}}}{\Delta x^2} 
		-\frac{\theta_{i-\frac{1}{2},j-\frac{1}{2}} v_{i-1,j-\frac{1}{2}}}{\Delta x^2} + u_{i+\frac{1}{2},j-1}\frac{\theta_{i+\frac{1}{2},j-\frac{1}{2}}-\theta_{i,j-1}}{\Delta x \Delta y} - u_{i-\frac{1}{2},j-1}\frac{\theta_{i-\frac{1}{2},j-\frac{1}{2}}-\theta_{i,j-1}}{\Delta x \Delta y} -  \frac{\theta_{i,j-\frac{1}{2}}p_{i,j-1}}{\mu \Delta y} \\ \\
	\end{bmatrix}.
\end{equation}

\end{document}